\pdfoutput=1
\documentclass[PAPER, cernpreprint,atlasdraft=false, USenglish, texmf, orcidlogo]{atlasdoc}
\usepackage{atlaspackage}
\usepackage{atlasbiblatex}

\usepackage{atlasphysics}
\usepackage{atlasheavyion}
\usepackage{bm}
\usepackage{placeins}
\usepackage{afterpage}
\usepackage{xcolor}
\usepackage{comment}
\usepackage{enumitem}
\graphicspath{{logos/}{figures/}}

\usepackage{ANA-HION-2025-04-PAPER-defs}

\AtlasTitle{Measurements of charged-particle
pseudorapidity and transverse momentum distributions
in O+O and Ne+Ne collisions at $\sqrt{s_{_\text{NN}}} = 5.36$ TeV with the ATLAS detector}

\AtlasAbstract{%
Measurements of charged-particle transverse
momentum spectra, multiplicity, and mean transverse momentum
are presented as a function of pseudorapidity and collision centrality in
O+O and Ne+Ne collisions at $\sqrt{s_{_\text{NN}}}= 5.36$~TeV
using $27.7~\mu\text{b}^{-1}$ and $53.1~\mu\text{b}^{-1}$ data sets recorded by the ATLAS experiment at the LHC.
The collision centrality is characterized by the total
transverse energy measured in the ATLAS forward calorimeters.
The kinematics of charged particles are reconstructed with the inner detector
over the fiducial pseudorapidity and transverse momentum ranges $|\eta|<2.5$
and $0.27 < p_{\text{T}} < 5$~GeV using data from the
ATLAS inner detector.
The per-event charged-particle pseudorapidity density $dn/d\eta$
and mean transverse momentum $\langle p_{\text{T}}\rangle$
are measured over this fiducial range as a function of $\eta$.
The results are reported in 5\%-wide centrality intervals
covering the 5--80\% centrality range, and in 1\%-wide intervals
covering the 0--5\% centrality range.
Invariant per-event yields are evaluated as a function of $\eta$
and $p_{\text{T}}$.
Their $p_{\text{T}}$ dependence is fitted to estimate extrapolated
$dn/d\eta$ and $\langle p_{\text{T}}\rangle$
values over $0 < p_{\text{T}} < 5$~GeV. To evaluate the impact of
using pseudorapidity instead of rapidity, measurements are also
performed as a function of rapidity computed using a pion mass hypothesis.
The fiducial and extrapolated results are compared with
hydrodynamic calculations.
Ratios of the $p_{\text{T}}$-differential yields between centrality
intervals, and between O+O and Ne+Ne collisions, are used to assess
the centrality- and system-dependence of radial flow developed in the collisions.
}

\AtlasRefCode{HION-2025-04}

\PreprintIdNumber{CERN-EP-2026-154}

\AtlasJournal{Phys.\ Rev.\ C.}
\hypersetup{pdftitle={ATLAS document},pdfauthor={The ATLAS Collaboration}}

\begin{document}

\maketitle

\section{Introduction}

Heavy-ion collisions at the Large Hadron Collider (LHC) are understood
to create highly out-of-equilibrium strongly interacting matter at
energy densities well in excess of $10~\text{GeV/fm}^3$
(see Ref.~\cite{Busza:2018rrf}).
This matter subsequently evolves within ${\lesssim}1$~fm/$c$~\cite{Schlichting:2019abc}
into a ``hydrodynamized'', although not yet completely thermalized,
quark--gluon plasma (QGP)~\cite{Baier:2000sb,Berges:2020fwq}.
Viscous hydrodynamic calculations~\cite{Heinz:2024jwu,Schenke:2010nt}
have been very successful in describing
a variety of soft single- and multi-hadron measurements at the
Relativistic Heavy Ion Collider (RHIC) and the LHC, but only if the
ratio of shear viscosity to entropy density ($\eta/s$)
in units of $\hbar$ is $\sim$0.1--0.2, close to the conjectured
lower bound on $\eta/s$~\cite{Kovtun:2004de,Romatschke:2007mq,Teaney:2009qa,Romatschke:2007mq}.
Such small values require the mean free path of the microscopic
constituents of the plasma
to be much smaller than the transverse size of the plasma which is
determined by the size of the colliding nuclei.

Extensive theoretical advances in the last two decades (see
e.g.\ Refs.~\cite{Schlichting:2019abc,Kurkela:2018vqr})
have provided a sound footing for understanding the rapid
transition from the out-of-equilibrium initial state to the
hydrodynamically evolving plasma.
The results of these advances also provide a rigorous connection between
certain final-state observables and the initial state.
In particular, the density per unit rapidity and per unit transverse area of
charged particles produced in a heavy-ion collision is found to be
closely related to the initial energy density and to the entropy
density in the plasma at the hydrodynamization time~\cite{Giacalone:2019ldn}.
Such a connection was originally proposed by Bjorken in the context of
a specific hydrodynamic model~\cite{Bjorken:1982qr}, but it is found to apply even for
the strongly out-of-equilibrium initial conditions of ultrarelativistic
heavy-ion collisions~\cite{Berges:2013fga,Giacalone:2019ldn,Strickland:2017kux}.
In addition, the mean transverse momentum of the produced particles
is sensitive to the collective transverse velocity or ``radial flow''
developed during the hydrodynamic expansion~\cite{PhysRevC.93.044908,Gardim:2019brr}.

A topic of great interest in the study of the QGP is whether there exists
some transverse size scale below which the above-described paradigm
for the creation, hydrodynamization, and subsequent evolution of
the QGP becomes invalid.
This question is made particularly urgent by the appearance of trends in
single- and multi-hadron data obtained from proton--proton~\cite{HION-2015-09,CMS-HIN-16-010,ALICE:2018pal,} and
proton--nucleus collisions~\cite{HION-2012-13,ALICE:2012eyl,ALICE:2016fzo,ALICE:2013wgn,} that are nearly identical to those observed
in heavy-ion collisions~\cite{Nagle:2018nvi, Weller:2017tsr}, suggesting that QGP can be created with a transverse
size comparable to the size of a proton, i.e.\ comparable to the QCD
confinement distance scale.
Systematic studies of multiple observables in collisions of nuclei of
different size -- since the nuclear size also determines the maximum transverse
size of the plasma -- may provide important insight into whether the paradigm
described above for the creation and evolution of the QGP evolves
smoothly with transverse size down to the size of a proton.
Considerations such as this have motivated interest in studying collisions
of nuclei much smaller than lead nuclei at the LHC~\cite{Citron:2018lsq}.

In July of 2025, the LHC carried out a long-anticipated set of special
runs that included \OO and \NeNe collisions at a center-of-mass energy
of 5.36~\TeV per nucleon pair.
The inclusion of \NeNe\ data-taking in the special runs arose from recent
strong interest in the heavy-ion community in studying the impact of the
shape of the colliding nuclei on the initial state, the subsequent
hydrodynamic evolution of the QGP, and on observables sensitive to
that evolution~\cite{Giacalone:2019pca,Giacalone:2020awm,Jia:2022ozr}.
The \Ox and \Ne nuclei have similar sizes, but are expected to have different
shapes, due to the role of alpha clustering~\cite{Frosini:2021sxj} in
the ground-state wave functions.
In particular, the \Ne nucleus is thought to acquire a ``bowling pin'' shape
due to the presence of an additional alpha cluster compared to the
\Ox nucleus.
Hydrodynamic calculations~\cite{Giacalone:2024luz} that start from
\emph{ab initio} calculations of the nuclear
geometry~\cite{Frosini:2021fjf,Frosini:2021sxj,Frosini:2021ddm,Lee:2025req} and
use the Trento model~\cite{Moreland:2014oya} to describe the initial
particle production and the energy-density distribution at
the onset of hydrodynamic evolution, predict measurable
effects from nuclear-shape differences.
Observables such as the
mean transverse momentum (\meanpt) and the Fourier coefficients
of the azimuthal harmonics~\cite{Voloshin:1994mz} (\vn), especially \vt, show sensitivity
to the different geometries of the \Ox\ and \Ne\ nuclei.
Measurements of the azimuthal Fourier coefficients of charged particles
produced in \OO and \NeNe collisions~\cite{HION-2025-02,CMS-HIN-25-009,ALICE:2025luc}
have yielded results consistent with predictions arising from the
hydrodynamic calculations.
However, tests of other predictions from the model, including the charged-particle
pseudorapidity density and \meanpt, are needed to verify both the initial
conditions of the calculations and their description of the radial
flow developed during the evolution of the plasma.

This \paper presents measurements of charged-particle pseudorapidity
density, transverse momentum, and mean transverse
momentum in \OO\ and \NeNe\ collisions, using data
collected by ATLAS during the aforementioned  special runs of the LHC.
The centrality of the \OO and \NeNe collisions is characterized by
the total transverse energy measured in the ATLAS forward calorimeters,
and results are presented over the centrality range 0--80\% in both collision systems.
Charged particles are reconstructed from hits in the ATLAS inner
detector over the ``fiducial'' pseudorapidity
and transverse momentum intervals, $|\eta| < 2.5$ and $0.27 < \pT <
5$~\GeV, respectively.
The measurements are performed as a function of \peta and
rapidity, where the latter is computed assuming the charged-pion
mass hypothesis and denoted by \ypi.
The use of pion rapidity provides results that are closer to those obtained using
true rapidity, which is the natural variable for describing
particle production in hadronic interactions but is not accessible
to ATLAS due to its limited ability to identify hadrons.
Results are presented for \dndeta,\footnote{Throughout
this \paper, when discussing differential distributions of particle yields
(versus \pT\ or \peta) the lower case ``\textit{n}'' denotes yield per event.
Thus, e.g.
$\dif n/\dif\eta \equiv (1/N_{\text{evt}}) \dif N/\dif\eta$.
}
\dndypi, and \meanpt evaluated as a function of \peta
and \ypi. Results are also presented for the double-differential
distributions, \invpTdndpTdeta and \invpTdndpTdypi in
different \peta or \ypi intervals.
Results for \dndeta and \meanpt are obtained both in the fiducial kinematic
range and with extrapolation to \pteqo\ using fits to the \pT
dependence of  \invpTdndpTdeta and \invpTdndpTdypi.
The measured \pT spectra are  compared
between different centrality intervals within the same collision
system and between \OO\ and \NeNe collisions.  The results of the
presented measurements are compared with predictions from hydrodynamic
calculations.

The remainder of this paper is structured as follows:
Section~\ref{sec:detector} describes the ATLAS experiment;
Section~\ref{sec:datasets} describes the data sets used in the
analysis and the object reconstruction; Section~\ref{sec:analysis}
presents the analysis methods; Section~\ref{sec:systematics}
discusses systematic uncertainties; Section~\ref{sec:result} presents
a subset of the results; and Section~\ref{sec:conclusion} summarizes the paper
and discusses conclusions drawn from the measurements. The appendix
presents more results that could not be included in
Section~\ref{sec:result} without detracting from its readability.


%

%
\vspace{-0.2in}
\section{ATLAS detector}
\vspace{-0.2in}
\label{sec:detector}

The ATLAS experiment~\cite{PERF-2007-01,GENR-2019-02} at the LHC comprises
a nearly hermetic central detector surrounding the collision
point\footnote{ATLAS uses a right-handed coordinate system with
its origin at the nominal interaction point (IP) in the center of
the detector and the \(z\)-axis along the beam pipe.
The \(x\)-axis points from the IP to the center of the LHC ring,
and the \(y\)-axis points upwards.
Polar coordinates \((r,\phi)\) are used in the transverse plane,
\(\phi\) being the azimuthal angle around the \(z\)-axis.
The pseudorapidity is defined in terms of the polar angle \(\theta\)
as \(\eta = -\ln \tan(\theta/2)\) and is equal to the rapidity
$ y = (1/2) \ln[(E + p_z )/(E - p_z)] $
in the relativistic limit.
Angular distance is measured in units of \(\Delta R \equiv \sqrt{(\Delta y)^{2} + (\Delta\phi)^{2}}\).
}
and special-purpose forward detectors including the zero-degree
calorimeters (ZDCs) that are described below.
The central detector consists of an inner tracking detector surrounded
by a thin superconducting solenoid, electromagnetic and hadronic
calorimeters covering the region \(|\eta|< 3.2\) and forward
calorimeters (FCals) covering \(3.2 < |\eta|< 4.9\), and a muon spectrometer
incorporating three large superconducting air-core toroidal magnets.

The inner-detector system (ID) is immersed in a \qty{2}{\tesla} axial
magnetic field and provides charged-particle tracking in the range
\(|\eta| < 2.5\).
A high-granularity silicon pixel detector covers the vertex region
and typically provides four measurements per track,
the first hit generally being in the insertable B-layer (IBL)
installed before Run~2~\cite{ATLAS-TDR-19,PIX-2018-001}.
It is surrounded by a silicon microstrip tracker (the SCT), which usually
provides eight measurements per track.
These silicon detectors are complemented by a transition radiation
tracker (TRT), which enables radially extended track reconstruction
up to \(|\eta| = 2.0\).

The FCals~\cite{Artamonov_2008} measure transverse energy in the
forward region and are used to characterize the centrality of
heavy-ion collisions.
Each FCal consists of three modules longitudinal in shower depth, one electromagnetic
and two hadronic, with dense absorber  matrices and narrow
liquid-argon gaps, optimized to measure the large particle flux
in the forward region.

Two ZDCs~\cite{Avoni:2025lmu} are located at
$\Delta z = \pm 140$~m from the center of ATLAS,
within the LHC ``Target Absorbers for Neutrals'' (TANs) shielding components.
The ZDCs measure neutral particles produced with $|\eta| \gtrsim 8.4$.
In heavy-ion collisions, these are primarily spectator neutrons that
did not participate in the collision and were emitted at very
forward rapidities.

The luminosity is measured mainly by the LUCID-2~\cite{LUCID2} detector,
which records Cherenkov light produced in the quartz windows of
photomultipliers located close to the beam pipe.

ATLAS uses a two-level trigger system~\cite{TRIG-2022-01} consisting of
a Level-1 (L1) trigger that is implemented using dedicated electronics
and a software-based ``high-level trigger'' (HLT) running algorithms
similar to those used in offline analysis.
The Athena software suite~\cite{ATL-SOFT-PUB-2026-001} is used in data simulation,
in the reconstruction and analysis of real and simulated data,
in detector operations, and in the trigger and data acquisition
systems of the experiment.


%

\vspace{-0.2in}
\section{Data sets and object reconstruction}
\label{sec:datasets}

The \OO data used in this analysis were collected during a dedicated period of
low-pileup operation, where the mean number of interactions
per bunch crossing was $\langle\mu\rangle \sim0.01$.
The \NeNe data were collected in a single LHC fill with
$\langle\mu\rangle$ varying between 0.04 and 0.16.
Minimum-bias events were selected using an L1 trigger based on the
TRT ``FastOR'' algorithm~\cite{ATL-INDET-PUB-2009-002},
together with an HLT requirement of at least one reconstructed
charged-particle track with $\pt > 200$~\MeV.
The events used in the analysis were recorded during stable LHC
running conditions and have no detector hardware or
readout errors~\cite{DAPR-2018-01}.
The integrated luminosities of \OO\ and \NeNe\ collisions sampled
by the TRT trigger and passing event-cleaning requirements (described
below) are $27.7~\mu\text{b}^{-1}$ and $53.1~\mu\text{b}^{-1}$, respectively.

Charged-particle tracks and collision vertices are reconstructed from
hits in the ID using standard methods~\cite{PERF-2015-08,ATL-PHYS-PUB-2015-026,ATL-PHYS-PUB-2019-015,IDTR-2021-01}.
For the nominal analysis, the reconstructed tracks are required to
have $|\eta| < 2.5$ and at least one pixel hit,
with an additional requirement of a hit in the IBL or a hit in the
next-to-innermost pixel layer when one is expected.\footnote{
A hit is expected if the extrapolated track crosses an active
region of a silicon-sensor module (pixel or SCT) that has not
been disabled.
A hit is said to be ``missing'' when it is expected but not found.
A missing hit is called a hole.
The hole counting does not consider layers before the first hit
or after the last hit.
If a track crosses a disabled module, then for the
purposes of hit counting, the disabled module is counted as a hit.
If a hit is used in reconstructing more than one track trajectory,
it is called a ``shared hit''.
}
The tracks are required to have at least six SCT hits.
To suppress secondary-track contributions, the transverse impact parameter of the track with
respect to the beamline, $d_0$, and the longitudinal impact parameter
of the track relative to the primary vertex~\cite{ATL-PHYS-PUB-2015-026,ATL-PHYS-PUB-2019-015},
$z_0\sin(\theta)$, are required to satisfy
$|d_0| < 1.5$~mm and $|z_0\sin(\theta)| < 1.5$~mm.
An alternative set of track selections is used to evaluate
systematic uncertainties in the measurement.
Like the nominal selection, the alternative selection requires
$|\eta| < 2.5$, $|d_0|<1.5$~mm, and $|z_0\sin(\theta)| < 1.5$~mm,
but its hit requirements are markedly different.
The track is required to have at least eight silicon-detector hits and at most
two silicon-detector holes, with at most one hole in the pixel detector,
at most one shared pixel hit or at most two shared SCT hits.
If a track has shared hits, then the number of required silicon-detector hits
is increased to ten.

The $z$-position of the primary vertex~\cite{ATL-PHYS-PUB-2015-026,ATL-PHYS-PUB-2019-015} is required to lie within
10~cm of the center of ATLAS.
In \OO and \NeNe collisions, backgrounds from ultraperipheral collisions (UPCs)
are suppressed relative to \PbPb\ collisions, but they still occur at an observable rate.
The dominant UPC background in \OO and \NeNe collisions is
inclusive photonuclear ($\gamma+A$) scattering,
which produces  asymmetric events with little energy emitted
in the hemisphere of the photon-emitting nucleus.
A selection that requires at least one topocluster~\cite{PERF-2014-07} having $\pt> 400$~\MeV\
within the acceptance of each FCal is used to reject these events.
An additional requirement rejects events in which the
fraction of all reconstructed tracks that satisfy the aforementioned
nominal selections is less than 50\%.
Such events are mostly associated with upstream interactions
and collisions involving ``satellite'' ions.
In the recorded \OO\ and \NeNe\ collision data, there are contributions
from pileup events, in which two or more inelastic collisions
occur in the same bunch crossing.
Most of the pileup events are removed by requiring exactly one
high-quality reconstructed vertex per event.
A high-quality vertex is defined as one that is well-constrained in position, having
a $z$-position variance of less than $0.02~\mathrm{mm}^2$.
Pileup events are further suppressed by exploiting the correlation
between energy recorded in the FCal (\sumET) and the energy
deposited in the ZDCs.
The residual pileup after these selections is estimated to be
negligible.

As in earlier ATLAS studies of heavy-ion collisions, events are categorized
into centrality percentiles based on the
\sumET~\cite{HION-2025-02,HION-2011-01,HION-2011-05,HION-2019-03}.
To relate the \sumET\ distribution to the sampled fraction of the
total inelastic \OO\ and \NeNe\ cross-sections,
a Glauber-model-based calculation~\cite{Miller:2007ri,Loizides:2025ule}
is used to fit the data, to extract the fraction of events selected
above a minimum \sumET\ threshold, and to estimate the systematic
uncertainty of that fraction~\cite{HION-2011-01,HION-2011-05,HION-2019-03}.
Additionally, the Glauber model is used to extract primary collision
characteristics, such as the average number of participating nucleons,
\npart, for each centrality interval.
The \npart\ values and their uncertainties corresponding to the
centrality intervals used here are shown in Table~\ref{tab:npart}
in the Appendix.
The Glauber analysis used in this paper represents an evolution
of that used in the ATLAS measurement of collective flow in
\OO and \NeNe collisions~\cite{HION-2025-02}.
It explicitly accounts for potential $\alpha$-particle clustering in
\Ox and \Ne nuclei by using nucleon density distributions obtained from
the PGCM~\cite{Frosini:2021ddm,Frosini:2021sxj} and
NLEFT~\cite{nleft} initial-state models as implemented in the
Trajectum hydrodynamic simulation code~\cite{Nijs:2021clz} (\texttt{TR\_OXYGEN\_V15}
and \texttt{TR\_NEON\_V14} with $\omega = 0.3$).
The \sumET values corresponding
to boundaries of \OO and \NeNe centrality intervals have changed
relative to Ref.~\cite{HION-2025-02} by
less than 1\% in the most central collisions and as much as
5\% in the most peripheral collisions.

The detector performance is studied using a sample of 50~million
minimum-bias \OO\ Monte Carlo (MC) events, corresponding to inclusive
inelastic collisions without a bias toward a specific hard process,
modeled with the HIJING\,v1.3b event generator~\cite{Wang:1991}.
The generated events were passed through a full simulation of the ATLAS
detector~\cite{SOFT-2010-01} using \GEANT~\cite{Agostinelli:2002hh},
and reconstructed using the same algorithms which are applied to the data.
Three alternative MC samples were produced using geometry variations that
contain additional material in either the IBL, the remainder of the ID, or the
pixel-detector services in the forward region.
These are used to
assess systematic uncertainties in the measurement that result from
the limited accuracy of the detector-material model
used in the simulation.


%
\section{Analysis}
\label{sec:analysis}
\vspace{-0.2in}

The measurements presented in this \paper include:
\begin{enumerate}[label=\roman*]
\item computation of the charged-particle \dndeta (or \dndypi) and $\meanpt(\eta)$
(or $\meanpt(\ypi)$) in the fiducial range of the analysis, i.e.\
$0.27 < \pT< 5$~\GeV\ and $|\peta| < 2.5$ (or $|\ypi| < 2.25$);
\item measurement of the invariant\footnote{
Here ``invariant'' refers to invariance under longitudinal boosts.
Strictly speaking, only the  \invpTdndpTdy %
distribution is invariant, whereas \invpTdndpTdeta
and \invpTdndpTdypi are not.
}
double-differential per-event  charged-particle yields,
\begin{align}
\Yeta \equiv \frac{\dd^2n}{\pT\,\dd \pT\,\dd\eta} \quad \text{and} \quad \Yypi\equiv\frac{\dd^2n}{\pT\,\dd \pT\, \dd\ypi},
\label{eq:observables}
\end{align}

\item extrapolation of the yields in Eq.~\eqref{eq:observables} to $\pt=0~\GeV$ by fitting the invariant
yields to smooth functions that are integrated analytically or numerically,
and estimation of the total or ``extrapolated'' \dndeta (or \dndypi)
and $\meanpt(\eta)$ (or $\meanpt(\ypi))$.
\end{enumerate}
The measurements performed using \ypi provide important physics
results as well as an essential calibration of the distortion of the \pT
spectra that results from the use of pseudorapidity. That distortion
is demonstrated using the HIJING Monte Carlo sample in
Section~\ref{sec:rapidity_bias}, and is  accounted for when performing
the \pteqo extrapolation of the \peta-based measurements using the
ratio of the measured \invpTdndpTdypi and \invpTdndpTdeta
distributions.

\subsection{Fiducial measurements}

In this analysis, primary charged particles are those with a mean lifetime
greater than $0.3\times10^{-10}$~s, produced either directly in the nucleus--nucleus
interaction or in decays of particles with shorter lifetimes~\cite{STDM-2010-06}.
For primary charged particles, the efficiency (\eps) of the track reconstruction procedure
and subsequent selections is determined using the HIJING MC sample.
The same sample is used to determine the misreconstructed-track\footnote{These are tracks
that do not correspond to true charged particles but arise
from combinatorial hits in the detector.}
(``fake-track'') fraction (\ff) and the secondary-track fraction
(\fs) among the selected tracks.
The latter arise mainly from weak decays of hadrons or interactions
with detector material.
The \eps, \ff, and \fs values are obtained after using standard
ATLAS procedures~\cite{PERF-2015-08,IDTR-2021-01} to match reconstructed
tracks to primary charged particles generated by HIJING.
Because of potential sensitivity of the reconstruction to detector occupancy,
the \eps, \ff, and \ff values are determined separately in
intervals of reconstructed charged-particle multiplicity \nchrec
computed over $|\peta| < 2.5$ and
$0.4 < \pt< 5$~\GeV.\footnote{A lower bound is placed on
\pT\ in the multiplicity calculation to match
the definition in previous ATLAS measurements~\cite{HION-2025-02}.}
The use of multiplicity instead of centrality allows the same corrections
to be applied to \OO and \NeNe data and avoids potential complications
in attempting to match HIJING and \OO or \NeNe centralities.

When evaluating \eps, \ff, and \fs, per-particle weights are applied
to MC events to correct for \pT-spectrum differences between data
and simulation and to adjust the kaon and proton fractions of the
total particle yield as a function of \pT\ to match measurements by
ALICE in \XeXe collisions~\cite{ALICE:2021lsv}.
These ``flavor weights'' are applied because the tracking efficiency depends
on the particle-species composition.
The flavor reweighting is performed in a centrality-independent manner,
using the ALICE measurements of the kaon-to-pion and proton-to-pion
ratios in the 50--60\% \XeXe centrality interval to correct the
corresponding HIJING fractions.
The 50--60\% centrality \XeXe\ collisions are chosen because the multiplicity
in this interval roughly corresponds to that in 0--10\% centrality
\OO\ collisions.
To evaluate the systematic uncertainties associated with the flavor reweighting,
alternative weights derived from the 70--90\% and 40--50\% \XeXe centrality
intervals measured by ALICE are also applied.
The reweighting factor depends on \pt\ and is $\sim$1--1.6 for kaons
and $\sim$0.5--1.4 for protons.
To account for higher secondary-track rates due to enhanced
strange hadron production~\cite{ALICE:2021lsv}, a \pT-independent
weight of 1.5, motivated by the maximum reweighting factor for kaons,
is applied to secondary tracks from $K_\text{s}$, $\Lambda$,
and $\Sigma$ decays.

A data-driven procedure is used to verify the fake-track fraction and
improve the secondary-track fraction estimates obtained from the MC simulation.
This procedure takes advantage of the fact that the distributions of
distance of closest approach of a track to the beamline, \dz,
are observed in MC events to have significantly different shapes for
reconstructed primary particles, fake tracks, and secondary tracks,
especially at low \pT\ where the fake-track fractions are largest.
Template fits to measured \dz distributions are performed in three \peta
intervals and in the \pT\ bins used in the analysis.
These yield \ff values within a few tenths of a percent of those obtained
from simulation, but \fs values larger than those obtained
from simulation. The ratios of the template-fit and MC-evaluated
secondary-track rates obtained in the three \peta intervals are then
applied as per-secondary-particle weights in the simulation. These weights
vary from near unity at low \pT to a maximum value between 2 and 2.5 near
$\pT = 2$~\GeV, and then decrease with increasing \pT.
No significant centrality dependence is observed in the template-fit
results, so the applied secondary-track reweighting factors are
the same for all \OO and \NeNe centralities.
Separate weights are obtained for the nominal and alternative track selections.
The systematic uncertainty associated with secondary particles is estimated
from the difference between results obtained with the two selections, each
using its own set of weights.

\begin{figure}[!tb]
\centering
\includegraphics[width=0.95\linewidth]{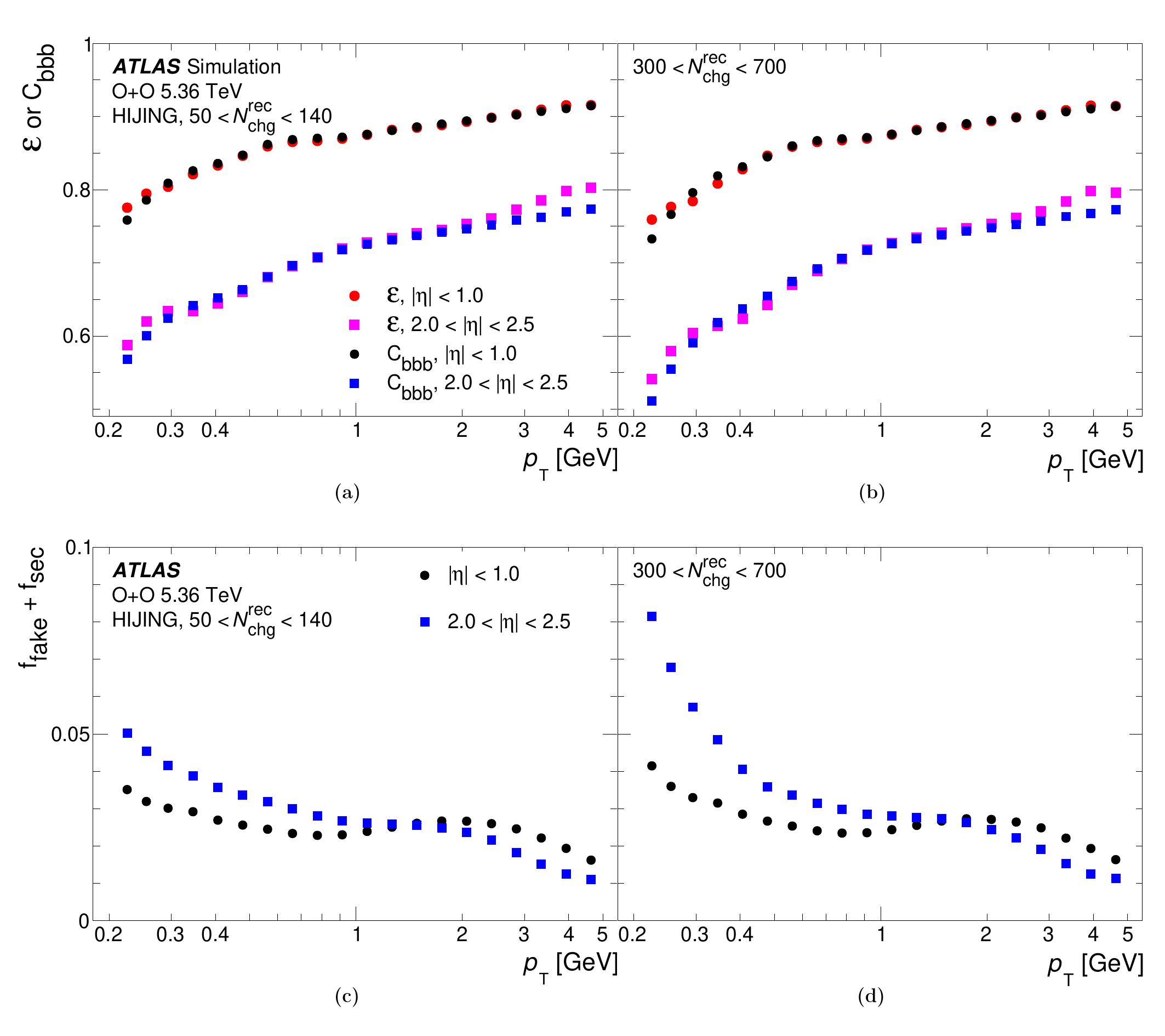}
\caption{
Example (a, b) reconstruction efficiencies, $\eps$, and bin-by-bin
correction factors, $C_{\mathrm{bbb}}$, together with (c, d)
fake- and secondary-track fractions, $\ff + \fs$.
The plots correspond to the nominal track-selection working
point and are shown for multiplicity intervals (a, c)
$50 < \nchrec < 140$ and (b, d) $300 < \nchrec < 700$,
and for pseudorapidity intervals $|\eta|<1.0$ and $2.0<|\eta|<2.5$.
}
\label{fig:corrections}
\end{figure}

In the data analysis, a per-track correction weight that accounts
for efficiency, fake tracks, and secondary particles is computed
according to
\begin{equation}
\wc = \frac{1 - (\ff + \fs)}{\eps}.
\label{eq:eff}
\end{equation}
Example results for the reconstruction efficiency $\eps$ and the combined
fake- and secondary-track fraction $\ff + \fs$, obtained for the
nominal track-selection working point are shown in Figure~\ref{fig:corrections} as
functions of \pT\ for two \peta\ intervals and two \nchrec\ intervals.
Figure~\ref{fig:corrections} also shows the bin-by-bin correction factor
$C_\text{bbb}$, which is discussed later and corrects for
migration of the reconstructed tracks in \pT\ and \peta.
The reconstruction efficiency increases significantly with increasing
\pT\ at low \pT, but increases slowly with \pT\ for $\pT\gtrsim0.7$~\GeV.
The combined fake- and secondary-track rates fall rapidly
with increasing \pT\ at low \pT, primarily due to the \pT\ dependence
of the fake-track fractions.
At higher \pT, the $\ff + \fs$ value is dominated by secondary tracks,
whose fraction varies slowly with \pT.

Aside from the effects of reconstruction efficiency, fake tracks,
and secondary tracks, the measurements are affected by migration of
the measured tracks in \pT\ and \peta.
The \pT\ resolution of the ID is dominated by multiple scattering
in the \pT\ range of the measurement, so  $\sigma_{\pT}/\pT$
varies slowly with \pT.
For $|\eta|<1$, $\sigma_{\pT}/\pT$ decreases from 3\% at $\pT= 0.2$~\GeV\
to 2\% at 0.4~\GeV\ and is nearly constant for higher \pT.
At the most forward pseudorapidities ($|\eta|>2$), $\sigma_{\pT}/\pT$ decreases
from ${\sim}6\%$ at $\pT= 0.2$~\GeV\ to 5\% at $\pT= 0.4$~\GeV,
above which it remains constant.
The (absolute) \peta\ resolution, $\sigma_{\eta}$, decreases with
increasing \pT.
For $|\eta|<1$, $\sigma_{\eta}$ decreases from 0.012 at $\pT=0.2$~\GeV\
to 0.002 at $\pT=5$~\GeV, while for $2<|\eta|<2.5$ it decreases from
0.018 at $\pT=0.2$~\GeV\ to 0.002 at $\pT=5$~\GeV.
Given the quoted \pT\ and \peta resolutions and the chosen \pT\ and
\peta intervals for the measurement, bin-by-bin corrections for
migration in \pT\ and \peta are sufficient for the measurement.
Thus, instead of applying the corrections in Eq.~\eqref{eq:eff}
and unfolding the measurements, simple bin-by-bin correction weights
\begin{equation*}
\wcbbb = \frac{1 - (\ff + \fs)}{C_\text{bbb}}
\end{equation*}
are applied to each particle in the data analysis.
Here, $C_{\text{bbb}}$ is the bin-by-bin correction factor
that accounts for both reconstruction efficiency and
migration in \pT and \peta, and is evaluated as a
function of \pt, \peta, and \nchrec.
Example results for $C_{\text{bbb}}$ are shown in
Figure~\ref{fig:corrections}.

The measured \peta-dependent quantities are defined for the fiducial region
in the following equations:
\begin{equation*}
\frac{{\dif n_{\text{fid}}^{\text{cent}}}}{\dif\eta} \equiv
\frac{1}{N^{\text{cent}}_{\text{evt}}} \dfrac{\sum\limits_{i\in \text{cent}}
\sum\limits_{j \in \Delta\eta} \wcbbbj}{\Delta \eta}\quad \text{and} \quad  \meanpt_{\text{fid}}^{\text{cent}} \equiv \frac{\sum\limits_{i \in \text{cent}}
\sum\limits_{j\in\Delta\eta} {\pT}_j \, \wcbbbj}{\sum\limits_{i \in \text{cent}}
\sum\limits_{j\in \Delta\eta} \wcbbbj},
\end{equation*}
where the labels ``cent'' and ``fid'' indicate the centrality interval
and the fiducial region of the measurement, the
index $i$ runs over events satisfying the event and
centrality selections, and the index $j$ runs over particles
within a given $\deta$ interval. Similar expressions with \peta replaced by \ypi define
the \ypi-dependent observables
The double-differential invariant yields are computed according to
\begin{equation}
\frac{\dif^2{n}}{\pT \dif\pT \dif\eta} \equiv
\frac{1}{N^{\text{cent}}_{\text{evt}}} \displaystyle \frac{\sum\limits_{i \in \text{cent}}
\sum\limits_{j \in (\Delta\eta,\Delta\pT)} \frac{ \wcbbbj}{{p_{\mathrm{T}}}_j}}{\Delta \eta \Delta \pT}.
\end{equation}
The measurements use 20 linear intervals and 36 logarithmic intervals in \peta
and \pT, respectively, spanning the fiducial ranges.
The measurements as a function of \ypi use the same intervals as
\peta, but results are not reported in the first and last interval,
owing to the loss of \ypi acceptance.

Applications of the above-described analysis procedure to
the reconstructed HIJING MC sample show closure of the \dndeta, \meanpt,
and double-differential yields to ${\ll}1$\% except in the 0--5\%
centrality interval, where the \dndeta yield exhibits a
${\sim}0.5\%$ degree of \peta-dependent nonclosure.
This nonclosure arises from the finite multiplicity binning of the
correction factors and is included in the systematic uncertainties
of the \dndeta measurements.

\subsection{Rapidity variables and impact on measurements\label{sec:rapidity_bias}}
The physics of particle production in hadronic or nuclear collisions
is most naturally expressed in terms of particle
rapidity (\rap).
The difference between \peta and \rap depends on \pT such that
\rap values migrate ``outward'', to larger $|\peta|$ values, at
low \pT, with the migration larger for increasing particle masses.
The migration of particles away from mid-rapidity leads to the well-known
depletion near
$\peta \sim 0$ in measurements of \dndeta in symmetric hadron--hadron
or nucleus--nucleus collisions. Since the migration is largest at
low \pT, it also has a significant impact on \meanpt.
Because most of the produced particles in hadronic collisions are
pions, the use of \ypi as an alternative to \peta removes much of
the distortion present in \dndeta and $\meanpt(\eta)$.
\begin{figure}[!b]
\centerline{
\includegraphics[width=0.95\textwidth]{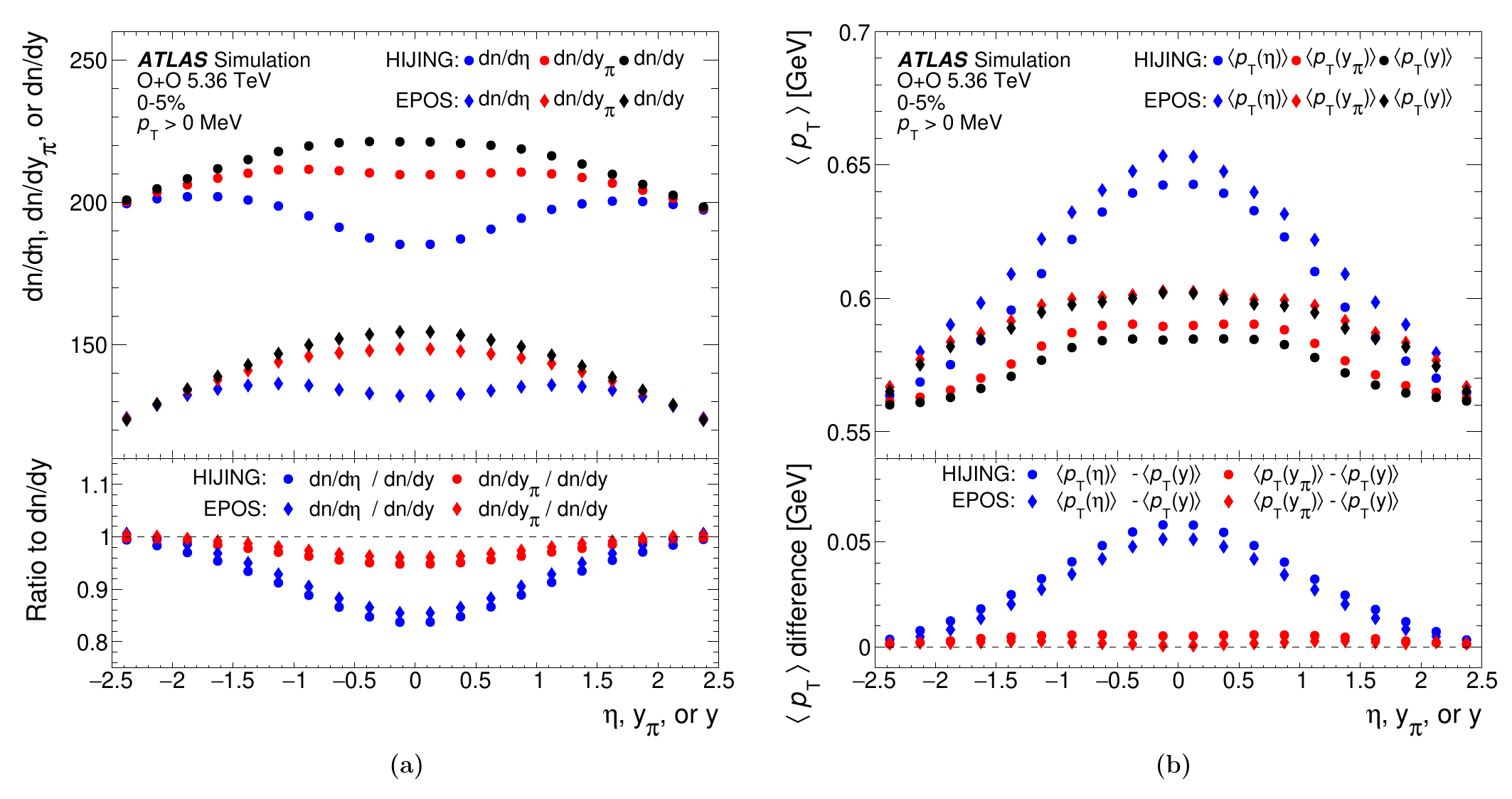}
}
\caption{
Comparison of (a) \dndeta, \dndypi, and \dndy and
(b) \meanpt computed in intervals of \peta, \ypi, and \rap
obtained from stable charged particles
with $|\rap| < 2.7$ and $0 < \pT < 5$~\GeV\
generated by
HIJING (circles) or EPOS (diamonds) in events satisfying the 0--5\% \OO centrality
selection. The bottom panels show (a) the ratio
of \dndeta and \dndypi to \dndy and (b) the difference between
$\meanpt(\peta)$ or $\meanpt(\ypi)$ and $\meanpt(y)$.
}
\label{fig:truthdndetameanpTRapVar}
\end{figure}

Figure~\ref{fig:truthdndetameanpTRapVar} demonstrates the above considerations
and the impact of the differences among the three rapidity variables,
\peta, \ypi, and \rap, based on comparisons using the HIJING sample and
a sample generated with the EPOS event generator~\cite{Pierog:2013ria}.
The figure compares $\dd n/\dd\eta$, $\dd n/\dd \ypi$ and $\dd n/\dd y$
for primary charged particles with $0<\pt<5~\GeV$,
produced by the generators in 5.36~\TeV\ \OO\ collisions in the 0--5\% centrality interval.
Comparisons between the \meanpt\ values as a function of \peta, \ypi and $y$ are also shown.
As the figure demonstrates, \dndeta is
smaller than \dndy everywhere, but especially so near $\peta =
0$. The \dndypi values are also smaller than \dndy, but the maximum
reduction is only ${\sim}5\%$ compared to a ${\sim}20\%$ reduction
in \dndeta. The \meanpt evaluated as a function of \peta is
significantly higher -- by as much as 50~\MeV\ -- than that
evaluated using true rapidity owing to the
migration of low-\pT particles away from $\peta \sim 0$.
However, the \meanpt evaluated as a function
of \ypi is much closer to $\meanpt(y)$, differing by less than 5~\MeV.
These observations justify the use of \ypi for some of the
measurements presented in this \paper.
\begin{figure}[!b]
\centerline{
\includegraphics[width=0.85\textwidth]{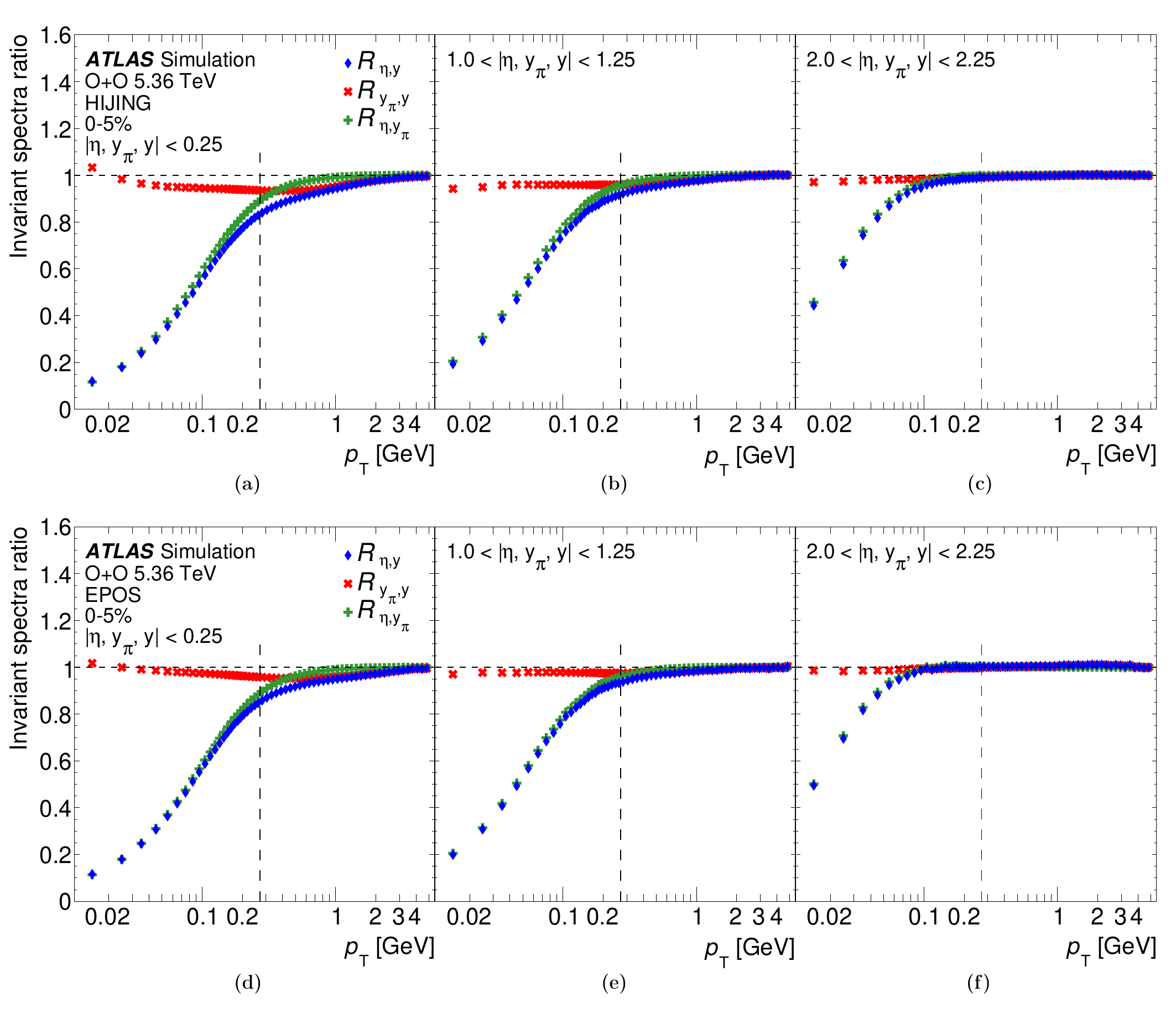}
}
\caption{
Ratios of invariant differential \pT spectra: \Yeta/\Yy (blue),
\Yypi/\Yy (red), and \Yeta/\Yypi (green)
for primary charged particles generated in 5.36~\TeV\
\OO collisions by (a, b, c) HIJING and (d, e, f) EPOS.
Results are shown for three different intervals of \peta, \ypi, or $y$,
with the interval varying from left to right.
The dashed vertical line indicates $\pT  = 0.27$~\GeV,
the nominal minimum \pT\ selection used in this analysis.
}
\label{fig:truthinvpTRatioRapVar}
\end{figure}

The impact of using different rapidity variables when evaluating the
invariant double-differential yields is illustrated in
Figure~\ref{fig:truthinvpTRatioRapVar}, which show ratios of
(i) \Yeta\ to $\Yy\equiv\invpTdndpTdy$,
(ii) \Yypi\ to $\Yy$, and
(iii) \Yeta\ to \Yypi.
Results are shown for HIJING at the top and EPOS at the bottom for three intervals of \peta,
\ypi, or \rap.
The effect of the migration in \peta can be clearly
observed at low \pT, where the \Yeta distribution is strongly
suppressed relative to \Yy. The \Yeta-to-\Yypi ratio shows a similar suppression at low \pT, which
means that the pions -- or the effects of the small pion mass --
dominate the migration responsible for the suppression
of \Yeta at low \pT. That conclusion has important
implications for this measurement, namely that the bulk of the distortion of
\Yeta can be obtained from the data by using the {\em measured} ratios
\begin{equation}
\Retaypi (\pT) = \frac{\Yeta}{\Yypi} \equiv \frac{\invpTdndpTdeta}{\invpTdndpTdypi},
\label{eq:Retaypi}
\end{equation}
where the numerator and denominator are evaluated using the
intervals of  \peta or \ypi.
The residual distortion of the \Yeta spectra, not included
in \Retaypi, is contained in the ratio
\begin{equation*}
\Rypiy (\pT) = \frac{\Yypi}{\Yy} \equiv \frac{\invpTdndpTdypi}{\invpTdndpTdy},
\end{equation*}
which shows a maximum deviation from unity of about 7\%.

The distortion of the \pT spectra seen in
Figure~\ref{fig:truthinvpTRatioRapVar} must be accounted for when
performing the \pteqo extrapolation. Fortunately, the bulk of that
distortion can be evaluated using the {\em measured} \Retaypi, which
greatly reduces potential sensitivity of the measurement to
mismodeling of the data -- particularly the flavor distributions -- in
the HIJING MC simulation.

\subsection{Extrapolation to zero transverse momentum~\label{sec:extrapolation}}
As described in the introduction, in addition to measuring
\dndeta and \meanpt within the fiducial acceptance, this measurement
also includes an extrapolation to \pteqo.
That extrapolation is performed by fitting the \pT dependence of
\Yeta or \Yypi to a function that can be integrated either
analytically or numerically.
Several functions, ($G(\pT)$) based on either the Hagedorn function
\begin{align}
G^{\text{H}}(\pT) & \equiv A/\left(1 + \frac{\pt}{p_0}\right)^n,
\label{eq:hagedorn}
\end{align}
with $A$, $p_0$ and $n$ being fit parameters,
or the Tsallis form $G^{\text{T}}(\pT)$, in which \pT\ in Eq.~\eqref{eq:hagedorn}
is replaced by the transverse mass $\mT = \sqrt{\pt^2 + m_{\pi}^2}$,
calculated under the charged-pion mass hypothesis,
were considered as part of the analysis.
The Hagedorn function was originally proposed~\cite{Hagedorn:1983wk}
because it could interpolate between negative exponential behavior at
low \pT or \mT and power-law evolution at high \pT.
Even so, neither the Hagedorn nor Tsallis function provides an adequate
description of the data.
However, augmenting the Hagedorn and Tsallis functions with an
additional negative exponential (NE) term in \pT\ or \mT, respectively:
\begin{align}
G^{\text{H+NE}}(\pT) & \equiv A/\left(1 + \frac{\pt}{p_0}\right)^n + Be^{-\pt/T},
\,\, \text{and} \,\,  G^{\text{T+NE}}(\pT) \equiv A/\left(1 + \frac{\mT}{p_0}\right)^n +Be^{-\mT/T},
\label{eq:Hne}
\end{align}
where $B$ and $T$ are additional fit-parameters, provides a
significantly better fit to the data.
Among the studied parameterizations,\footnote{``Blast-wave''
fits~\cite{Schnedermann:1993ws} were also considered but did not
provide an improved description of the data.}
these two forms provide the best description of the data over the
full \pT\ range of the measurement.
Although the Tsallis plus negative exponential function,
$G^{\text{T+NE}}$, yields better closure when applied to the HIJING-
and EPOS-generated primary charged-particle distributions in \OO
and \NeNe collisions, the $G^{\text{H+NE}}$
function provides better fits to the data, so it is used as the nominal model for
extrapolating to \ptEqualO, while results obtained with
$G^{\text{T+NE}}$ are used for evaluating systematic uncertainties.

\begin{figure}[!tb]
\centerline{
\includegraphics[width=0.9\textwidth]{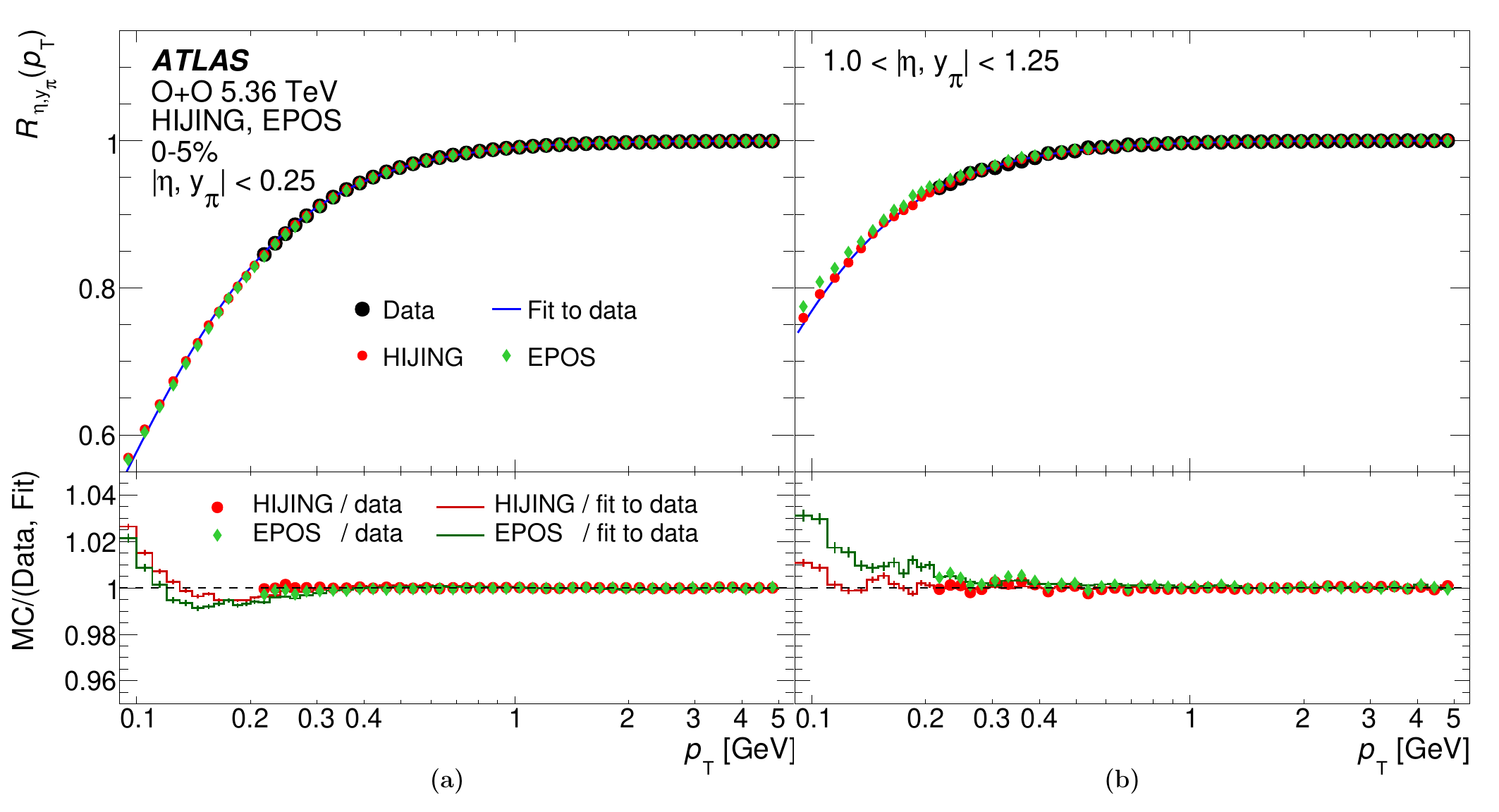}
}
\caption{
Results for $\Retaypi (\pT)$ (Eq.~\eqref{eq:Retaypi}) in data
compared with results from HIJING (red circles) and EPOS (green diamonds)
in the 0--5\% centrality interval.
Results are shown for (a) $|\eta|,|\ypi| < 0.25$ and
(b) $1 < |\eta|,|\ypi| < 1.25$.
}
\label{fig:ptspectRatioDataMC}
\end{figure}

Fits to Eq.~(\ref{eq:Hne}) are used to extrapolate the \Yypi\ distributions
to \ptEqualO.
However, in order to properly fit the above functions to the \Yeta
distributions, the modification of the \pT spectra resulting from the
migration between $y$ and \peta  illustrated in
Figure~\ref{fig:truthinvpTRatioRapVar} must be addressed. As suggested
above, nearly all of the modification can be measured using
\Retaypi.  Figure~\ref{fig:ptspectRatioDataMC} shows results
for this ratio in \OO collisions in the 0--5\% centrality interval in
two $|\peta|$ intervals. The ratios are evaluated over an extended \pT
range, $0.21 < \pT < 5$~\GeV in order to best constrain \Retaypi at
lower \pT.
The results are compared with similar ratios constructed using
primary charged particles generated by HIJING (red circles) and EPOS (green diamonds).
HIJING describes the measured \Retaypi in both $|\peta|$ intervals.
EPOS describes the data well for $|\eta|<0.25$ but
systematically deviates from the measured \Retaypi at
larger $|\peta|$, as evidenced by the small differences between EPOS
and data in the $1 < |\eta| < 1.25$ interval.
The variations of the measured $\Retaypi$ with \pT are
well described by functions of the form
\begin{equation}
\Retaypi^{\text{fit}}(\pT) =  \frac{h(\pT)}{C + h(\pT)},~\text{with}~h(\pT) = \left(\pT\right)^a \left[1 + \gamma \left(\pT\right)^a\right],
\label{eq:Rfit}
\end{equation}
where $C$, $a$ and $\gamma$ are fit parameters.
The values of the parameterized functions in Eq.~(\ref{eq:Rfit}) decrease
as $\pT \rightarrow 0$ and approach unity at large \pT.
The data cannot constrain the term involving $\gamma$ in Eq.~\eqref{eq:Rfit}, but the MC
$\Retaypi^{\text{fit}}(\pT)$ distributions, evaluated
over $0 < \pT < 5$~\GeV, are better described with a small, nonzero
value for $\gamma$ that is used when fitting the data.
Resulting fits to the measured $\Retaypi(\pT)$ are shown with
the curves in Figure~\ref{fig:ptspectRatioDataMC}.
The ratios of the MC \Retaypi values to those obtained from the data
are shown by the curves in the lower panels.
The incorporation of the MC-determined $\gamma$ term in $h(\pT)$
ensures that $\Retaypi^{\text{fit}}(\pT)$ remains
compatible with both HIJING and EPOS at low \pT, except at larger
$|\peta|$,  where EPOS shows a systematic deviation from the data.
In order to assess the potential bias resulting from the use of MC simulation
to constrain $\gamma$, the analysis is repeated with $\gamma = 0$,
and the difference is included as a systematic uncertainty.

The parameter
$C$ in Eq.~\eqref{eq:Rfit} is small and consistent with zero at large
$|\peta|$ and is largest at $\peta = 0$. For all centrality
intervals in both \OO and \NeNe collisions, the \peta dependence of $C$
is well described by a Gaussian function centered at zero. In the application
of $\Retaypi^{\text{fit}}(\pT)$ in the analysis described
below, Gaussian fits to $C(\peta)$ are performed separately for each
collision system, centrality interval, and analysis variation, and the
results of the fit are used when evaluating
$\Retaypi^{\text{fit}}(\pT)$. Statistical variations of the
Gaussian fit functions at the 68\% confidence level are used to assess systematic
uncertainties associated with the smoothing of $C(\peta)$.
\begin{figure}[!tb]
\centerline{
\includegraphics[width=0.95\textwidth]{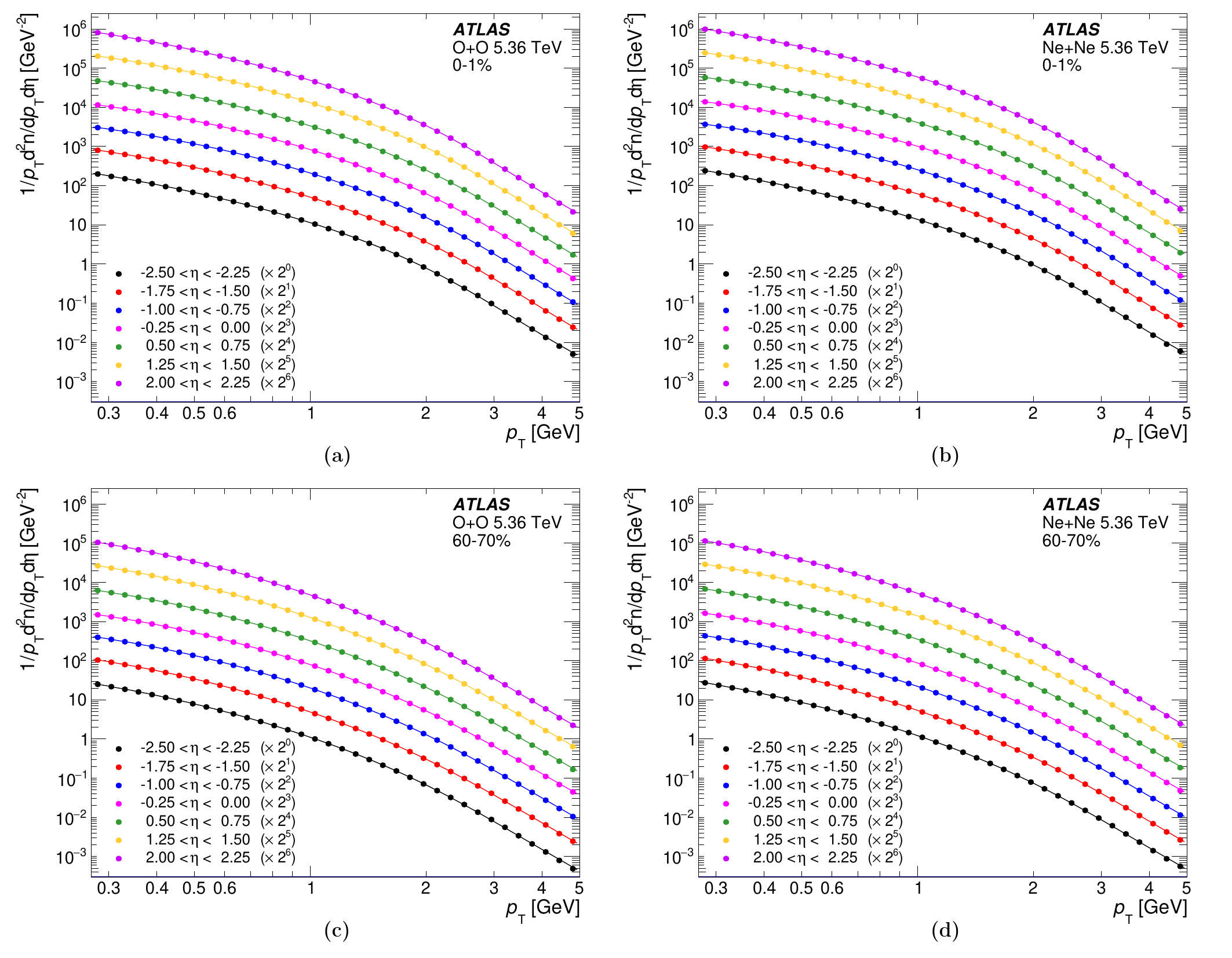}
}
\caption{
Measured \invpTdndpTdeta for a subset of the \peta intervals used in the \dndeta and
\meanpt measurement for \OO (left) and \NeNe (right) collisions in the 0--1\% (top)
and 60--70\% (bottom) centrality intervals. The spectra are each scaled by a factor of $2^{n}$,
as indicated in the legend, to improve visibility between data measured in different \peta
intervals. Statistical errors are shown but are smaller than the size of the data points.
The solid lines show the results of fits to the data using the expression in
Eq.~(\ref{eq:ptFitMethod}) with $G = G^{\text{H+NE}}$.
}
\label{fig:pTSpectraEtaDep}
\end{figure}

\begin{figure}[!tb]
\centerline{
\includegraphics[width=0.9\textwidth]{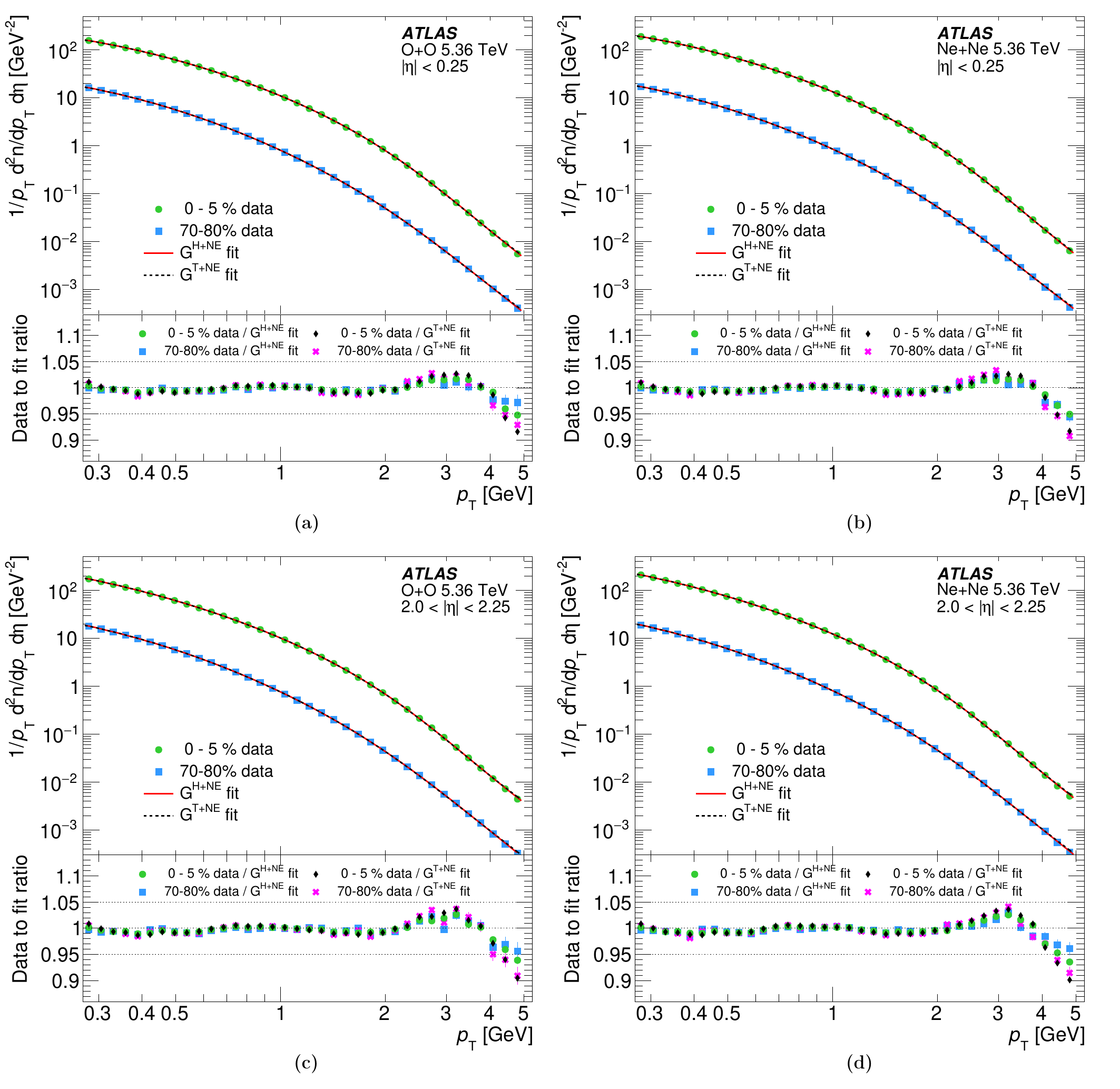}
}
\caption{
Measured \Yeta and their ratios to fit functions (see
Eq.~\eqref{eq:ptFitMethod}) in the pseudorapidity intervals
(a, b) $|\peta|< 0.25$ and (c, d) $2.0 < |\peta| < 2.25$
for (a, c) \OO collisions and (b, d) \NeNe collisions in
the 0--5\% and 70--80\% centrality intervals. Solid red and
dashed black lines in each panel represent fits of the functions
$G^{\text{H+NE}}(\pT)$ and $G^{\text{T+NE}}(\pT)$, respectively, to the data.
The error bars on the data points indicate statistical uncertainties only.
The dotted horizontal lines in the lower section of each panel
indicate unity and unity $\pm 5\%$.
}
\label{fig:pTFitRatios}
\end{figure}

The complete parameterization of the \pT dependence of \Yeta is
expressed as
\begin{equation}
\ffitpt = \Retaypi^{\mathrm{fit}} (\pT) \, \times \Rypiy(\pT) \, \times
G^{\text{model}}(\pT),
\label{eq:ptFitMethod}
\end{equation}
where $G^{\text{model}}(\pT)$ is either $G^{\text{H+NE}}$ or $G^{\text{T+NE}}$.
This function, with $\Retaypi^{\mathrm{fit}}$ obtained from fits to data (Eq.~\eqref{eq:Rfit})
and \Rypiy obtained from MC simulation, is fitted to the measured data in each \peta interval
using both $G^{\text{H+NE}}(\pT)$ and $G^{\text{T+NE}}(\pT)$.
The nominal fits are performed over the transverse momentum range $0.27 < \pT < 5$~\GeV.
Results of such fits are illustrated in
Figure~\ref{fig:pTSpectraEtaDep}, which shows the
measured \invpTdndpTdeta distributions versus \pT in a subset of the
analyzed \peta intervals for \OO and \NeNe collisions in the 0--1\%
and 60--70\% centrality intervals. The fits obtained using Eq.~(\ref{eq:ptFitMethod})
with the $G^{\text{H+NE}}$ functional form are also shown. The fits
are capable of describing the full shape of the
measured \invpTdndpTdeta, although small deviations of the fit from the data would not
be visible in the figure. To better illustrate the quality of the fits,
Figure~\ref{fig:pTFitRatios} shows results for \invpTdndpTdeta
in \OO and \NeNe collisions in 0--5\% and 70--80\% centrality
intervals using ``symmetrized'' pseudorapidity intervals $|\peta| <
0.25$ (top) and $2.0 < |\peta| < 2.25$ (bottom) and ratios of the
fit function to the data. The ratios differ
from unity by ${<}2\%$ at low \pT, but show a systematic deviation
from the data above 3~GeV. That deviation, however, has negligible
impact on the extrapolated \dndeta and \meanpt results.

While the extrapolated \dndeta and \meanpt can, in principle, be computed
directly from $f^{\text{fit}}_{\eta}(\pT)$, for the purposes of this
measurement the focus is on correcting the fiducial measurements.
Thus, instead, the fit results are used to compute the two quantities
\begin{align}
\Fextr & = \frac{\int_0^{5~\GeV}  {\dd\pT} \; \pT\,\ffit }{\int_{0.27~\GeV}^{5~\GeV} {\dd\pT} \; \pT\, \ffit}
\quad \mathrm{and}\quad
\Dextr  = \frac{\int_0^{5~\GeV} \dd\pT \; \pT^2\, \ffit}{\int_0^{5~\GeV} \dd\pT \; \pT\, \ffit}
- \frac{\int_{0.27~\GeV}^{5~\GeV} \dd\pT \; \pT^2\, \ffit}{\int_{0.27~\GeV}^{5~\GeV} \dd\pT \; \pT\, \ffit}.
\label{eq:Fextr}
\end{align}
where the \pT factors in the integral are needed to convert from
invariant differential yields to, e.g., $\dd^2n/\dd\pT\, \dd \eta$.
The quantity \Fextr represents the multiplicative factor needed to
correct the fiducial \dndeta to the full, extrapolated \dndeta,
while \Dextr is an additive correction needed to convert the
fiducial \meanpt to the full extrapolated result. In both \Fextr
and \Dextr, the normalization of \ffit cancels out in the ratio. Systematic
variations of \Fextr and \Dextr associated with the extrapolation
procedure are made by including the MC-motivated $\gamma$ term in
$h(\pT)$ (see Eq.~\eqref{eq:Rfit}) or not, by varying the $a$ parameter within statistical
uncertainties, by varying $C(\eta)$ within
the uncertainty in the Gaussian smoothing function as described above, and by including
the $\Rypiy(\pT)$ obtained from HIJING or setting that term to unity
in Eq.~\eqref{eq:ptFitMethod}. These variations are found to
change the  \Fextr values by at most a few tenths of a percent and
\Dextr by less than a few MeV. The inclusion of \Retaypi in
Eq.~\eqref{eq:ptFitMethod} reduces \dndeta and \Fextr by
an \peta-dependent fraction that at $\peta = 0$ is ${\lesssim}5\%$ and
increases \meanpt and \Dextr by a \peta-dependent shift that is
${\lesssim}25$~\MeV.

\begin{figure}[!tb]
\centering
\includegraphics[width=0.75\textwidth]{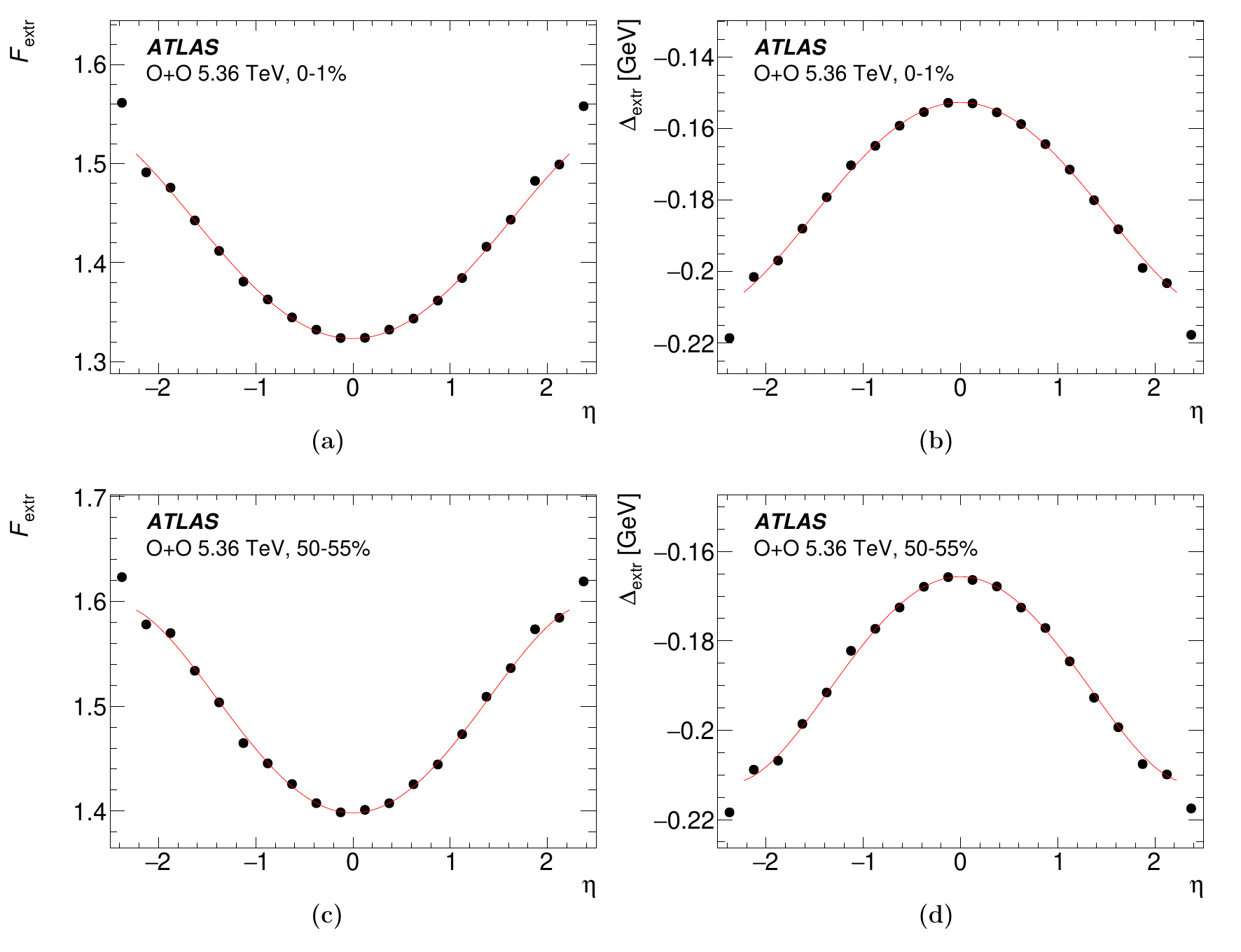}
\caption{
Example results for (a, c) \Fextr and (b, d) \Dextr obtained by
fitting measured \Yeta distributions to the form in
Eq.~\eqref{eq:ptFitMethod} and applying Eqs.~\eqref{eq:Fextr}. %
Results are shown for the (a, b) 0--1\% and (c, d) 55--60\%
centrality intervals in \OO collisions.
The solid lines show the fits used to smooth and ensure that
\Fextr and \Dextr are symmetric around $\peta = 0$.
The data with $|\eta|>2.25$ are excluded from the
smoothing, as the tracking is poorer at forward rapidities.
}
\label{fig:FextrDextr}
\end{figure}

Because \OO\ and \NeNe\ collision systems are intrinsically
forward-backward symmetric, the corrections to the fiducial
measurements of \dndeta and \meanpt, the \Fextr and \Dextr values,
should be symmetric in rapidity.
To ensure that symmetry, and to remove statistical fluctuations,
\Fextr and \Dextr are fitted to even polynomial functions of $\peta$.
However, the data with $|\eta|>2.25$ are excluded from the
smoothing procedure because the tracking performance is poorer
at forward rapidities, leading to larger variations in
\Fextr and \Dextr that are not well described by
the even-polynomial fits.
Examples of the results for \Fextr and \Dextr obtained from the
aforementioned procedures, and the fits obtained to smooth the
values, are shown in Figure~\ref{fig:FextrDextr}.

%

%
%

\section{Systematic uncertainties}
\label{sec:systematics}
\vspace{-0.2in}

Systematic uncertainties in the measurements
arise primarily from potential inaccuracies in the corrections for
reconstruction efficiency, fake- and secondary-track rates, and bin migration.
Such effects can originate from mismodeling of the detector geometry
and response in the MC simulations, imperfect modeling of the true particle kinematic
distributions in \OO\ or \NeNe collisions, differences between the detector
conditions during data-taking and those assumed in the
MC simulation, or from the procedures used to evaluate and apply
the corrections.
In addition to MC-based effects, the measurements can be affected by
uncertainties in the definition of the centrality intervals, biases
associated with the trigger selection, or background events passing
the event selection.
For results extrapolated to \pteqo, additional uncertainties arise
from the modeling of the extrapolation procedure.
The evaluation of the systematic uncertainties associated with these
sources is described below.

\begin{enumerate}[label=\roman*]
\item \textbf{Track selection:}
The track selection criteria determine the relative contributions of
primary charged particles, secondary particles, and fake tracks
entering the analysis.
The results obtained with the nominal selection are compared with
those using the alternative track selections described in
Section~\ref{sec:datasets}.
By varying the tracking requirements, the relative contributions of primary,
secondary, and fake tracks are modified, and the associated systematic
uncertainty therefore covers the likely spread of the contributions
from secondary and fake tracks in the data.
This uncertainty varies only weakly with centrality and is taken to
be symmetric about the nominal measurement.

\item \textbf{Centrality definition:}
The uncertainty in the centrality determination over the 0--80\% centrality range is
estimated to be $\pm2.2\%$ ($\pm2.5\%$) for the \OO\ (\NeNe) data.
This uncertainty reflects variations in the nuclear density profiles
used in the Glauber fits, as well as additional sources such as
changes to the \sumET\ thresholds above which the fits are performed.
The systematic uncertainty associated with the centrality definition
is evaluated by repeating the full analysis using alternative
centrality thresholds obtained from the Glauber study.
This uncertainty exhibits a strong dependence on centrality: it is
smallest in the most central collisions and increases
monotonically for increasingly peripheral collisions.
For the \dndeta\ measurements, this is the dominant systematic
uncertainty affecting the most peripheral centrality intervals.

\item \textbf{Tracking efficiency:}
As mentioned in Section~\ref{sec:analysis}, the measurements
use corrections to adjust the tracking efficiencies and
fake- and secondary-track rates that are obtained using MC simulations.
The sensitivity of the results to the modeling of the detector
material and geometry is evaluated using previously described alternative MC samples.
The difference between the results obtained with the alternative and
nominal simulations is taken as the systematic uncertainty.
The point-to-point variations are assumed to be symmetric about the
nominal result and are combined in quadrature.
The resulting fractional uncertainties are largely independent of
centrality, but show a strong dependence on $\eta$ for the
\dndeta\ measurement.
At forward pseudorapidity ($|\eta|>2$), this source becomes one of
the dominant contributors to the total uncertainty in \dndeta.

\item \textbf{MC reweighting:}
As described in Section~\ref{sec:analysis}, residual mismodeling of
the particle kinematic distributions in \HIJING\ is addressed through
\pt-dependent and flavor-dependent reweighting of the MC-generated
charged-particle distributions.
The reweighting factors are obtained from comparisons with ALICE measurements of
$K/\pi$ and $p/\pi$ \pt\ spectra in \XeXe\ collisions~\cite{ALICE:2021lsv},
and are applied to kaons and protons in the \HIJING\ simulation.
The nominal analysis uses reweighting factors obtained from the
50--60\% centrality interval in \XeXe\ collisions.
This \XeXe\ multiplicity roughly corresponds to 0--10\%
centrality \OO\ collisions.
Alternative reweighting
factors are derived from the 40--50\% and 60--90\% centrality intervals,
and the resulting changes in the measured observables are included
as a systematic uncertainty.

\item \textbf{MC nonclosure:}
A closure test is performed by comparing the values of \dndeta\ and
\meanpt\ obtained directly from generated particles in the MC
simulation with those obtained by applying the full reconstruction
and analysis procedure to the simulated events.
Closure is found to hold within statistical uncertainties over
the 5--80\% centrality range.
However, for the \dndeta\ measurement in the most central collisions,
a small degree of nonclosure is observed.
This nonclosure amounts to approximately 1\% in the 0--1\%
centrality interval.
A systematic uncertainty is therefore assigned to the \dndeta\
measurement; it is zero above 5\% centrality and increases
linearly from 0\% at 4--5\% centrality to 1\% at 0--1\% centrality.
No nonclosure effect is observed for the \meanpt\ measurement, and no
corresponding systematic uncertainty is assigned.

\item \textbf{Fit model for \pteqo extrapolation:}
This uncertainty applies only to results extrapolated to \pteqo.
It accounts for the dependence of the extrapolated measurements on
the functional form used to describe the \pt\ spectra.
As discussed in Section~\ref{sec:analysis} the nominal results
are obtained using a ``Hagedorn + negative exponential'' parameterization.
The change in the results when using an alternative
``Tsallis + negative exponential'' parameterization is assigned as a
systematic uncertainty.
This uncertainty is taken to be symmetric about the nominal value.
It is one of the leading uncertainties in the results after extrapolation to \pteqo.

\item \textbf{Fit range for \pteqo extrapolation:}
This uncertainty also applies only to results extrapolated to \pteqo.
It accounts for the dependence of the extrapolation on the \pt\
range used in the fit.
The nominal extrapolation uses fits to the \pt\ spectra over the
range 0.27--5.0~\GeV.
Alternative fit ranges of 0.24--5.0~\GeV\ and 0.27--3.9~\GeV\ are
used to assess this uncertainty.

\item \textbf{$y_{\pi}{-}\eta $ weight:}
This uncertainty is another that applies only to results extrapolated to \pteqo.
The value of $R_{\eta,y{\pi}}^{\text{fit}}(\pT)$ (Eq.~\eqref{eq:Rfit})
is varied within its uncertainties, and the effect is propagated
through the analysis.
This includes setting of the $\gamma$ term in Eq.~\eqref{eq:Rfit} to zero,
as discussed in Section~\ref{sec:extrapolation}.
Additionally, checks are done where the MC-determined \Rypiy\ term
is removed from \ffitpt\ (Eq.~\eqref{eq:ptFitMethod})
and the analysis is repeated.
The change in the final results is found to be negligible.
\end{enumerate}

\begin{figure}[p]
\centering
\includegraphics[width=0.9\textwidth]{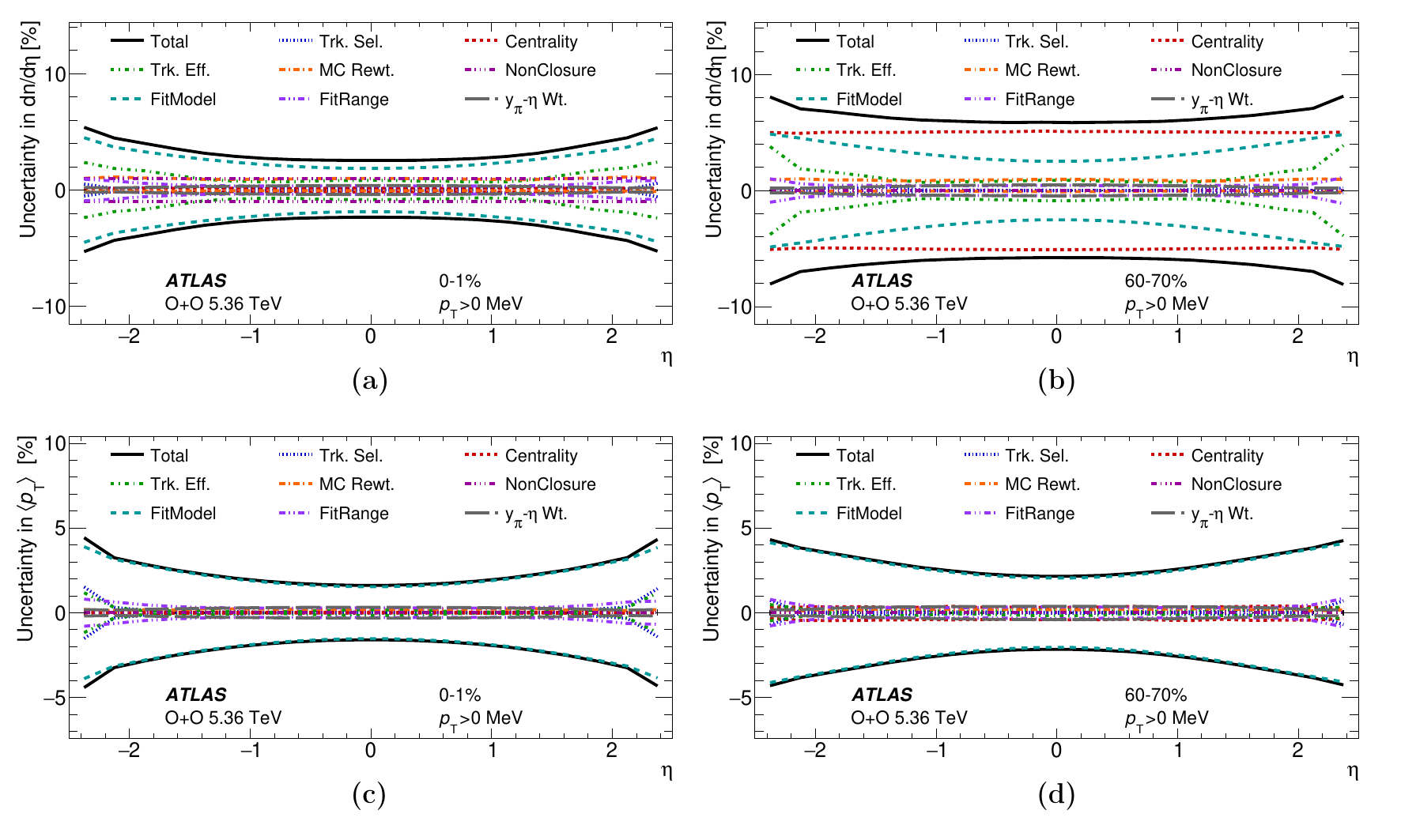}
\caption{
Summary of the systematic uncertainty estimates for (a, b) \dndeta\
and (c, d) \meanpt\ in \OO\ collisions, for the (a, c)  0--1\%
and (b, d) 60--70\% centrality intervals.
Both the total uncertainty and its components are shown.
}
\label{fig:dndetaSystSummary}
\end{figure}

\begin{figure}[p]
\centering
\includegraphics[width=0.9\textwidth]{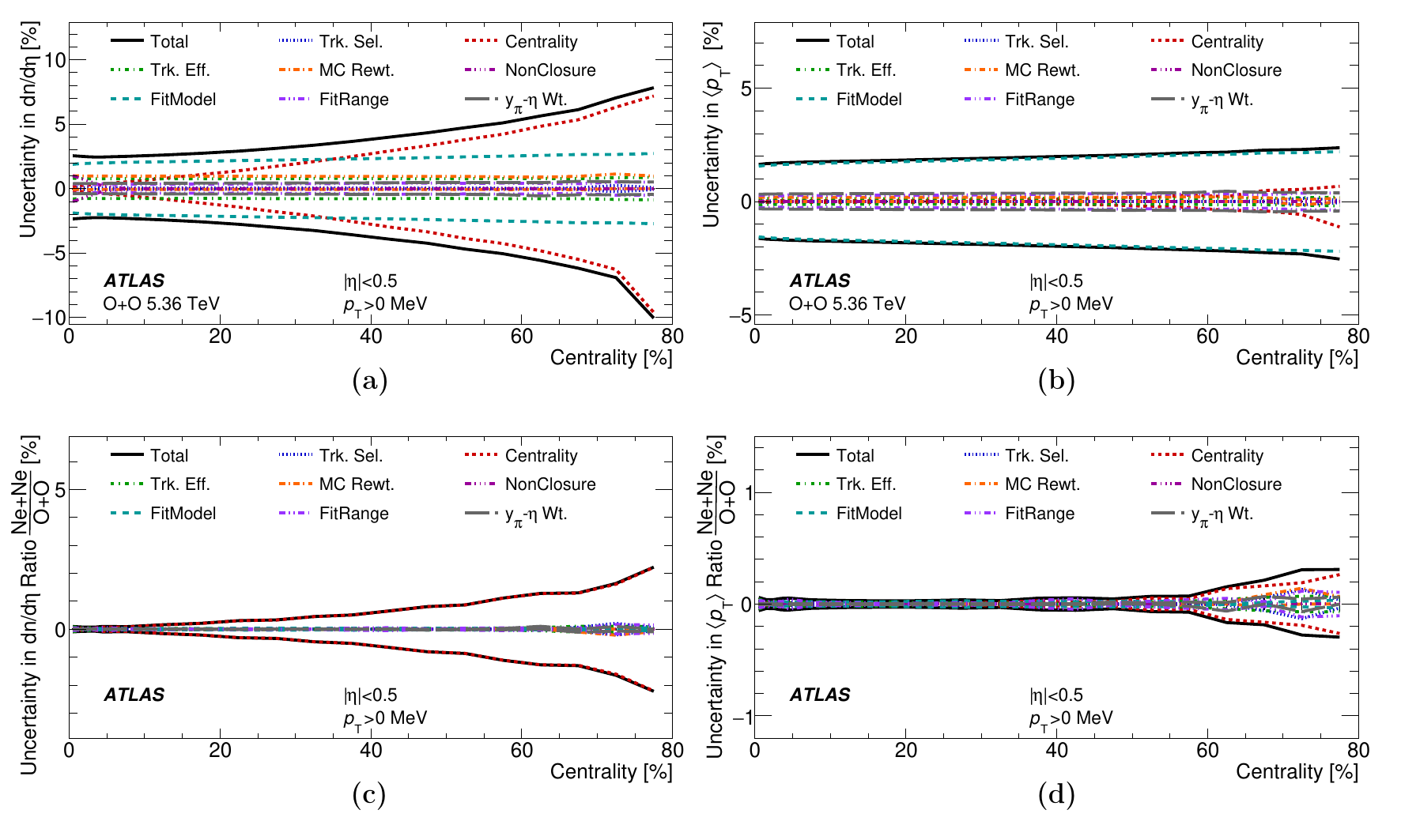}
\caption{
Summary of the systematic uncertainty estimates for (a) \dndeta\
and (b) \meanpt\ at mid-rapidity ($|\eta|<0.5$), as a function
of centrality in \OO\ collisions.
The uncertainties for the ratios of the \NeNe\ to \OO\ measurements
for (c) \dndeta\ and (d) \meanpt\ are also shown.
The plots show the total uncertainty and the contributions from
the individual components.
}
\label{fig:SystSummaryCentdep}
\end{figure}
The total systematic uncertainty is obtained as the quadrature sum
of the individual contributions.
Figure~\ref{fig:dndetaSystSummary} summarizes the systematic
uncertainties for the \dndeta\ and \meanpt\ measurements in \OO\
collisions for a central and a peripheral centrality interval.
The uncertainty associated with the chosen functional form of the
fit model is typically the leading systematic uncertainty.
However, in peripheral collisions, the uncertainty from the
centrality determination also contributes significantly.
Figure~\ref{fig:SystSummaryCentdep} shows the centrality dependence
of the systematic uncertainties at mid-rapidity.
Figure~\ref{fig:SystSummaryCentdep} also shows the systematic
uncertainties in the \NeNe/\OO ratios of the \dndeta\ and \meanpt spectra,
where correlated uncertainties cancel out.
Uncertainties not associated with the extrapolation also apply to the
$\pt>0.27$~\GeV\ fiducial measurements.


%
\FloatBarrier
\section{Results}
\label{sec:result}
\vspace{-0.2in}

%
This section presents a subset of the results
-- in most cases for a limited set of centrality intervals -- of the measurements performed
using the procedures described in the preceding sections.
Results not included here are presented in the appendix.

\subsection{Theoretical calculations}
The measurements presented in this paper are compared with theoretical
calculations from Ref.~\cite{Mantysaari:2025tcg}
(labeled as ``IPGlasma'') and from Ref.~\cite{Giacalone:2024luz}
(labeled as ``Trento''). They are also compared with events generated with the
EPOS model~\cite{Pierog:2013ria}. The following paragraphs describe
each calculation in more detail.

The ``IPGlasma'' calculation combines the Projected Generator Coordinate
Method (PGCM)~\cite{Lee:2008fa}, an \textit{ab initio} nuclear-structure
model for oxygen and neon, with the IPGlasma framework~\cite{Schenke:2012wb}
to compute the initial energy densities of the
colliding nuclei.
These energy densities are evolved using viscous relativistic
hydrodynamics implemented in MUSIC~\cite{Schenke:2010nt}.
A hadronic afterburner based on UrQMD~\cite{Bleicher:1999xi,Bass:1998ca}
is subsequently applied to obtain the final particle spectra
and multiplicities.
The estimated theory uncertainties in the IPGlasma calculations are
purely statistical.

The ``Trento'' calculations are described in Ref.~\cite{Giacalone:2024luz},
and use either the PGCM~\cite{Lee:2008fa} or the Nuclear Lattice Effective
Field Theory (NLEFT) simulations~\cite{Lee:2008fa,nleft} to determine
the nuclear configurations of oxygen and neon.
These configurations are used to generate initial-state energy
densities using the Trento model~\cite{Moreland:2014oya}.
The resulting events are evolved using the Trajectum hydrodynamic
framework~\cite{Nijs:2020roc,Nijs:2021clz} to obtain the final particle
spectra and multiplicity distributions.
The estimated theory uncertainties in the Trento calculations include statistical uncertainties
from finite event sampling and systematic uncertainties obtained by
sampling multiple parameter sets from a Bayesian posterior constrained
by \PbPb\ data for several observables (see Refs.~\cite{Giacalone:2024luz,Giacalone:2023cet}),
as well as variations arising from finite grid spacing and nuclear-structure modeling.

The EPOS calculations are performed using version 4.0.3 of the EPOS model in its
core--corona configuration~\cite{Werner:2024fwk, Werner:2023zvo, Werner:2023fne, Werner:2023mod, Werner:2023jps} including hydrodynamic modeling.
Initial particle production is described within a parton-based
Gribov--Regge multiple scattering framework, with parton ladders
hadronizing via string fragmentation.
The produced matter is separated into a dense core, which undergoes
hydrodynamic evolution, and a dilute corona, which hadronizes
without significant final-state interactions.
Final-state hadronic rescatterings are included through a hadronic cascade.
The estimated theory uncertainties in the EPOS calculations are
purely statistical.

In the figures shown in this section, the statistical and
systematic uncertainties of the Trento model calculations
are combined in quadrature and shown as total uncertainties
for clarity.
Since the IPGlasma and EPOS model calculations include only
statistical uncertainties, their displayed total uncertainties
are purely statistical.

The Trento calculations are available only for the
\pteqo\ extrapolated case, and are not shown when
comparing with the $\pt>270~\MeV$ fiducial measurements,
except when comparing ratios where the effect of the fiducial
acceptance largely cancels.
Similarly, IPGlasma calculations for the \pt\ spectra are
not available in 1\%-wide centrality intervals and are not
shown when comparing with the \pt\ spectra measured in 1\%-wide
centrality intervals.

\subsection{Transverse momentum distributions}
\begin{figure}[!htb]
\centering
\includegraphics[width=0.99\linewidth]{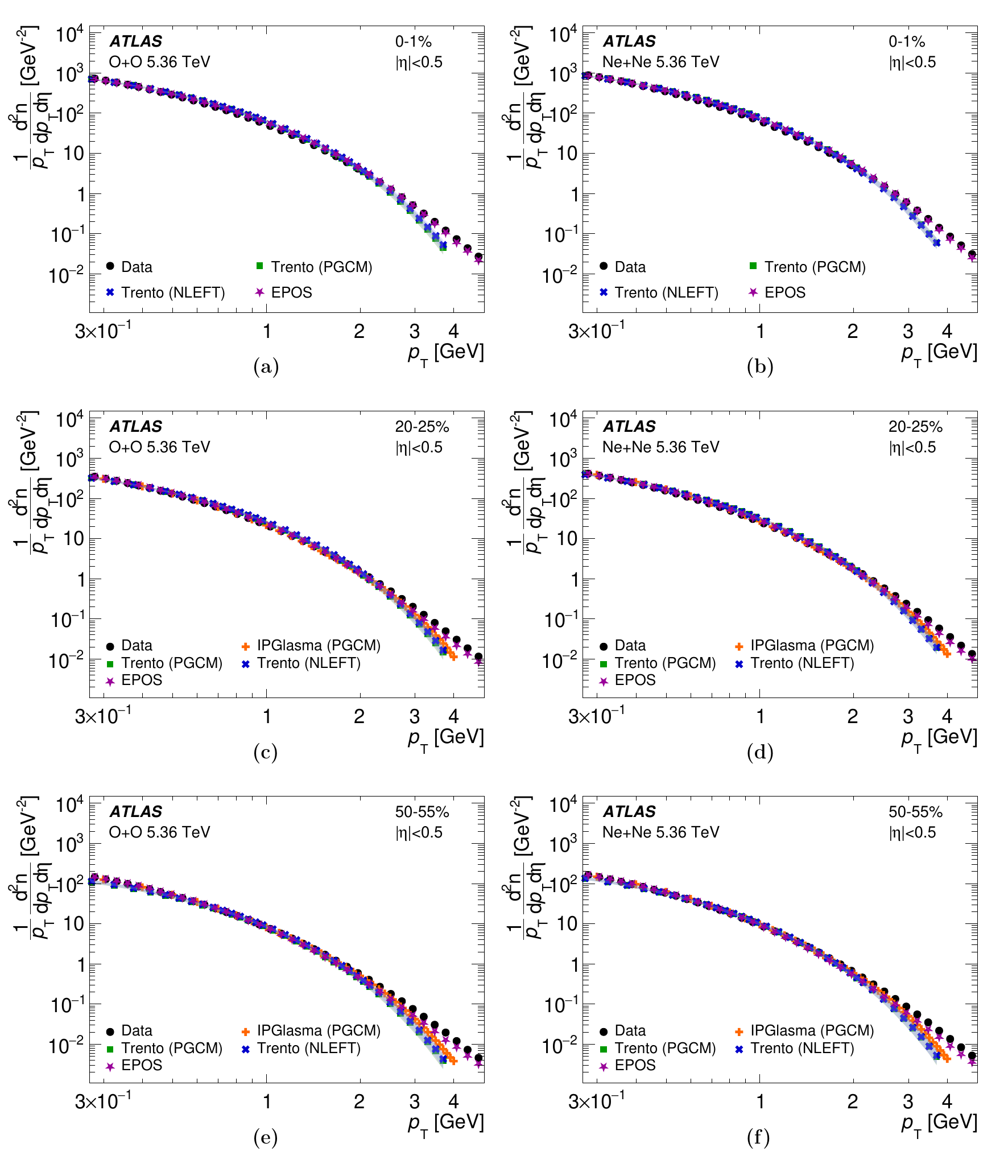}
\caption{
The measured \pt\ spectra in (a, c, e) \OO\ and
(b, d, f) \NeNe\ collisions at mid-rapidity
for the (a, b) 0--1\%, (c, d) 20--25\%
and (e, f) 50--55\% centrality intervals.
For comparison, predictions from models described in Ref.~\cite{Mantysaari:2025tcg}
(labeled ``IPGlasma'') and Ref.~\cite{Giacalone:2024luz} (labeled ``Trento''),
as well as the EPOS model, are also shown.
For the data, the vertical lines and shaded bars indicate statistical and
systematic uncertainties, respectively, and in most cases are smaller
than the size of the data points.
For the theory calculations, the shaded bands indicate total uncertainties.
}
\label{fig:results_ptspec_Theory}
\end{figure}

\begin{figure}[p]
\centering
\includegraphics[width=0.99\linewidth]{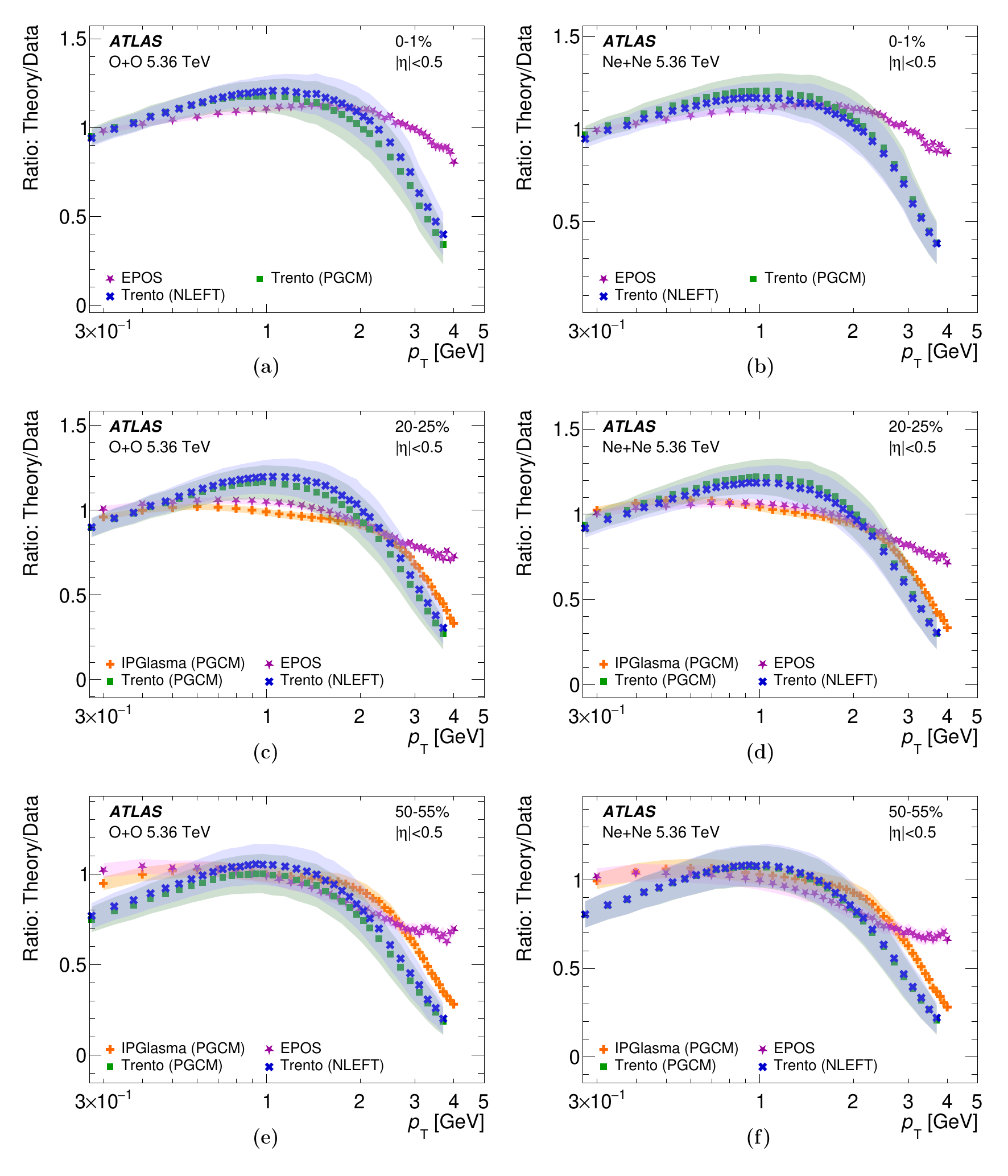}
\caption{
Ratios of the theory predictions to the measured \pt\ spectra in (a, c, e) \OO\ and
(b, d, f) \NeNe\ collisions at mid-rapidity
for the (a, b) 0--1\%, (c, d) 20--25\%
and (e, f) 50--55\% centrality intervals.
Ratios based on predictions from models described in
Ref.~\cite{Mantysaari:2025tcg} (labeled ``IPGlasma'')
and Ref.~\cite{Giacalone:2024luz} (labeled ``Trento''),
as well as the EPOS model, are shown.
The shaded bands indicate total uncertainties.
}
\label{fig:results_ptspec_TheoryRatio}
\end{figure}

Figure~\ref{fig:results_ptspec_Theory} presents the
double-differential invariant yields, \invpTdndpTdeta, measured in \OO
(left) and \NeNe (right) collisions in three centrality intervals and
also shows the
corresponding theoretical distributions.
Figure~\ref{fig:results_ptspec_TheoryRatio}
presents the ratios of the theory predictions to the data.
Figures~\ref{fig:results_ptspec_OO_Theory_aux}--\ref{fig:results_ptspec_NeNe_Theory_aux}
and~\ref{fig:results_ptspec_OO_TheoryRatio_aux}--\ref{fig:results_ptspec_NeNe_TheoryRatio_aux}
in the Appendix respectively show invariant yields and ratios of theory predictions to data
in additional centrality intervals.
In general, the \pt\ spectra in the data are qualitatively reproduced by the
theory calculations, especially at low \pT.
For $\pt<2$~\GeV, the Trento calculations are within about $\pm15$\% of the data,
but show a systematic deviation with increasing \pT\ that is more pronounced in the more
peripheral collisions.
In contrast, the IPGlasma and EPOS ratios to data are nearly
\pT-independent and within about $\pm5$\% of unity for $\pT \lesssim 1$~GeV.
Both the Trento and IPGlasma calculations underestimate the data for
$\pT \gtrsim 2$~GeV, as does EPOS but to a lesser degree.
The failure of the Trento and IPGlasma calculations to describe the
data at high \pT presumably arises because the hydrodynamic models
do not include hard processes. Hard processes are included in EPOS,
but the hard-scattering rates may be insufficient to describe the data.

\begin{figure}[!tb]
\centering
\includegraphics[width=0.9\linewidth]{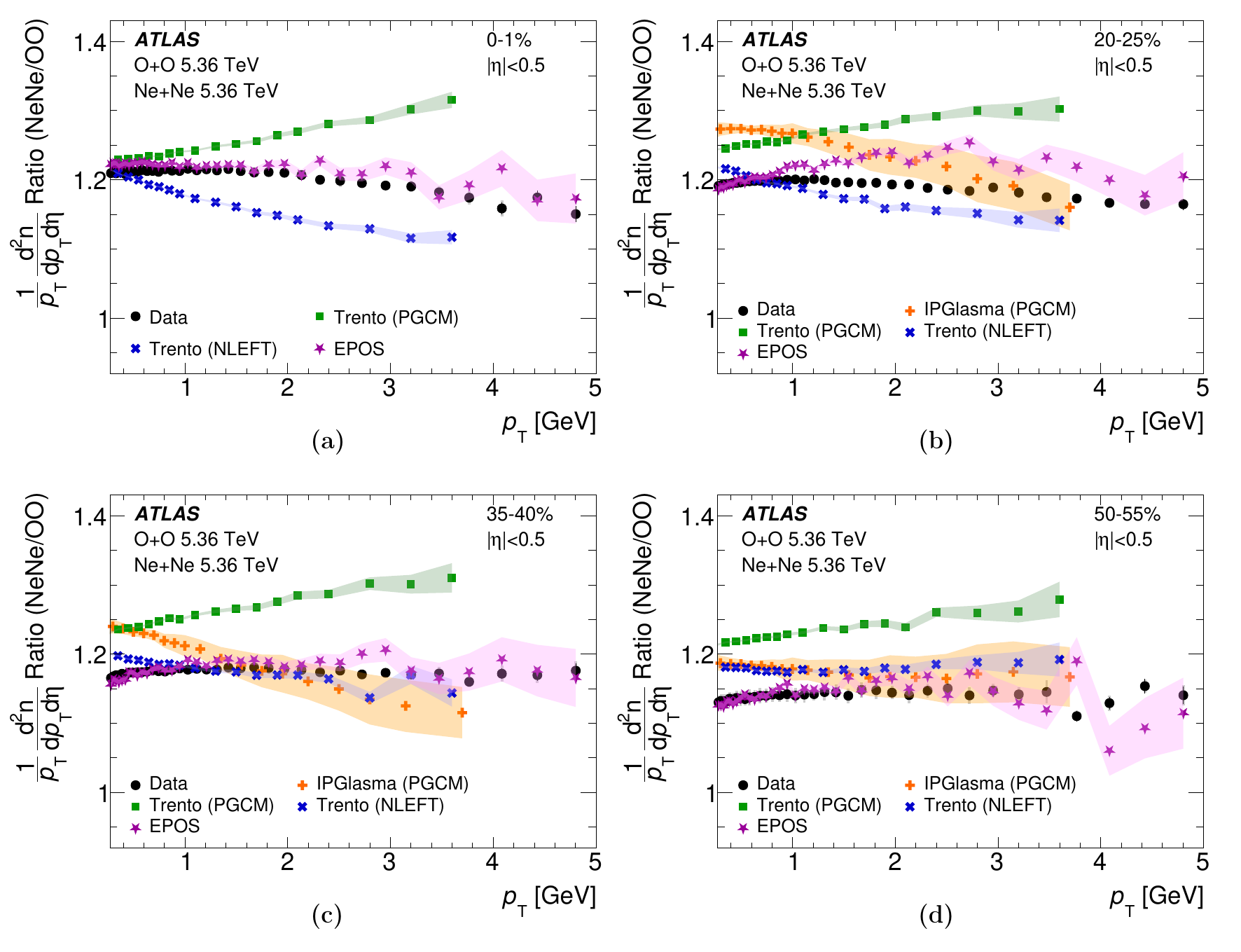}
\caption{
The ratio of the \pt\ spectra measured in \NeNe\ and \OO\ collisions
at mid-rapidity ($|\eta|<0.5$),
for the (a) 0--1\%, (b) 20--25\%, (c) 25--40\%, and (d) 50--55\%
centrality intervals.
Comparisons with calculations from models described in
Ref.~\cite{Mantysaari:2025tcg} (labeled ``IPGlasma'')
and Ref.~\cite{Giacalone:2024luz} (labeled ``Trento''),
as well as the EPOS model, are also shown.
For the  data, the vertical lines and shaded bars indicate statistical and
systematic uncertainties, respectively.
For the theory points, the shaded bands indicate total uncertainties.
}
\label{fig:results_ptspec_Ratio_Theory}
\end{figure}

Figure~\ref{fig:results_ptspec_Ratio_Theory} shows the ratio of the
\NeNe\ to \OO\ \pT\ spectra, together with comparisons with
theoretical calculations.
Figure~\ref{fig:results_ptspec_Ratio_Theory_aux}
in the Appendix shows similar results for additional
centrality intervals.
Clear differences between data and the
theoretical predictions are seen in the magnitude of the ratios,
indicating deficiencies in the description of the (relative)
overall particle yield. These are
discussed further below.  The measured \NeNe to \OO ratios are found
to be nearly independent of \pT, with the ratios varying by ${<}10\%$ over the measured range.
However, in the more central collisions -- i.e.\ in the top panels --
a slow decrease in the ratio with increasing \pT is seen,
while in the more peripheral intervals -- bottom panels --
a slight increase in the ratio with increasing \pT is observed at
low \pT. The EPOS calculations generally reproduce the trends seen in the
measurements, except in the 20--25\% centrality interval. In contrast,
the IPGlasma- and Trento-based hydrodynamic model calculations
exhibit a much stronger \pt\ dependence in the \NeNe to \OO ratios
than is observed in the data, with the NLEFT and PGCM Trento
calculations showing opposite trends. In central collisions, the data
lie within the range spanned by the NLEFT and PGCM Trento predictions,
suggesting that neither of the Trento models provides the correct initial
conditions for the hydrodynamic evolution.
In the most peripheral centrality interval shown in the figure,
the hydrodynamic calculations disagree with the measurements
but exhibit a weak \pT\ dependence similar to that observed
in the data.
\begin{figure}[!tb]
\centering
\includegraphics[width=0.95\linewidth]{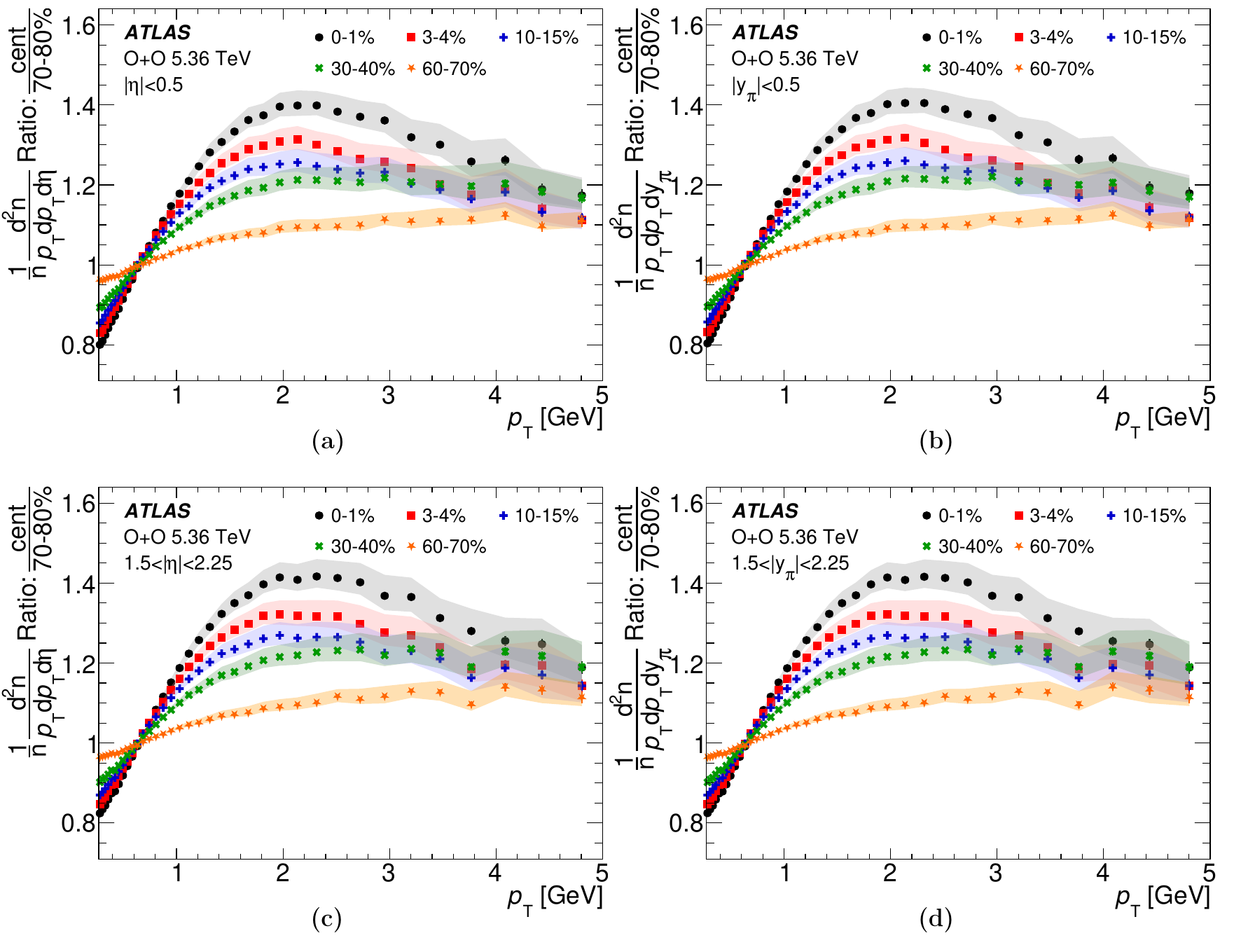}
\caption{
Ratios of the \pt\ spectra measured in \OO\ collisions in
different centrality intervals, as indicated in the legends,
to those measured in the 70--80\% interval.
The results are shown for (a) $|\eta|<0.5$, (b) $|\ypi|<0.5$,
(c) $1.5<|\eta|<2.25$ and (d) $1.5<|\ypi|<2.25$.
The vertical lines and shaded bands indicate statistical and
systematic uncertainties, respectively.
}
\label{fig:results_RatioPeriph_OO}
\end{figure}

\begin{figure}[!tb]
\centering
\includegraphics[width=0.95\linewidth]{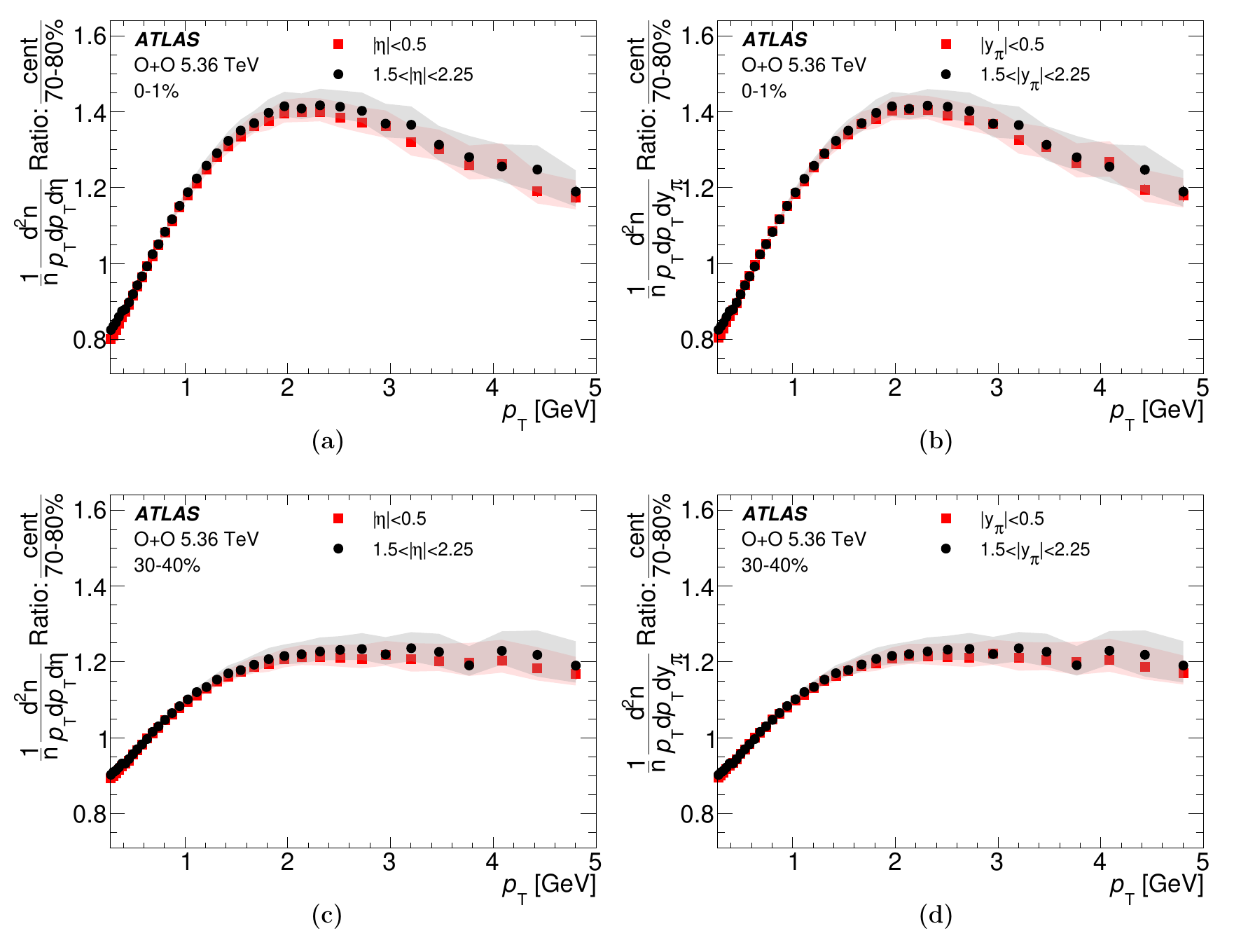}
\caption{
Ratios of the \pt\ spectra in (a, b) 0--1\% and
(c, d) 30--40\% centrality \OO\ collisions
to those measured in the 70--80\% centrality interval.
Each panel compares the ratios at mid-rapidity and forward rapidity.
The results are shown for (a, c) \peta\  and  (b, d) \ypi\ differential measurements.
The vertical lines and shaded bands indicate statistical uncertainties
(often smaller than the size of the data points) and
systematic uncertainties, respectively.
}
\label{fig:results_RatioPeriph_OO_fwd}
\end{figure}

\begin{figure}[!tb]
\centering
\includegraphics[width=0.99\linewidth]{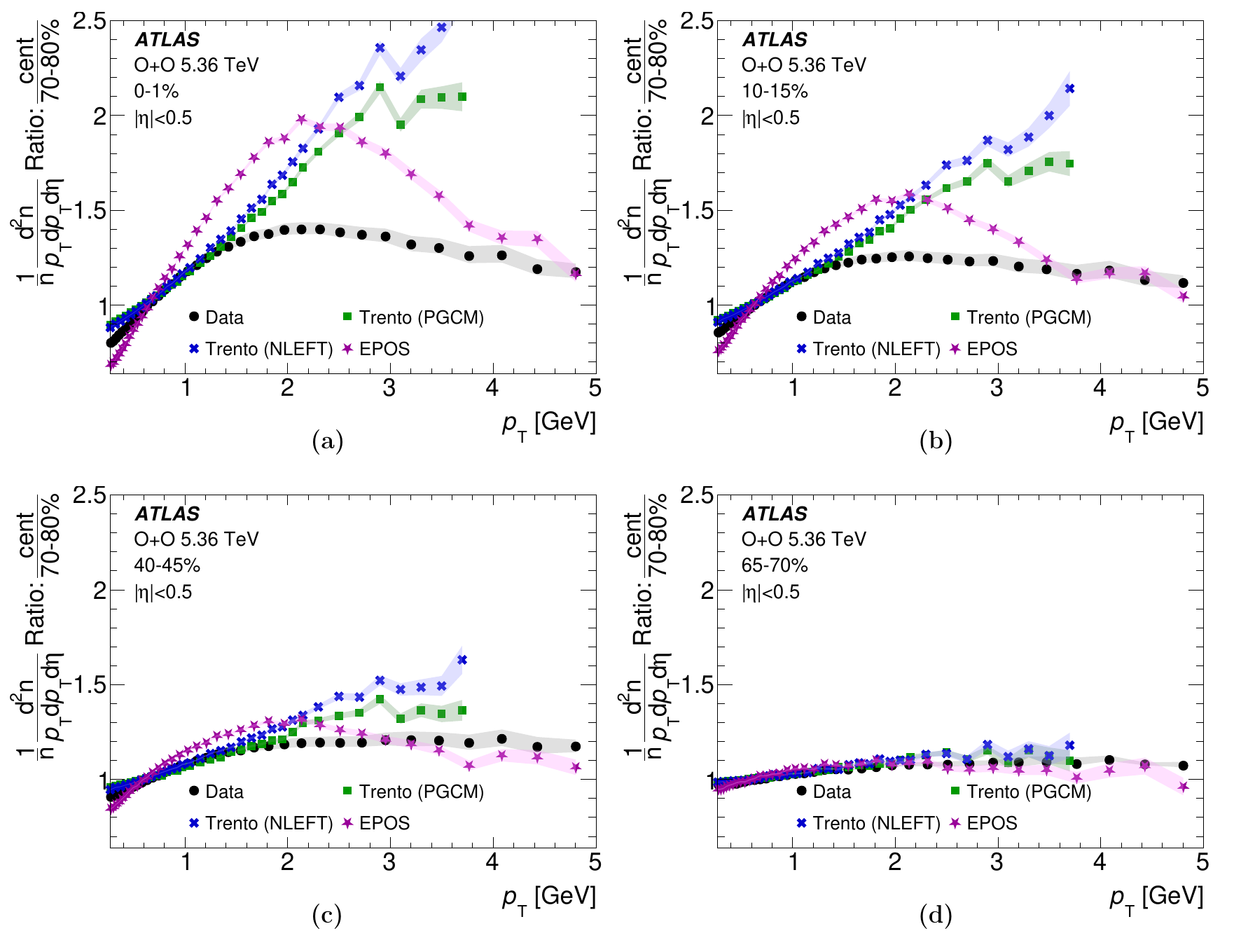}
\caption{
Ratios of the \pt\ spectra in (a) 0--1\%, (b) 10--15\%,
(c) 40--45\% and (d) 65--70\% centrality \OO\ collisions
to those measured in the 70--80\% centrality interval.
Comparisons with the EPOS model and hydrodynamic models based
on IPGlasma and Trento calculations are also shown.
For the data, the vertical lines and shaded bands indicate
statistical uncertainties (always smaller than the size
of the data points) and systematic uncertainties, respectively.
For the theory points, the shaded bands indicate total uncertainties.
}
\label{fig:results_RatioPeriph_OO_TheoryCompare}
\end{figure}

Figure~\ref{fig:results_RatioPeriph_OO} shows the ratios of the \invpTdndpTdeta
spectra in \OO\ collisions at various centralities to that measured
in the 70--80\% peripheral centrality interval.
For these ratios, the spectra are first normalized by the \pt-integrated multiplicity
in the corresponding \peta\ or \ypi\ interval.
This normalization removes the overall change in particle production
between the centrality intervals, allowing the ratios to emphasize changes
in the shape of the \pt\ spectra.
Figures~\ref{fig:results_RatioPeriph_OO_aux} and~\ref{fig:results_RatioPeriphYpi_OO_aux}
in the Appendix show similar results for additional
centrality intervals in \OO\ collisions.
Figures~\ref{fig:results_RatioPeriph_NeNe_aux} and~\ref{fig:results_RatioPeriphYpi_NeNe_aux}
in the Appendix show similar results for \NeNe\ collisions.
From these ratios, it is observed that when going from peripheral to central
collisions, the spectra become
progressively harder, with the yields below $ \pT \sim 0.5$~GeV
suppressed in the more central collisions, and the yields at higher \pT
enhanced. However, the enhancement then decreases with increasing \pT such
that by 5~GeV the ratios have little centrality dependence.
This trend is observed in both the $\eta$-binned measurements and the $\ypi$-binned
measurements, and at both mid-rapidity and forward rapidity.
The observed behavior is consistent with the development of stronger radial
flow in the more central \OO collisions combined with the dominance of
hard processes at the upper end of the measured \pT range.

To more directly test the rapidity dependence of the hardening
of the \pT spectra, Figure~\ref{fig:results_RatioPeriph_OO_fwd}
shows the ratios of the \invpTdndpTdeta (left) and \invpTdndpTdypi (right)
spectra at 0--1\% and 30--40\% centrality to those at 70--80\% centrality at mid-rapidity
($|\eta,\ypi|< 0.5$) and forward rapidity ($1.5 < |\peta,\ypi|
< 2.5$).
Results for additional centrality intervals are shown in the Appendix,
in Figures~\ref{fig:results_RatioPeriph_OO_etacompare_aux}--\ref{fig:results_RatioPeriphYpi_OO_etacompare_aux} for \OO\ collisions
and in Figures~\ref{fig:results_RatioPeriph_NeNe_etacompare_aux}--\ref{fig:results_RatioPeriphYpi_NeNe_etacompare_aux} for \NeNe\ collisions.
The \invpTdndpTdeta and \invpTdndpTdypi ratios are
consistent, indicating that the bias in the \peta distributions at low
\pT cancels out in the ratios. The mid-rapidity and forward-rapidity
ratios are consistent within uncertainties,
indicating that the radial flow developed in the \OO collisions is
empirically observed to be approximately boost-invariant over $|\eta|< 2.5$.

Figure~\ref{fig:results_RatioPeriph_OO_TheoryCompare} compares the
measured \invpTdndpTdeta ratios with those obtained from the model calculations.
Figures~\ref{fig:results_RatioPeriph_OO_TheoryCompare_aux}
and~\ref{fig:results_RatioPeriph_NeNe_TheoryCompare_aux}
in the Appendix show similar results for additional centrality
intervals for \OO\ and \NeNe\ collisions, respectively.
In general, the ratios in the model calculations show a stronger
dependence on \pt\ than observed in the data.
The EPOS ratios reach a maximum and then decrease
with increasing \pT as in the data, but the maximum values are substantially
higher than in the data except in the most peripheral (65--70\%)
centrality interval.
The hydrodynamic calculations keep increasing monotonically,
probably because of the lack of hard contributions.

\subsection{Particle yields and mean transverse momenta}

\begin{figure}[!tb]
\centering
\includegraphics[width=0.80\linewidth]{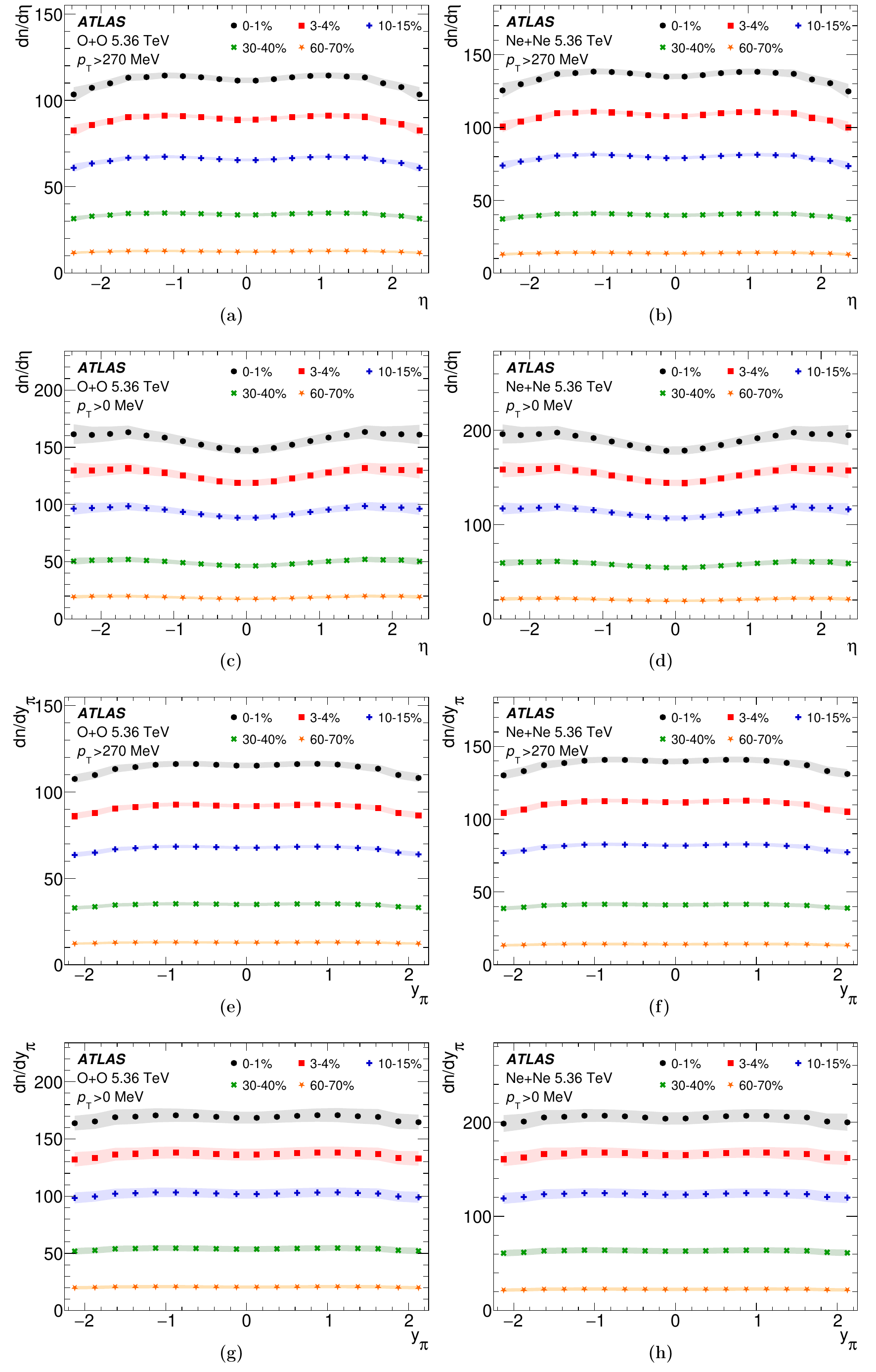}
\caption{
The \dndeta\ distributions in (a, c) \OO  and
(b, d) \NeNe collisions, as a function of $\eta$,
for several centrality intervals as indicated in the legends.
Panels (a, b) show the measurements in the $\pt>270~\MeV$ fiducial
acceptance, and panels (c, d) show the results for the
\pteqo\ extrapolation.
Also shown are the \dndypi\ distributions for (e, g) \OO\ and (f, h) \NeNe\
collisions for the (e, f) $\pt>270~\MeV$ fiducial acceptance and the (g, h) \pteqo\ extrapolation.
The vertical lines and shaded bands indicate statistical and
systematic uncertainties, respectively.
Statistical uncertainties in all cases are smaller than the size
of the data points.
}
\label{fig:results_dndeta_selected}
\end{figure}

\begin{figure}[!tb]
\centering
\includegraphics[width=0.80\linewidth]{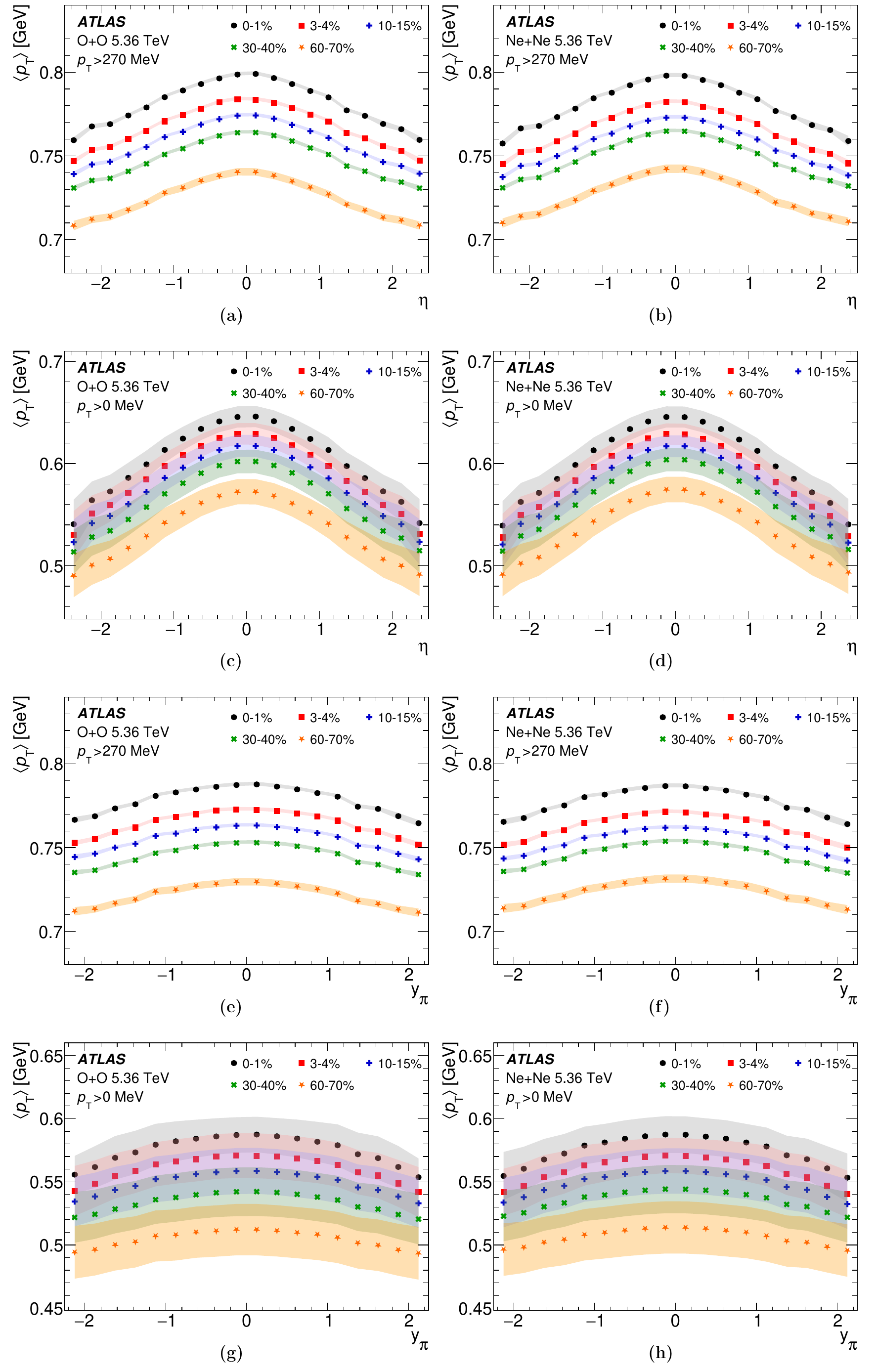}
\caption{
The \meanpt\ distributions as a function of \peta\ in (a, c) \OO\  and
(b, d) \NeNe\ collisions in several centrality intervals as indicated in the legends.
Panels (a, b) show the measurements in the $\pt>270~\MeV$ fiducial
acceptance, and panels (c, d) show the results for the
\pteqo\ extrapolation.
Also shown are the \meanpt\ distributions as a function of \ypi\ for (e, g) \OO\ and (f, h) \NeNe\
collisions for the (e, f) $\pt>270~\MeV$ fiducial acceptance and the (g, h) \pteqo\ extrapolation.
The vertical lines and shaded bands indicate statistical and
systematic uncertainties, respectively.
Statistical uncertainties in all cases are smaller than the size
of the data points.
For some of the centrality intervals the data points are
slightly shifted along the $x$-axis for clarity.
}
\label{fig:results_MeanpT_selected}
\end{figure}

\begin{figure}[!tb]
\centering
\includegraphics[width=0.99\linewidth]{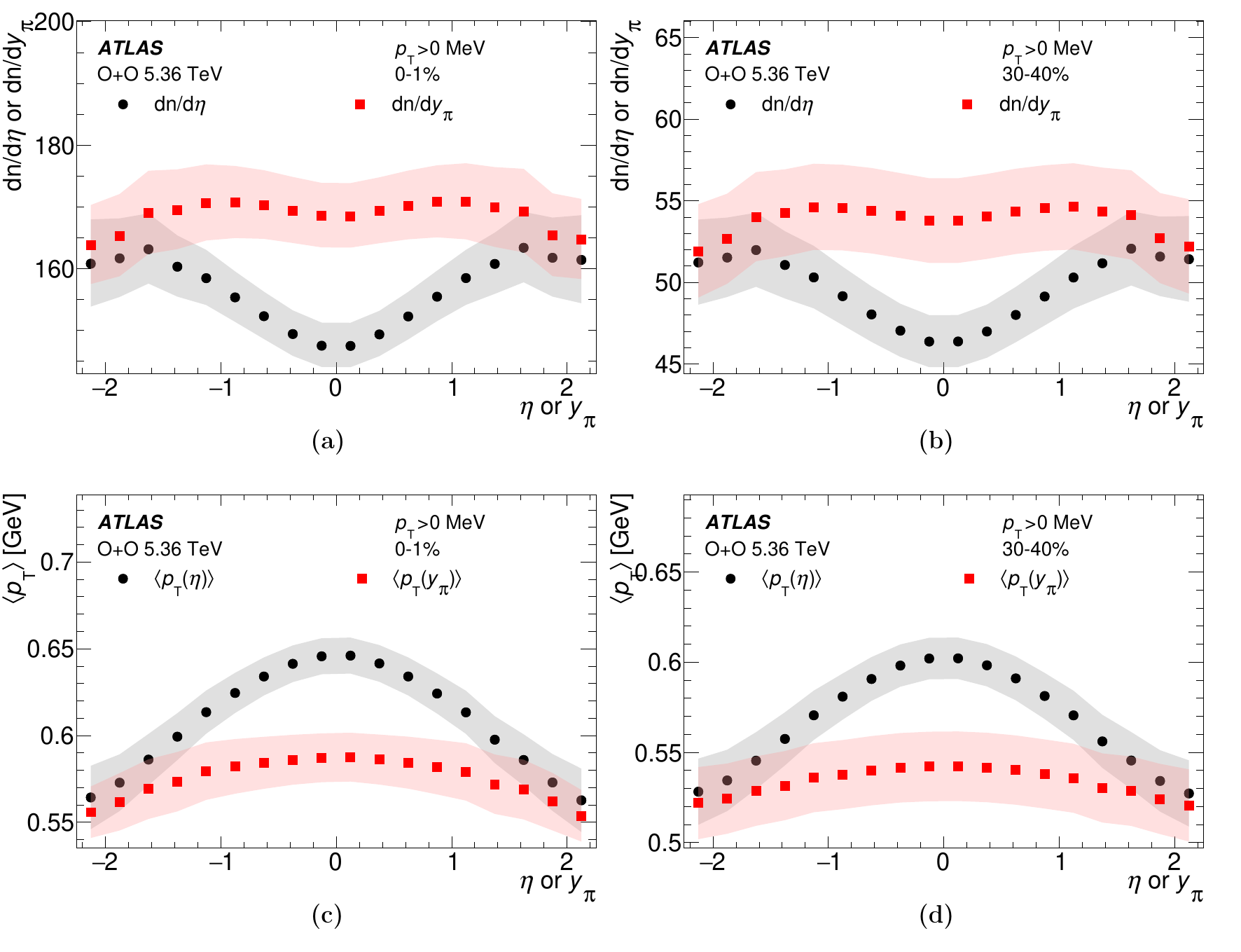}
\caption{
Comparison of the (a, b) \dndeta\ and \dndypi and
(c, d) $\meanpt(\eta)$ and $\meanpt(\ypi)$ distributions,
in the \pteqo\ extrapolation case,
for (a, c) 0--1\% and (b, d) 30--40\% central \OO\ collisions.
The vertical lines and shaded bands indicate statistical and
systematic uncertainties, respectively.
Statistical uncertainties in all cases are smaller than the size
of the data points.
The \ypi-dependence data points are slightly shifted along
the $x$-axis for clarity.
}
\label{fig:results_compare_ypi_eta_dep}
\end{figure}

\afterpage{\FloatBarrier}
Results of the \dndeta\ and \dndypi\ measurements are shown in
Figure~\ref{fig:results_dndeta_selected} for \OO\ and \NeNe\
collisions in several representative centrality intervals.
Results are presented for the fiducial region $\pt>270~\MeV$
as well as for the \pteqo\ extrapolation.
Results for additional centrality intervals are presented
in Figures~\ref{fig:results_dndetaOO_aux}--\ref{fig:results_dndypiOO_ptExtrap0_aux}
in the Appendix.
The \dndeta\ distributions exhibit the expected dependence
on pseudorapidity.
A dip is observed at mid-pseudorapidity, which arises from
presenting the distributions as a function of pseudorapidity
rather than rapidity (see Section~\ref{sec:rapidity_bias}).
This dip is more pronounced for the \pteqo\ extrapolation and
is largely removed when the measurements are plotted as a
function of \ypi.
The \meanpt\ measurements as a function of $\eta$ and \ypi\ are
shown in Figure~\ref{fig:results_MeanpT_selected} for both
the $\pt>270~\MeV$ fiducial region and the \pteqo\ extrapolation.
Results for additional centrality intervals are presented
in Figures~\ref{fig:results_MeanpTvsEtaOO_aux}--\ref{fig:results_MeanpTvsypiOO_ptExtrap0_aux}
in the Appendix.
The \meanpt\ values peak at mid-(pseudo)rapidity and decrease
toward forward rapidities.
In addition, \meanpt\ increases monotonically when going from peripheral
to central collisions.
The \dndeta\ values obtained from the \pteqo\ extrapolation are
approximately 1.4 times larger than the corresponding
fiducial measurements.
In contrast, the \meanpt\ values from the \pteqo\ extrapolation
are approximately 150--200~\MeV lower than the fiducial measurements,
reflecting the inclusion of low-\pt\ particles in the
extrapolated spectra.
Across the measured \peta\ range, the \meanpt\ variation is
$\sim$100~\MeV for the \pteqo\ case.
In comparison, the \ypi-dependent measurements shown in
Figures~\ref{fig:results_MeanpT_selected}(e) and~\ref{fig:results_MeanpT_selected}(f)
exhibit a much weaker dependence, varying by only $\sim$30~\MeV
over the measured \ypi\ range.
As discussed previously, this difference arises from the
kinematic mapping between rapidity and pseudorapidity.
Particles at a given $|\ypi|$ correspond to larger $|\eta|$ values,
with the migration being more pronounced at low \pt.
Consequently, low-\pt\ particles shift further outward when
plotted versus $\eta$, which increases \meanpt\ and reduces
\dndeta\ near $|\eta|\sim0$, leading to the stronger variation
in the results plotted as a function of \peta.
This effect is clearly shown in Figure~\ref{fig:results_compare_ypi_eta_dep},
where the \peta- and \ypi-dependent measurements are plotted together.
Similar comparisons for additional centrality intervals,
as well as for the $\pt>0.27$~\GeV fiducial measurements,
are included in the Appendix in Figures~\ref{fig:results_CompareYandEta_OO_dNdeta_aux}--\ref{fig:results_CompareYandEta_OO_meanpT_ptExtrap0_aux} for \OO\ collisions,
and in Figures~\ref{fig:results_CompareYandEta_NeNe_dNdeta_aux}--\ref{fig:results_CompareYandEta_NeNe_meanpT_ptExtrap0_aux} for \NeNe\ collisions.
\begin{figure}[!tb]
\centering
\includegraphics[width=0.99\linewidth]{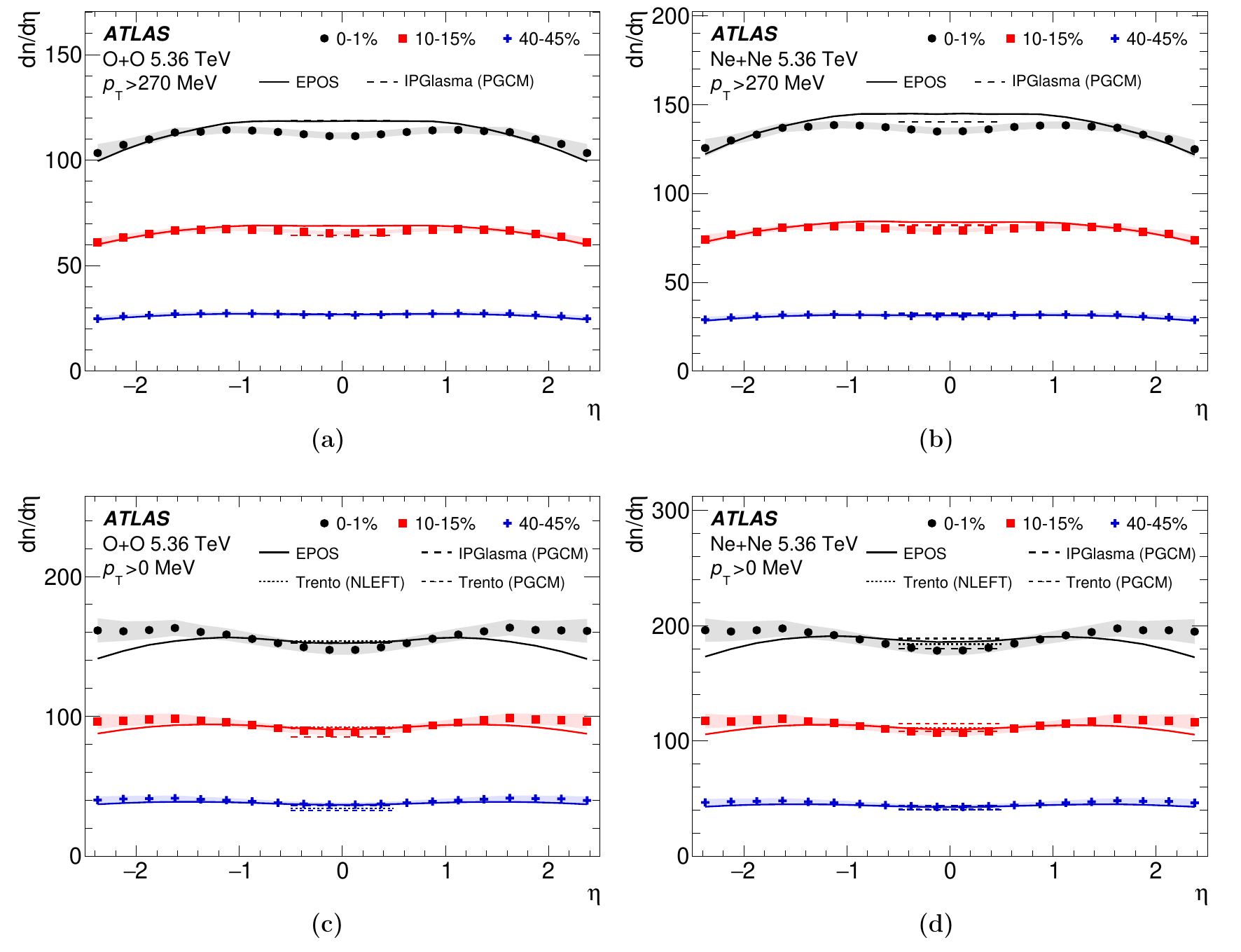}
\caption{
Comparison of the measured \dndeta\ with model calculations for
(a, c) \OO\ and (b, d) \NeNe\ collisions.
Panels (a, b) correspond to the $\pt>0.27$~\GeV fiducial region,
and panels (c, d) to the \pteqo\ extrapolation.
The EPOS calculations are shown over the full $\eta$ range, while
the IPGlasma- and Trento-based hydrodynamic calculations are
averaged over $|\eta|<0.5$ and displayed only in this region.
}
\label{fig:results_dndeta_selected_theory_compare}
\end{figure}

\begin{figure}[!tb]
\centering
\includegraphics[width=0.99\linewidth]{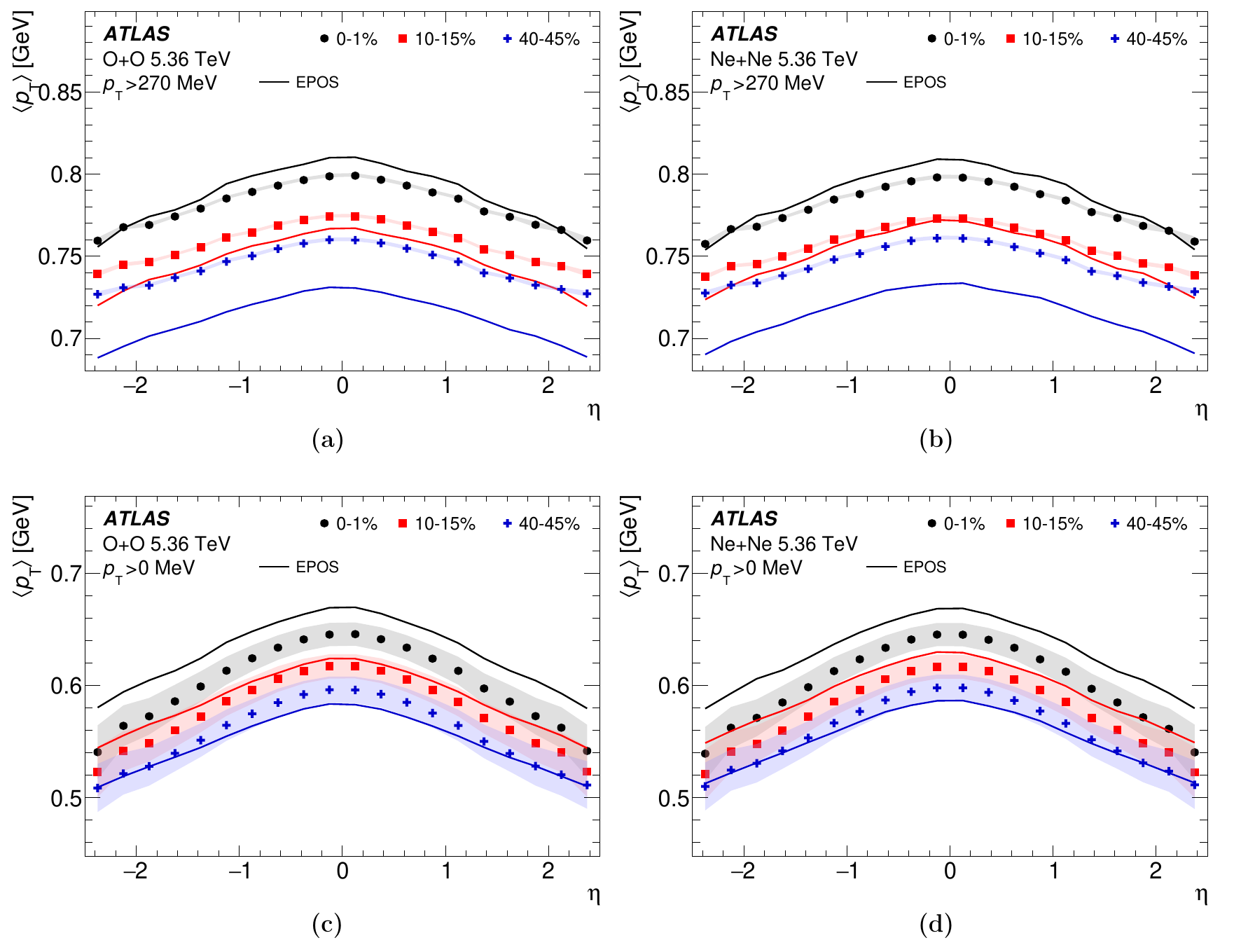}
\caption{
Comparison of the $\eta$ dependence of the measured \meanpt\
with calculations from the EPOS model for (a, c) \OO\ and (b, d) \NeNe\ collisions.
Panels (a, b) correspond to the $\pt>0.27$~\GeV fiducial region,
and panels (c, d) to the \pteqo\ extrapolation.
}
\label{fig:results_meanpt_selected_theory_compare}
\end{figure}

Figure~\ref{fig:results_dndeta_selected_theory_compare} shows
comparisons of the \dndeta\ measurements with calculations
from the EPOS model.
Comparisons with the IPGlasma- and Trento-based hydrodynamic
calculations, averaged over $|\eta|<0.5$ are also shown.
Similar comparisons for additional centrality intervals in \OO\ collisions
are shown in Figure~\ref{fig:results_OO_dNdeta_ptExtrap0_Theory_aux}
and for \NeNe\ collisions are shown in Figure~\ref{fig:results_NeNe_dNdeta_ptExtrap0_Theory_aux}
in the Appendix.
Figure~\ref{fig:results_meanpt_selected_theory_compare}
shows data--EPOS comparisons for \meanpt\ as a function of $\eta$.
Similar comparisons in additional centrality intervals are shown for \OO\ collisions
in Figure~\ref{fig:results_OO_meanpT_ptExtrap0_Theory_aux}
and are shown for \NeNe\ collisions in Figure~\ref{fig:results_NeNe_meanpT_ptExtrap0_Theory_aux}
in the Appendix.
In general, the calculations qualitatively reproduce the trends
seen in the measurements.
More detailed quantitative data--model comparisons for the centrality
dependence of \dndeta\ and \meanpt\ at mid-rapidity ($|\eta|<0.5$)
are discussed in Section~\ref{sec:centdep}.

\subsection{Centrality dependence of particle yields and mean
transverse momenta~\label{sec:centdep}}

\begin{figure}[!tb]
\centering
\includegraphics[width=0.99\linewidth]{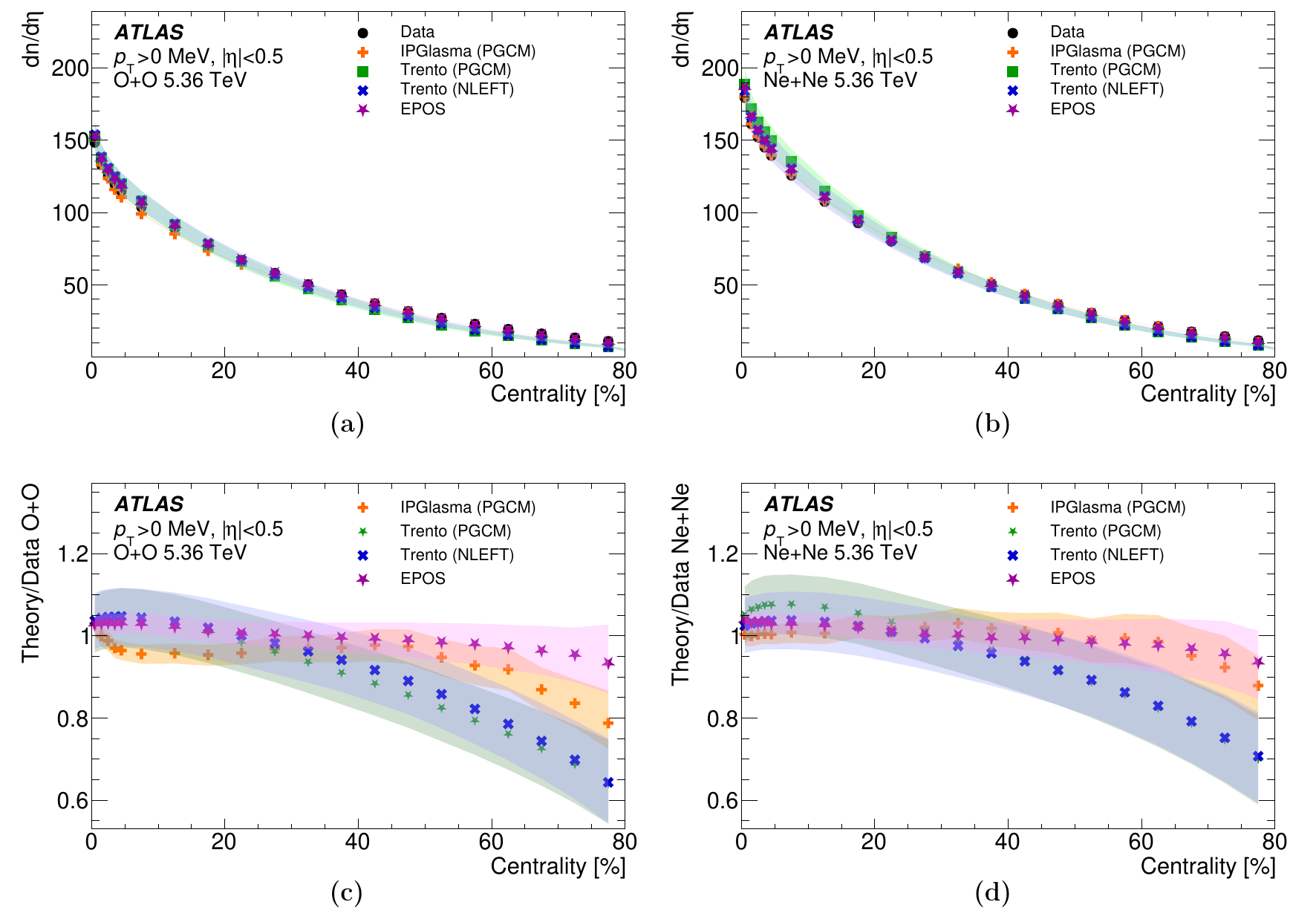}
\caption{
The centrality dependence of \dndeta\ measured at mid-rapidity
($|\eta|<0.5$) in (a) \OO\ and (b) \NeNe\ collisions.
Comparisons with calculations from models described in
Ref.~\cite{Mantysaari:2025tcg} (labeled ``IPGlasma'') and
Ref.~\cite{Giacalone:2024luz} (labeled ``Trento'') as well as the
EPOS model are also shown.
Theory-to-data ratios are shown for (c) \OO\ and (d) \NeNe\ collisions.
For the data in panels (a, b), the vertical lines and shaded bars
indicate statistical and systematic uncertainties, respectively,
and in all cases are smaller than the size of the data points.
For the theory points in panels (a, b) and the theory-to-data ratios
in panels (c, d), the shaded bands indicate combined
(statistical + systematic) uncertainties.
}
\label{fig:results_dndetavsCent_ptExtrap0}
\end{figure}

Figure~\ref{fig:results_dndetavsCent_ptExtrap0} shows the centrality
dependence of \dndeta\ measured at mid-rapidity
($|\eta|<0.5$) for the \pteqo\ extrapolation.
Comparisons with model calculations are also included.
As is typical in nucleus--nucleus collisions, the charged-particle
multiplicity rises most steeply toward the most central collisions.
Overall, the models reproduce the observed centrality dependence
reasonably well.
For both the \OO\ and \NeNe\ collisions, the Trento-based hydrodynamic
calculations agree with the measurements within $\sim$1$\sigma$
uncertainties over the 0--60\% centrality range.
The IPGlasma-based hydrodynamic calculations and the EPOS model
show even better agreement with the data.
This is illustrated by the theory-to-data ratios shown in the
lower panels of Figure~\ref{fig:results_dndetavsCent_ptExtrap0}.
In peripheral collisions, noticeable discrepancies between
data and the Trento- and IPGlasma-based model calculations emerge.
These differences are larger for \OO\ than for \NeNe\ collisions, and
are more pronounced for the Trento model than for the IPGlasma model.
Part of the discrepancy may arise from the different rapidity regions
used to define centrality in the data and in the Trento-
and IPGlasma-based hydrodynamic model calculations.
In the data, centrality is determined using forward-calorimeter
transverse energy.
For the EPOS model comparisons, the centrality was determined using
the generated final particles in the FCal acceptance.
However, the Trento and IPGlasma based model calculations shown here
determine centrality by using charged-particle multiplicity at mid-rapidity.
Similar comparisons for the $\pt>0.27$~\GeV\ fiducial region are
shown in Figure~\ref{fig:results_dndetavsCent_aux} in the Appendix.
The fiducial measurements show similar trends.
\begin{figure}[!tb]
\centering
\includegraphics[width=0.99\linewidth]{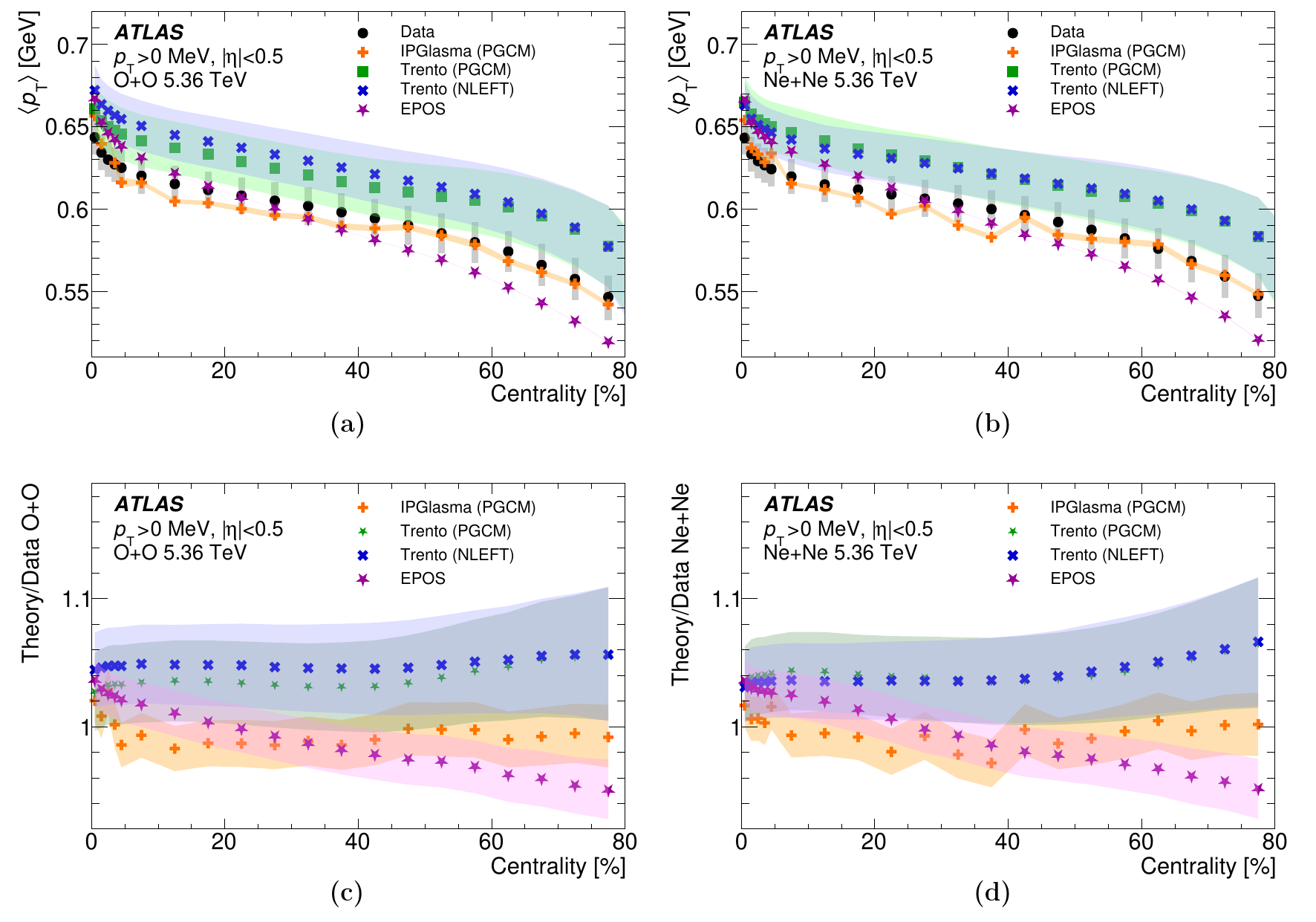}
\caption{
The centrality dependence of \meanpt\ measured at mid-rapidity
($|\eta|<0.5$) in (a) \OO\ and (b) \NeNe\ collisions.
Comparisons with calculations from the EPOS model and models described in
Ref.~\cite{Mantysaari:2025tcg} (labeled ``IPGlasma'') and
Ref.~\cite{Giacalone:2024luz} (labeled ``Trento'') are also shown.
Theory-to-data ratios are shown for (c) \OO\ and (d) \NeNe\ collisions.
For the data in panels (a, b), the vertical lines and vertical bands
indicate statistical and systematic uncertainties, respectively.
Statistical uncertainties in all cases are smaller than the size
of the data points.
For the theory points in panels (a, b) and the theory-to-data ratios
in panels (c, d), the shaded bands indicate combined
(statistical + systematic) uncertainties.
}
\label{fig:results_meanpTvsCent_ptExtrap0}
\end{figure}

Figure~\ref{fig:results_meanpTvsCent_ptExtrap0} shows the centrality
dependence of \meanpt\ at mid-rapidity ($|\eta|<0.5$) for the
\pteqo\ extrapolation, together with model comparisons.
In contrast to the \dndeta\ results, the \meanpt\ values are nearly
identical for \OO\ and \NeNe\ collisions.
At a given centrality, the differences between the two systems are
within 0.5\% (less than 5~\MeV) over the full measured centrality range.
The \meanpt\ decreases monotonically from central to peripheral
collisions, with a total variation of $\sim$100~\MeV\
across the centrality range.
The model calculations reproduce the observed centrality dependence
well and typically agree with the data within $\pm5\%$.
The Trento-based hydrodynamic calculations predict \meanpt\ values
slightly higher than those observed in the data.
The EPOS model calculations transition from higher values than in data in
central collisions to lower values in peripheral
collisions.
The IPGlasma-based hydrodynamic calculations agree best
with the measurements.
Similar comparisons for the $\pt>0.27$~\GeV\ fiducial region are
shown in Figure~\ref{fig:results_meanpTvsCent_aux} in the Appendix.
The fiducial measurements show similar trends.
\begin{figure}[!tb]
\centering
\includegraphics[width=0.99\linewidth]{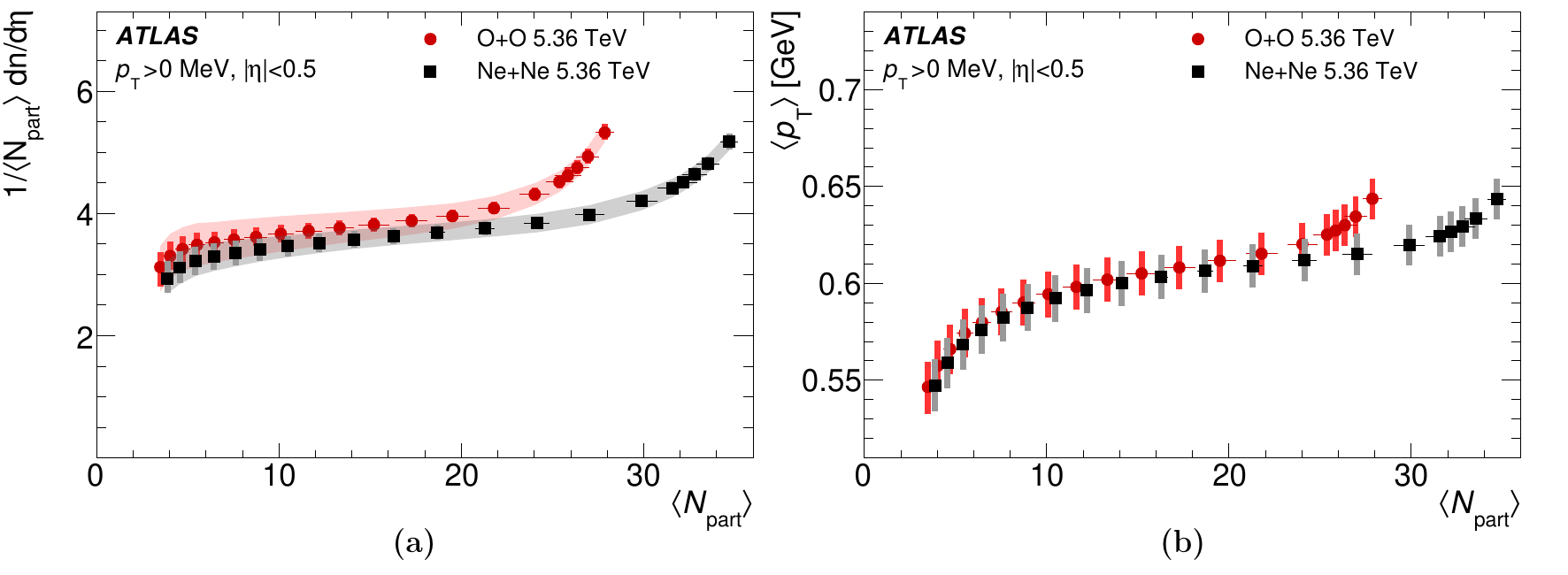}
\caption{
The \npart\ dependence of (a) $(1/\npart) \dndeta$ and (b) \meanpt\ measured
at mid-rapidity ($|\eta|<0.5$).
The vertical lines and shaded bars indicate statistical and
systematic uncertainties, respectively.
In panel (a), the shaded bands indicate the total systematic uncertainty,
which additionally includes the uncertainty in the \npart\ determination.
Statistical uncertainties in all cases are smaller than the size
of the data points.
}
\label{fig:results_npart_ptExtrap0}
\end{figure}
\begin{figure}[!tb]
\centering
\includegraphics[width=0.99\linewidth]{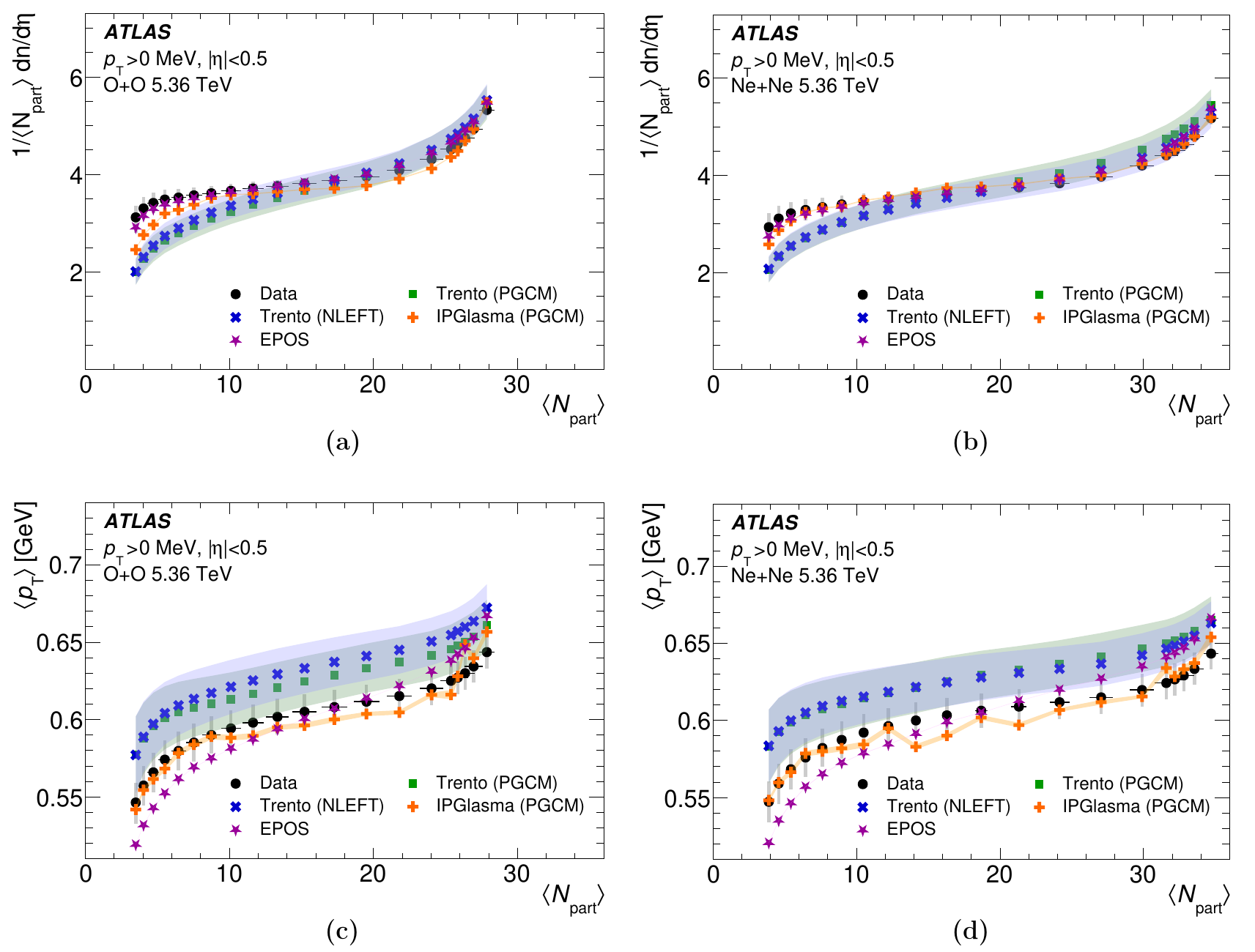}
\caption{
The \npart\ dependence of the (a, b) $(1/\npart) \;\!\dndeta$ and (c, d) \meanpt\ measured
at mid-rapidity ($|\eta|<0.5$) compared with model calculations,
for (a, c) \OO\ and (b, d) \NeNe\ collisions.
For the data, the vertical lines and shaded bars
indicate statistical and systematic uncertainties, respectively.
The statistical uncertainties in all cases are smaller than the size
of the data points.
For the theory points, the shaded bands indicate combined
(statistical + systematic) uncertainties.
}
\label{fig:results_npart_theorycompare_ptExtrap0}
\end{figure}

Figure~\ref{fig:results_npart_ptExtrap0} shows the dependence of
\dndeta\ and \meanpt\ measured at mid-rapidity on the number of
participating nucleons in the collisions, \npart.
For \npart\ values below approximately 25, corresponding to the 10\%
and more peripheral centrality intervals, the measured $(1/\npart) \;\!\dndeta$
in \OO\ and \NeNe\ are consistent, within uncertainties.
In this region, the \dndeta\ dependence is approximately linear in
\npart.
At larger \npart, the \dndeta and \meanpt values
rise more sharply as the maximum \Npart values are approached.
This rise is consistent with the behavior observed in ultracentral collisions in
\PbPb\ and \XeXe collisions~\cite{Gardim:2019brr,HION-2021-20}, where for small impact parameters the
transverse area of the medium becomes constant, but initial-state
fluctuations increase the energy density leading to enhanced radial
flow and, thus, higher \meanpt.
Figure~\ref{fig:results_npart_theorycompare_ptExtrap0} compares the
\npart\ dependence of the \dndeta\ and \meanpt\ measurements at
mid-rapidity with model calculations.
The trends observed here are similar to those in
Figures~\ref{fig:results_dndetavsCent_ptExtrap0} and~\ref{fig:results_meanpTvsCent_ptExtrap0}:
for \dndeta, the models are closer to the data in central collisions
but underestimate it in peripheral collisions, while for \meanpt,
the data lie between the different model predictions.
\begin{figure}[!tb]
\centering
\includegraphics[width=0.99\linewidth]{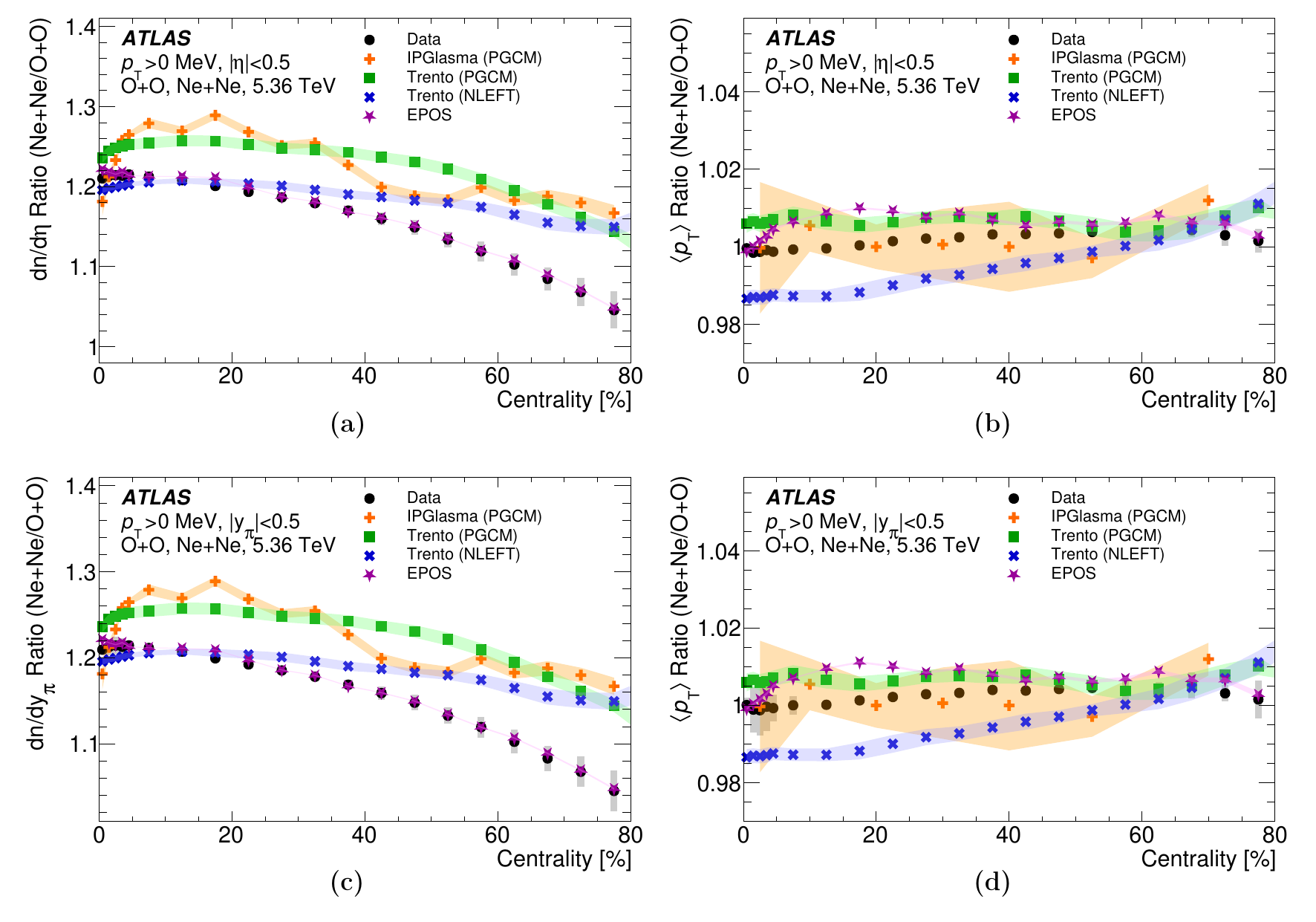}
\caption{
The centrality dependence of the ratio \NeNe/\OO for (a) \dndeta\
and (b) \meanpt, measured at mid-pseudorapidity ($|\eta|<0.5$).
Similar results are also shown for (c) \dndypi\ and (d) \meanpt,
measured at mid-rapidity ($|\ypi|<0.5$).
Comparisons with calculations from the EPOS model and models described in
Ref.~\cite{Mantysaari:2025tcg} (labeled ``IPGlasma'')
and Ref.~\cite{Giacalone:2024luz} (labeled ``Trento'') are also shown.
The model calculations in all cases are for the $|\eta|<0.5$ case,
and are identical in the upper and lower panels.
For the data, the vertical lines and shaded bars
indicate statistical and systematic uncertainties, respectively.
Statistical uncertainties in all cases are smaller than the size
of the data points.
For the theory points, the error bars indicate combined
(statistical + systematic) uncertainties.
}
\label{fig:results_ratio_ptExtrap0}
\end{figure}
Figure~\ref{fig:results_ratio_ptExtrap0} shows the \NeNe/\OO\ ratios
for \dndeta, \dndypi, and \meanpt\ measured at mid-rapidity,
for results extrapolated to \pteqo.
Figures~\ref{fig:results_dndeta_ratio_aux}--\ref{fig:results_meanpTvsypi_ratio_aux}
in the Appendix show similar comparisons
for the $\pt>0.27$~\GeV\ fiducial region, together with the
measured \dndeta, \dndypi, and \meanpt\ values, as a function of centrality.
For \dndeta, the ratio is approximately 1.2 in 0--1\% centrality
collisions, increases slightly over the 0--5\% centrality range, and
then decreases monotonically to approximately 1.04 by the 75--80\%
centrality interval.
Comparisons with model calculations are also shown in Figure~\ref{fig:results_ratio_ptExtrap0}.
In the IPGlasma- and Trento-based hydrodynamic model calculations,
the increase in the ratio over the 0--5\% centrality
collisions is significantly larger, particularly for the IPGlasma
model, and persists over the 0--20\% centrality range.
The subsequent decrease toward peripheral collisions is substantially
weaker than that observed in the data.
The EPOS model calculations are consistent with the data throughout
the measured multiplicity range.
For the \meanpt\ ratios, the data exhibit variations
typically less than $\pm1\%$ across the measured centrality range.
The measured ratios lie between the different model predictions.
The results for $|\ypi|<0.5$ are nearly identical to the
measurements for $|\eta|<0.5$.
For these ratios, the measurements for the $\pt>0.27$~\GeV\ fiducial region are also
nearly identical to the \pteqo\ extrapolation case as shown
in Figures~\ref{fig:results_dndeta_ratio_aux}--\ref{fig:results_meanpTvsypi_ratio_aux}.
The uncertainty in these measured ratios is smaller than the differences
between predictions obtained with different nuclear-structure models
(PGCM and NLEFT), indicating that these measurements are sensitive to
the modeling of the nuclear structure.
%


%

%
%
\FloatBarrier

\section{Conclusion}
\label{sec:conclusion}

This \paper presents measurements of charged-particle \pt\ spectra,
multiplicity, and mean transverse momentum in \OO\ and \NeNe\ collisions
at 5.36~\TeV per nucleon pair, using the ATLAS detector at the LHC.
The integrated luminosities of the \OO\ and \NeNe\ data sets
used in this measurement are $27.7~\mu\text{b}^{-1}$ and
$53.1~\mu\text{b}^{-1}$, respectively.
The measurements are performed using charged-particle tracks
reconstructed in the inner detector over the fiducial phase space
$0.27 < \pt < 5$~GeV and $|\eta|<2.5$.
Results are reported as a function of pseudorapidity and collision
centrality, where the latter is characterized by the total
transverse energy measured in the forward calorimeters.
Reconstruction efficiencies and fake- and secondary-track contributions
are corrected using Monte Carlo simulations, and a comprehensive
evaluation of systematic uncertainties is performed.
Systematic uncertainties in the measurements arise from the track
selection, detector material description, centrality determination,
and procedure for extrapolation to $\pteqo$.

Measurements are also performed as a function of rapidity \ypi computed
for the charged-pion mass hypothesis.
The use of \ypi\ significantly reduces effects arising from the
difference between pseudorapidity and true rapidity, particularly
near mid-rapidity, providing results closer to those that would be
obtained using true rapidity.
The dynamics of particle production are expected to be weakly
dependent on rapidity, a feature reflected in the measurements
presented as a function of \ypi.

Ratios of the \pt\ spectra between central and peripheral collisions are measured.
These indicate a hardening of the \pT\ spectra from peripheral to central
collisions for $\pT\lesssim2$~\GeV, and demonstrate the buildup of radial flow.
This buildup is found to be similar at mid-rapidity and forward rapidity,
consistent with the approximate boost invariance of radial flow.
Measurements of the \NeNe/\OO\ ratio of the spectra for a given centrality
are found to be nearly independent of \pT.

The charged-particle multiplicity and mean transverse momentum are
both measured within the fiducial acceptance and extrapolated to \pteqo,
using fits to the measured \pt spectra.
The results show a smooth evolution of \dndeta\ and \meanpt\ with
centrality and (pseudo-)rapidity in both collision systems.
For a given centrality interval, \NeNe\ collisions exhibit larger
charged-particle multiplicities than \OO\ collisions.
The \NeNe/\OO\ ratio of \dndeta\ values varies between approximately
1.05 and 1.20, while the corresponding ratio of \meanpt\ values varies
between approximately 0.99 and 1.01.
These ratios are consistent between the \peta- and \ypi-based measurements.
The \NeNe/\OO\ ratios are also found to be consistent between the fiducial
and \pteqo\ measurements.
The multiplicities and \meanpt\ exhibit a significantly weaker dependence
as a function of \ypi\ than as a function of \peta, reflecting
the approximate boost invariance of particle production.

The measurements are compared with hydrodynamic model calculations
in which the initial energy densities are computed using the
IPGlasma and Trento frameworks.
These frameworks incorporate state-of-the-art nuclear-structure
models for oxygen and neon to provide initial conditions for
simulations of the quark--gluon plasma evolution.
The IPGlasma calculations are coupled to the MUSIC hydrodynamic
package, while the Trento calculations use the Trajectum package.
Comparisons with the EPOS event generator are also made.
The model calculations generally reproduce the observed trends,
but exhibit some deviations from the data
that increase for peripheral collisions.
Among the models considered, EPOS provides the best overall
agreement with the data, while IPGlasma-based predictions
show slightly better agreement than those based on Trento.
Comparisons of the \NeNe/\OO\ ratios for the \pt\ spectra, \dndeta\ and \meanpt\ with
Trento calculations using the PGCM and NLEFT nuclear-structure models
demonstrate that the measurements presented here provide sensitivity to
differences between nuclear-structure models.

These measurements constitute the first detailed ATLAS results for
charged-particle production in \OO\ and \NeNe\ collisions at the
LHC.
They provide important experimental constraints on models of particle
production and collective behavior in small- and intermediate-sized
nuclear systems, and complement existing measurements in \pp, \pA,
and heavy-ion collisions.
Together, the \dndeta\ and \meanpt\ measurements reported here provide
a baseline for understanding particle production and collective
effects in light-ion collisions, and contribute to a more complete
picture of the onset and evolution of quark--gluon plasma dynamics
as a function of system size.


%

%
%
%
%
%
\section*{Acknowledgments}
%

%

%
%

%
%

We thank CERN for the very successful operation of the LHC and its injectors, as well as the support staff at
CERN and at our institutions worldwide without whom ATLAS could not be operated efficiently.

The crucial computing support from all WLCG partners is acknowledged gratefully, in particular from CERN, the ATLAS Tier-1 facilities at TRIUMF/SFU (Canada), NDGF (Denmark, Norway, Sweden), CC-IN2P3 (France), KIT/GridKA (Germany), INFN-CNAF (Italy), NL-T1 (Netherlands), PIC (Spain), RAL (UK) and BNL (USA), the Tier-2 facilities worldwide and large non-WLCG resource providers. Major contributors of computing resources are listed in Ref.~\cite{ATL-SOFT-PUB-2026-001}.

We gratefully acknowledge the support of ANPCyT, Argentina; YerPhI, Armenia; ARC, Australia; BMWFW and FWF, Austria; ANAS, Azerbaijan; CNPq and FAPESP, Brazil; NSERC, NRC and CFI, Canada; CERN; ANID, Chile; CAS, MOST and NSFC, China; Minciencias, Colombia; MEYS CR, Czech Republic; DNRF and DNSRC, Denmark; IN2P3-CNRS and CEA-DRF/IRFU, France; SRNSFG, Georgia; BMFTR, HGF and MPG, Germany; GSRI, Greece; RGC and Hong Kong SAR, China; ICHEP and Academy of Sciences and Humanities, Israel; INFN, Italy; MEXT and JSPS, Japan; CNRST, Morocco; NWO, Netherlands; RCN, Norway; MNiSW, Poland; FCT, Portugal; MNE/IFA, Romania; MSTDI, Serbia; MSSR, Slovakia; ARIS and MVZI, Slovenia; DSI/NRF, South Africa; MICIU/AEI, Spain; SRC and Wallenberg Foundation, Sweden; SERI, SNSF and Cantons of Bern and Geneva, Switzerland; NSTC, Taipei; TENMAK, T\"urkiye; STFC/UKRI, United Kingdom; DOE and NSF, United States of America.

Individual groups and members have received support from BCKDF, CANARIE, CRC and DRAC, Canada; CERN-CZ, FORTE and PRIMUS, Czech Republic; COST, ERC, ERDF, Horizon 2020 and Marie Sk{\l}odowska-Curie Actions, European Union; Investissements d'Avenir Labex, Investissements d'Avenir Idex and ANR, France; DFG and AvH Foundation, Germany; Herakleitos, Thales and Aristeia programmes co-financed by EU-ESF and the Greek NSRF, Greece; BSF-NSF and MINERVA, Israel; NCN and NAWA, Poland; La Caixa Banking Foundation, CERCA and AGAUR programs from Generalitat de Catalunya and PROMETEO and GenT Programmes Generalitat Valenciana, Spain; G\"{o}ran Gustafssons Stiftelse, Sweden; The Royal Society and Leverhulme Trust, United Kingdom; Eric and Wendy Schmidt Fund for Strategic Innovation, United States of America.

In addition, individual members wish to acknowledge support from Chile: Agencia Nacional de Investigaci\'on y Desarrollo (ANID FONDECYT reg. 1230987, FONDECYT 1230812, FONDECYT 1240864, Fondecyt 3240661, Fondecyt Regular 1240721); China: Chinese Ministry of Science and Technology (MOST-2023YFA1605700, MOST-2023YFA1609300), National Natural Science Foundation of China (NSFC 12275265, NSFC-W2543005); Czech Republic: Czech Science Foundation (GACR - 24-11373S), Ministry of Education Youth and Sports (ERC-CZ-LL2327, FORTE CZ.02.01.01/00/22\_008/0004632), PRIMUS Research Programme (PRIMUS/21/SCI/017); EU: H2020 European Research Council (ERC - 101002463); European Union: European Research Council (BARD No. 101116429, ERC - 101219398, ERC - 948254, ERC 101089007), European Regional Development Fund (HE COFUND GA No.101081355, ERDF), Marie Sklodowska-Curie Actions (GAP-101168829); France: Agence Nationale de la Recherche (ANR-21-CE31-0013, ANR-22-EDIR-0002, ANR-24-CE31-0504-01); Germany: Deutsche Forschungsgemeinschaft (DFG - 469666862); China: Research Grants Council (GRF); Italy: Istituto Nazionale di Fisica Nucleare (LHC-MIUR - 28003/2025), Ministero dell'Università e della Ricerca (NextGenEU 153D23001490006 M4C2.1.1, NextGenEU I53D23000820006 M4C2.1.1, SOE2024\_0000023); Japan: Japan Society for the Promotion of Science (JSPS KAKENHI  JP25H0063, JSPS KAKENHI JP22H01227, JSPS KAKENHI JP22H04944, JSPS KAKENHI JP22KK0227, JSPS KAKENHI JP24K23939, JSPS KAKENHI JP24KK0251, JSPS KAKENHI JP25H00650, JSPS KAKENHI JP25H01291, JSPS KAKENHI JP25K01011, JSPS KAKENHI JP25K01023); Poland: Polish National Science Centre (NCN 2021/42/E/ST2/00350, NCN OPUS 2023/51/B/ST2/02507, NCN OPUS nr 2022/47/B/ST2/03059, NCN UMO-2019/34/E/ST2/00393, UMO-2022/47/O/ST2/00148, UMO-2023/49/B/ST2/04085, UMO-2023/51/B/ST2/00920, UMO-2024/53/N/ST2/00869); Spain: Agència de Gestió d'Ajuts Universitaris i de Recerca. (AGAUR - 2023 BP 00141), Ministry of Science and Innovation (RYC2019-028510-I, RYC2020-030254-I, RYC2021-031273-I, RYC2022-038164-I), Ministerio de Ciencia, Innovación y Universidades/Agencia Estatal de Investigaci\'on (EU NextGenerationEU (PRTR-C17.I1), PID2022-142604OB-C22); Sweden: Carl Trygger Foundation (Carl Trygger Foundation CTS 22:2312), Swedish Research Council (Swedish Research Council 2023-04654, VR 2021-03651, VR 2022-03845, VR 2022-04683, VR 2023-03403, VR 2024-05451, VR 2025-05940), Knut and Alice Wallenberg Foundation (KAW 2023.0366); Switzerland: Swiss National Science Foundation (SNSF - PCEFP2\_194658); United Kingdom: The Binks Trust, Royal Society (NIF-R1-231091); United States of America: U.S. Department of Energy (ECA DE-AC02-76SF00515), John Templeton Foundation (John Templeton Foundation 63206), Neubauer Family Foundation.

%
%


%
\clearpage
\appendix
\part*{Appendix}
\addcontentsline{toc}{part}{Appendix}

\begin{figure}[h]
\centering
\includegraphics[width=0.99\linewidth]{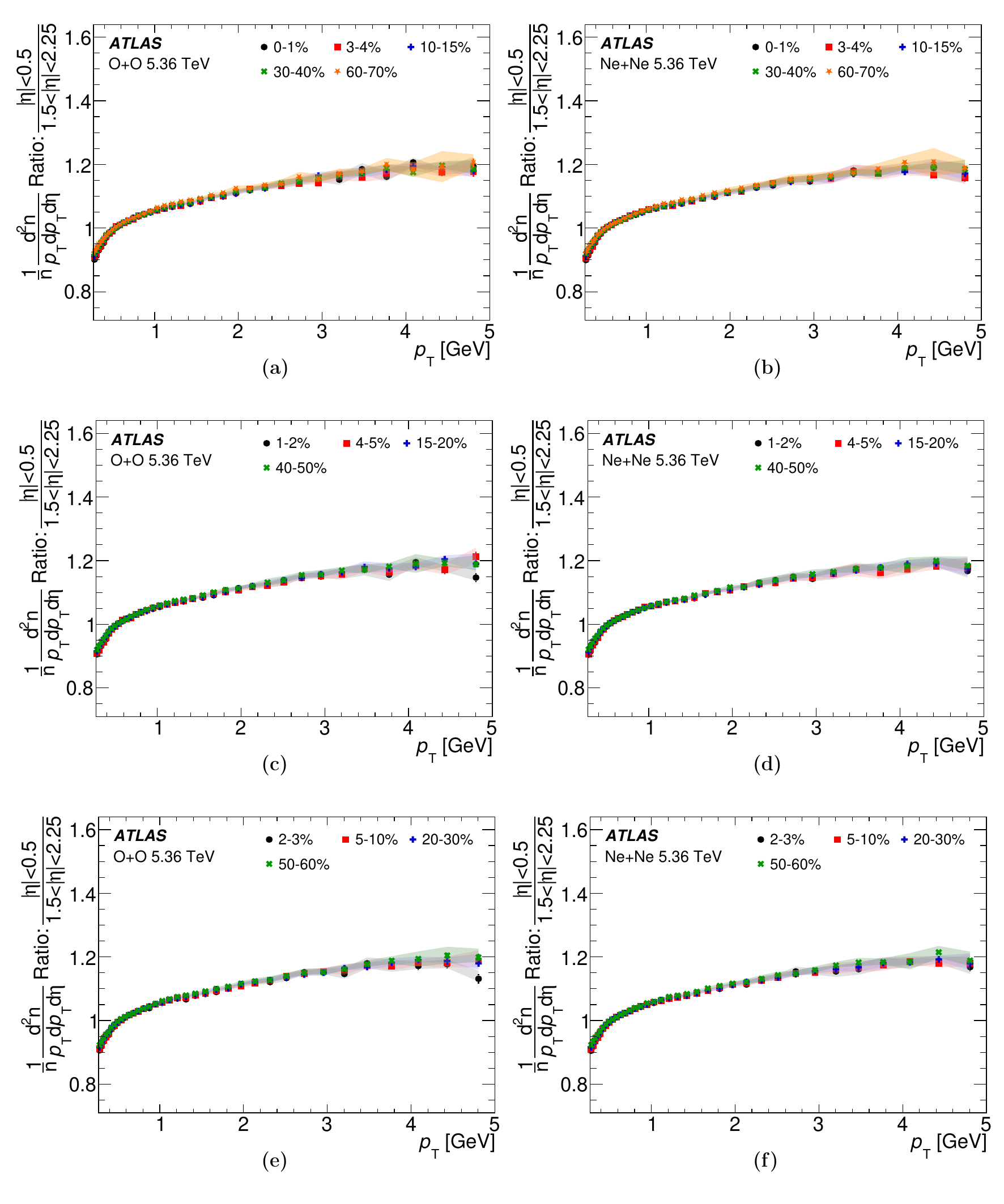}
\caption{
Comparisons between ratios of \pt\ spectra measured over $|\eta|<0.5$
to those measured over $1.5<|\eta|<2.25$,
for (a, c, e) \OO\ and (b, d, f) \NeNe\ collisions at various centralities.
The vertical lines and shaded bands indicate statistical and
systematic uncertainties, respectively.
The statistical uncertainties in most cases are smaller
than the size of the data points.
}
\label{fig:results_RefEtaRatio_aux}
\end{figure}

\begin{figure}[h]
\centering
\includegraphics[width=0.99\linewidth]{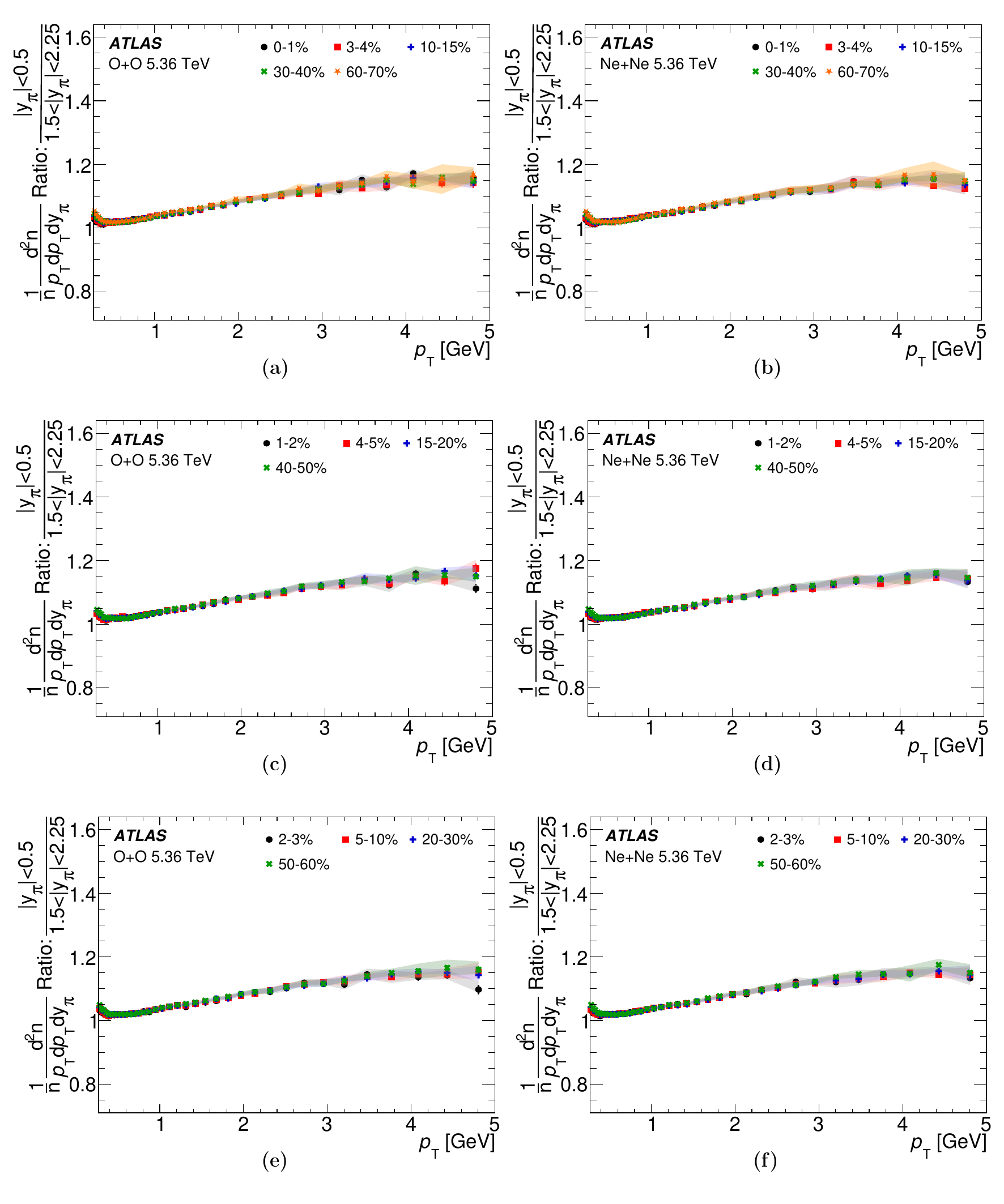}
\caption{
Comparisons between ratios of \pt\ spectra measured over $|\ypi|<0.5$
to those measured over $1.5<|\ypi|<2.25$ for (a, c, e) \OO\
and (b, d, f) \NeNe\ collisions at various centralities.
The vertical lines and shaded bands indicate statistical and
systematic uncertainties, respectively.
The statistical uncertainties in most cases are smaller
than the size of the data points.
This figure is similar to Figure~\ref{fig:results_RefEtaRatio_aux},
but compares the spectra in different intervals of $\ypi$.
}
\label{fig:results_RefEtaRatioYpi_aux}
\end{figure}

Figure~\ref{fig:results_RefEtaRatio_aux} shows the ratios of the \pt\ spectra measured
in $|\eta|<0.5$ to those measured in $1.5<|\eta|<2.25$,
for \OO\ and \NeNe\ collisions.
Figure~\ref{fig:results_RefEtaRatioYpi_aux} shows the corresponding ratios for spectra measured
in intervals of \ypi, comparing $|\ypi|<0.5$ with $1.5<|\ypi|<2.25$.
In Figure~\ref{fig:results_RefEtaRatio_aux}, a depletion is observed at $\pt\lesssim0.6~\GeV$,
which is absent in Figure~\ref{fig:results_RefEtaRatioYpi_aux}.
This depletion is an artifact of presenting the results in
intervals of $\eta$.
It arises from the $\eta$-migration discussed in Section~\ref{sec:rapidity_bias},
which is enhanced at low \pt\ and near $\eta\sim0$, and modifies the shape of the \pt\ spectra.

The remaining figures and tables in the Appendix support results shown in the
main body of the paper and are referenced there;
they are therefore not discussed further here.

\begin{table}[h]
\caption{
The values of \npart\ and their uncertainties, $\delta\npart$, for the
centrality intervals used in this analysis.
The values are obtained from a Glauber model analysis
detailed in the text.
\label{tab:npart}
}
\centering
\begin{tabular}{|c|| r|r|| r|r|}\hline
& \multicolumn{2}{c|}{\OO}       & \multicolumn{2}{c|}{\NeNe}    \\ \hline
Centrality  & $\npart$ & $\delta\npart$      & \npart    & $\delta\npart$    \\
$[\%]$    &          &                     &           &                   \\ \hline
0--1 & 27.88 & 0.52 & 34.71 & 0.50 \\
1--2 & 26.96 & 0.63 & 33.55 & 0.63 \\
2--3 & 26.37 & 0.68 & 32.81 & 0.70 \\
3--4 & 25.86 & 0.72 & 32.18 & 0.76 \\
4--5 & 25.39 & 0.75 & 31.59 & 0.80 \\
~~5--10 & 24.04 & 0.82 & 29.90 & 0.85 \\
10--15 & 21.80 & 0.88 & 27.05 & 0.81 \\
15--20 & 19.52 & 0.89 & 24.15 & 0.69 \\
20--25 & 17.29 & 0.88 & 21.32 & 0.54 \\
25--30 & 15.22 & 0.84 & 18.69 & 0.44 \\
30--35 & 13.34 & 0.78 & 16.30 & 0.35 \\
35--40 & 11.64 & 0.74 & 14.14 & 0.27 \\
40--45 & 10.11 & 0.69 & 12.22 & 0.21 \\
45--50 & 8.75 & 0.64 & 10.50 & 0.17 \\
50--55 & 7.54 & 0.57 & 8.97 & 0.16 \\
55--60 & 6.47 & 0.51 & 7.63 & 0.17 \\
60--65 & 5.52 & 0.47 & 6.45 & 0.15 \\
65--70 & 4.72 & 0.40 & 5.43 & 0.13 \\
70--75 & 4.04 & 0.34 & 4.58 & 0.11 \\
75--80 & 3.50 & 0.28 & 3.89 & 0.09 \\
20--30 & 16.26 & 0.86 & 20.01 & 0.49 \\
30--40 & 12.49 & 0.76 & 15.22 & 0.31 \\
40--50 & 9.43 & 0.67 & 11.36 & 0.19 \\
50--60 & 7.00 & 0.55 & 8.30 & 0.16 \\
60--70 & 5.12 & 0.43 & 5.94 & 0.13 \\
70--80 & 3.77 & 0.31 & 4.24 & 0.10 \\\hline
\end{tabular}
\end{table}

\begin{figure}[h]
\centering
\includegraphics[width=0.81\linewidth]{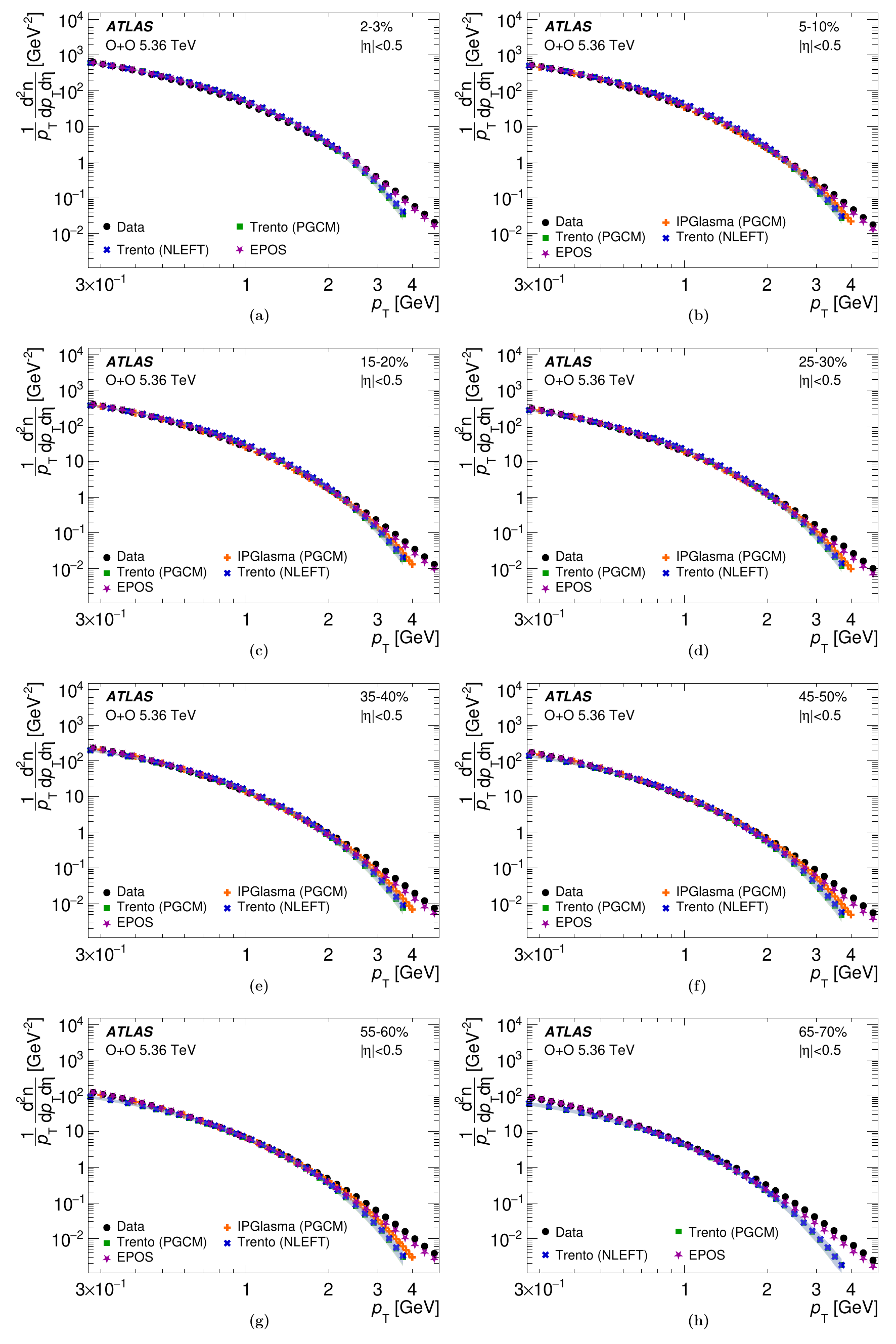}
\caption{
The measured \pt\ spectra in \OO\ collisions at mid-rapidity ($|\eta|<0.5$).
Each panel shows the results for a different centrality interval.
Comparisons with calculations from EPOS and models described in
Ref.~\cite{Mantysaari:2025tcg} (labeled ``IPGlasma'')
and Ref.~\cite{Giacalone:2024luz} (labeled ``Trento'') are also shown.
For the data, the vertical lines and vertical bars indicate statistical and
systematic uncertainties, respectively, and in most cases are smaller
than the size of the data points.
For the model calculations, the shaded bands indicate total uncertainties.
}
\label{fig:results_ptspec_OO_Theory_aux}
\end{figure}

\begin{figure}[h]
\centering
\includegraphics[width=0.81\linewidth]{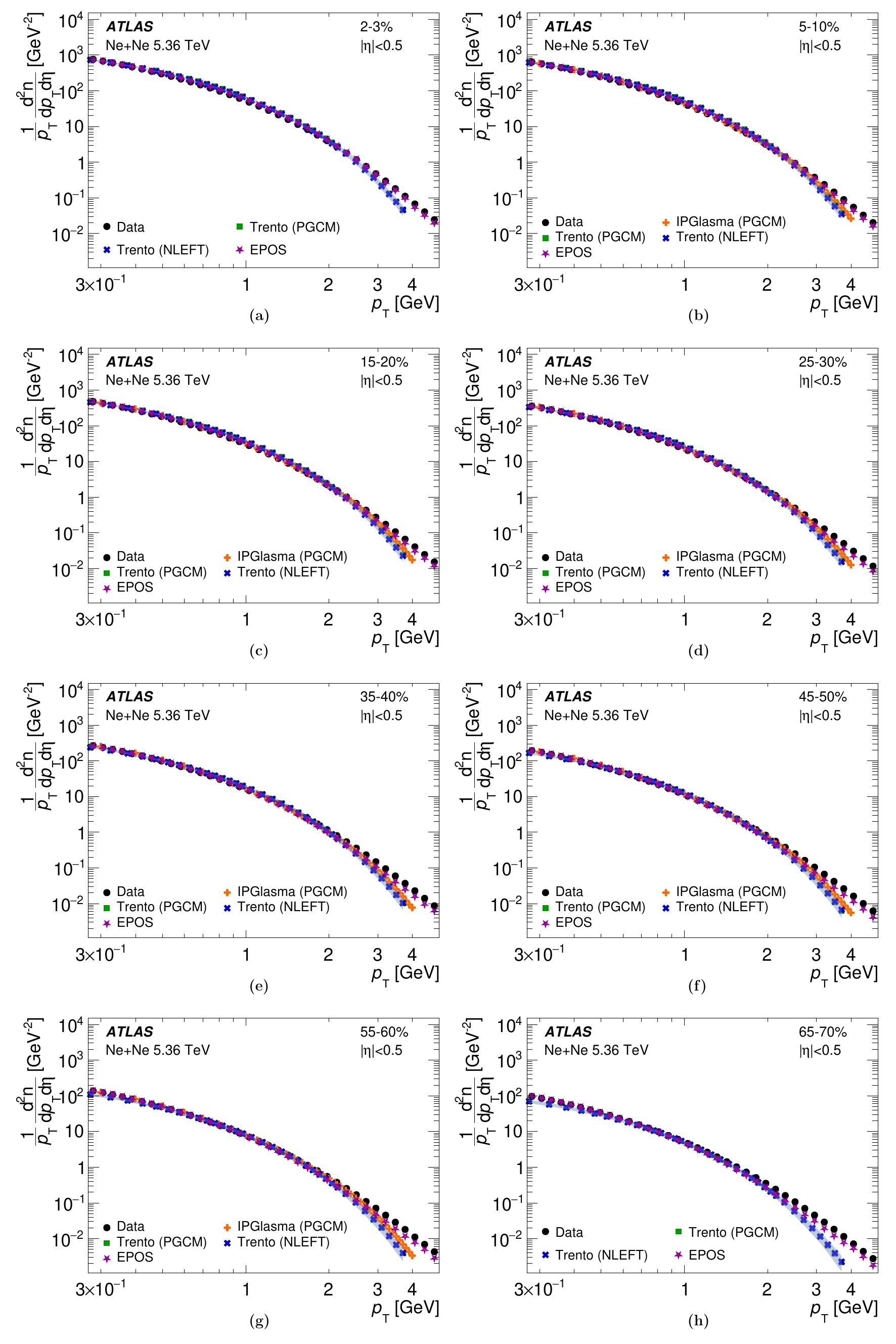}
\caption{
Same as Figure~\ref{fig:results_ptspec_OO_Theory_aux}, but for \NeNe\ collisions.
}
\label{fig:results_ptspec_NeNe_Theory_aux}
\end{figure}

\begin{figure}[h]
\centering
\includegraphics[width=0.81\linewidth]{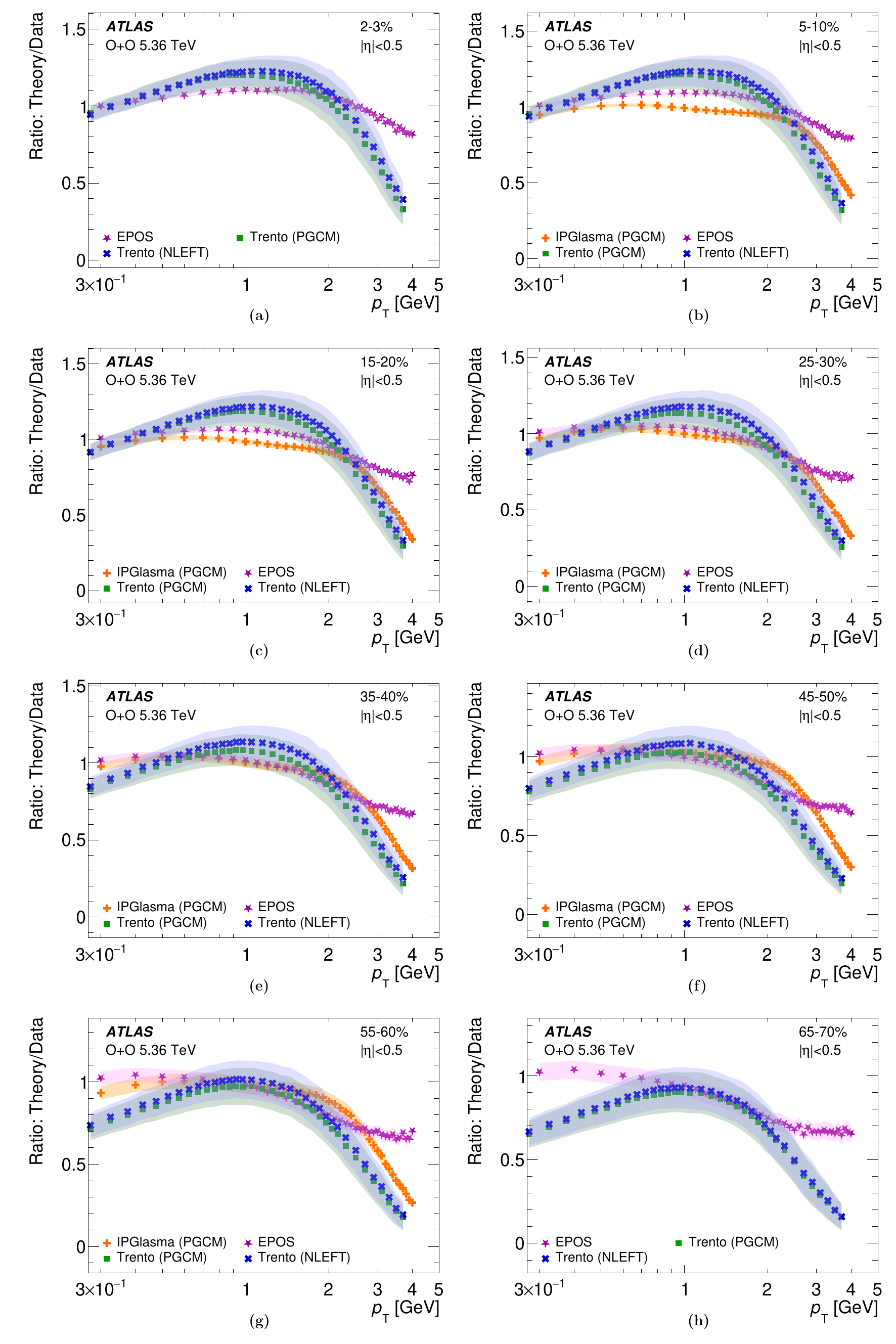}
\caption{
Ratios of theory predictions to measured \pt\ spectra
in \OO\ collisions at mid-rapidity ($|\eta|<0.5$).
Each panel shows the results for a different centrality interval.
The ratios are shown for EPOS and models described in
Ref.~\cite{Mantysaari:2025tcg} (labeled ``IPGlasma'')
and Ref.~\cite{Giacalone:2024luz} (labeled ``Trento'').
The shaded bands indicate total uncertainties.
}
\label{fig:results_ptspec_OO_TheoryRatio_aux}
\end{figure}

\begin{figure}[h]
\centering
\includegraphics[width=0.81\linewidth]{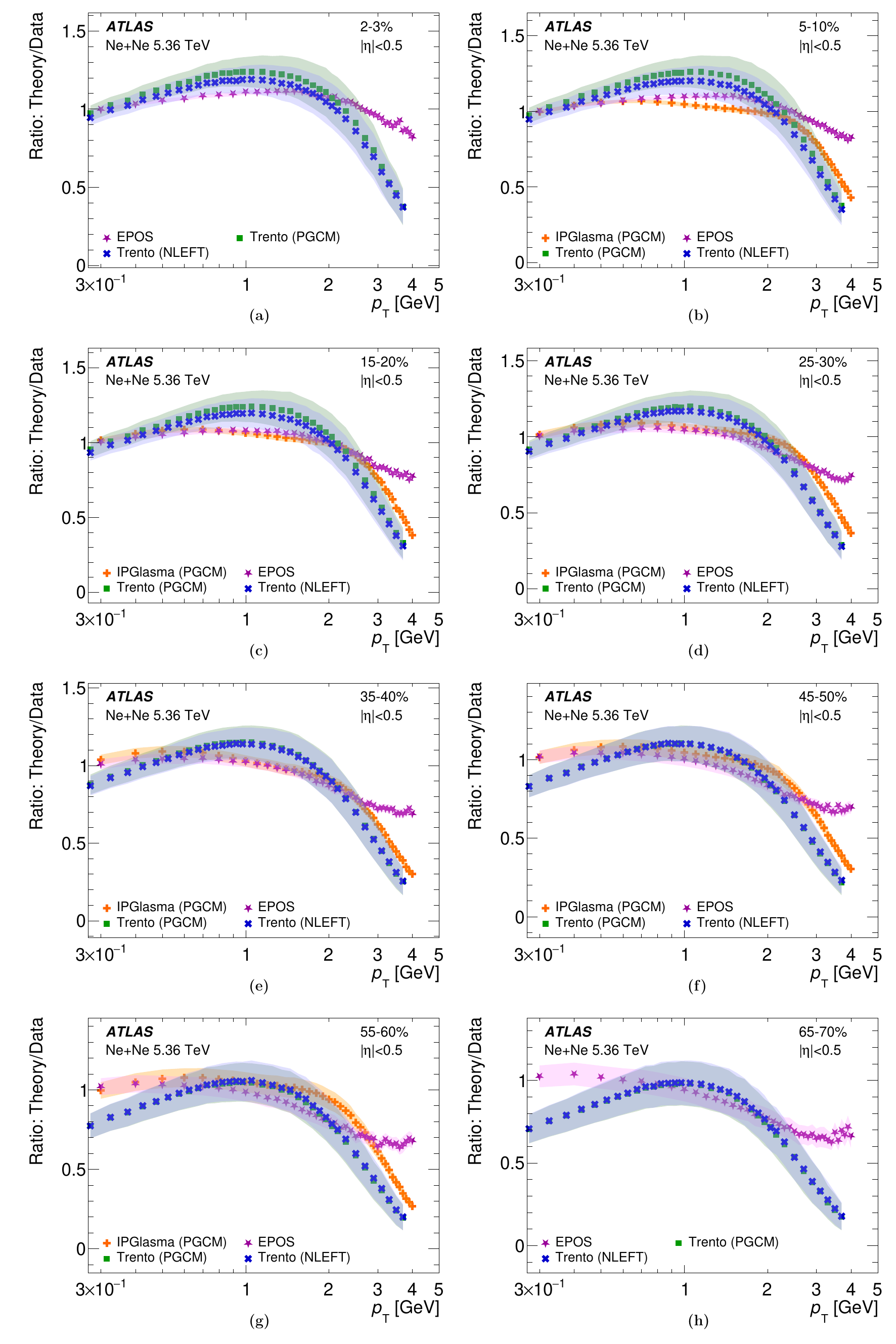}
\caption{
Same as Figure~\ref{fig:results_ptspec_OO_TheoryRatio_aux}, but for \NeNe\ collisions.
}
\label{fig:results_ptspec_NeNe_TheoryRatio_aux}
\end{figure}

\begin{figure}[h]
\centering
\includegraphics[width=0.85\linewidth]{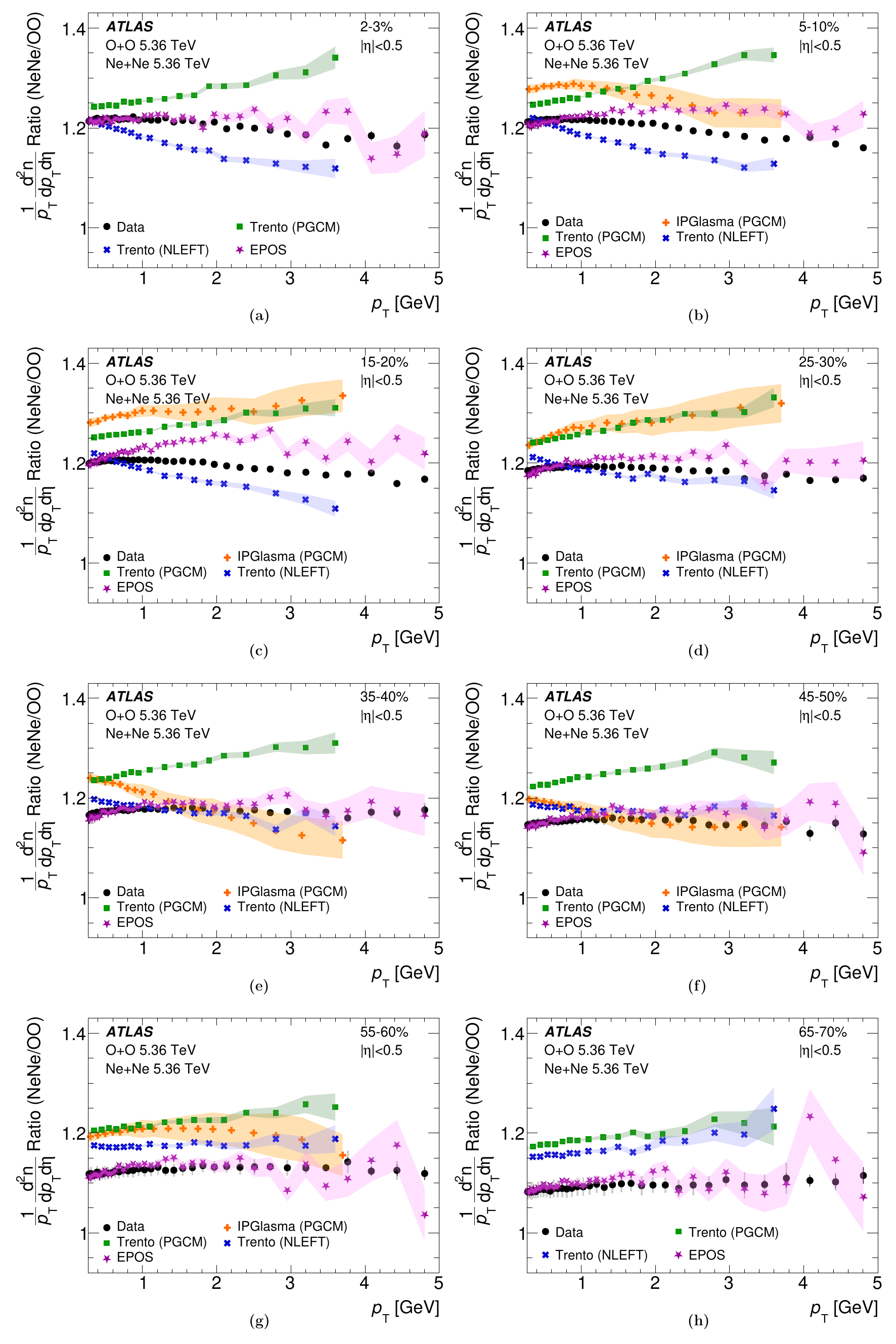}
\caption{
Ratios of \pt\ spectra measured in \NeNe\ collisions to those
measured in \OO\ collisions, at mid-rapidity ($|\eta|<0.5$).
Each panel shows the results for a different centrality interval.
Comparisons with calculations from EPOS and models described in
Ref.~\cite{Mantysaari:2025tcg} (labeled ``IPGlasma'')
and Ref.~\cite{Giacalone:2024luz} (labeled ``Trento'') are also shown.
For the data, the vertical lines and vertical bars indicate statistical and
systematic uncertainties, respectively, and in most cases are smaller
than the size of the data points.
For the model calculations, the shaded bands indicate total uncertainties.
}
\label{fig:results_ptspec_Ratio_Theory_aux}
\end{figure}

\begin{figure}[h]
\centering
\includegraphics[width=0.99\linewidth]{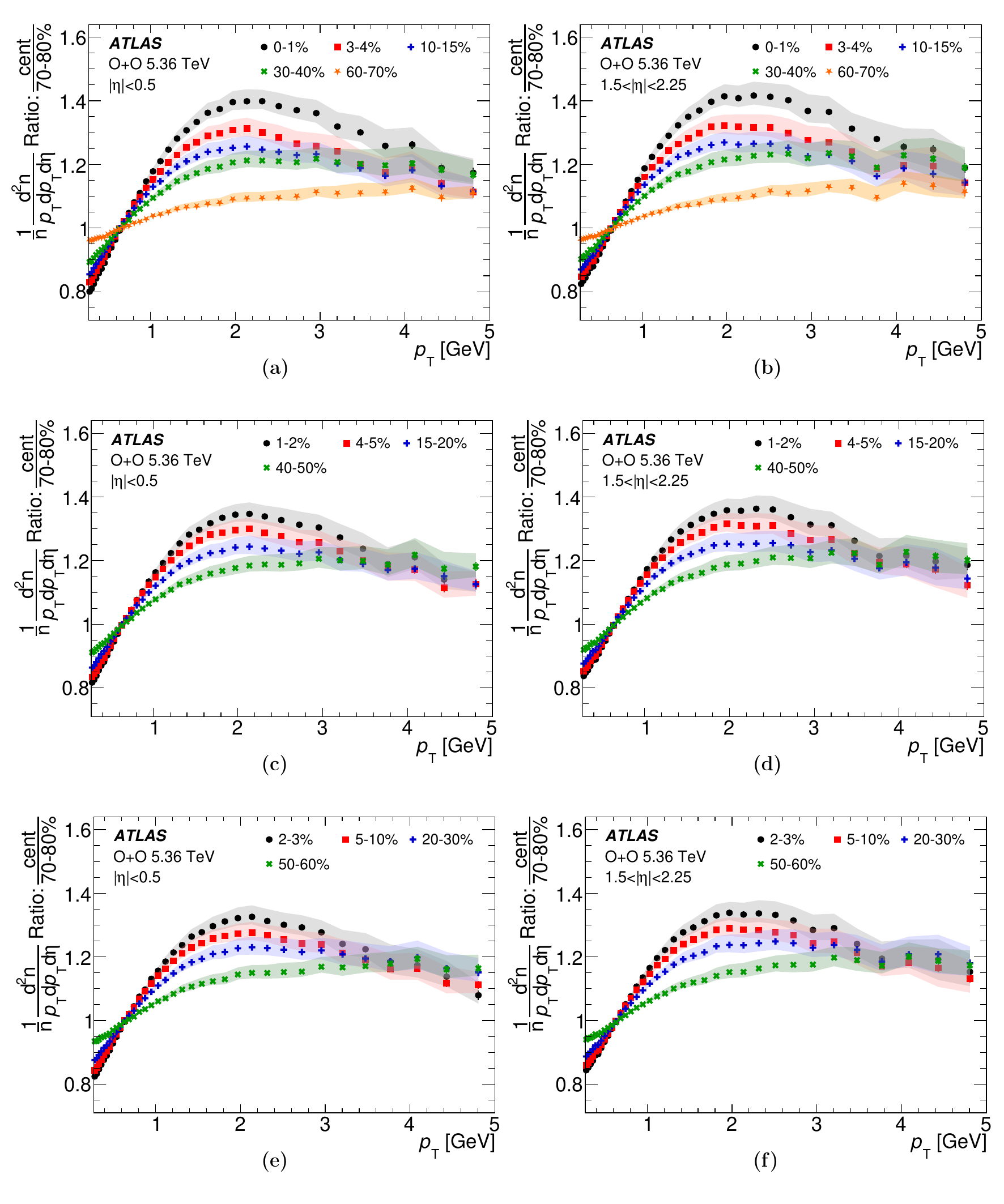}
\caption{
Comparisons between ratios of \pt\ spectra measured in
\OO\ collisions at various centralities to those
measured in the 70--80\% centrality interval.
Panels (a, c, e) show the measurements at mid-rapidity ($|\eta|<0.5$),
and panels (b, d, f) show the measurements at forward rapidity ($1.5<|\eta|<2.25$).
The vertical lines and shaded bands indicate statistical and
systematic uncertainties, respectively.
The statistical uncertainties in most cases are smaller
than the size of the data points.
Panels (a) and (b) are identical to Figures~\ref{fig:results_RatioPeriph_OO}a
and~\ref{fig:results_RatioPeriph_OO}c, respectively.
}
\label{fig:results_RatioPeriph_OO_aux}
\end{figure}

\begin{figure}[h]
\centering
\includegraphics[width=0.99\linewidth]{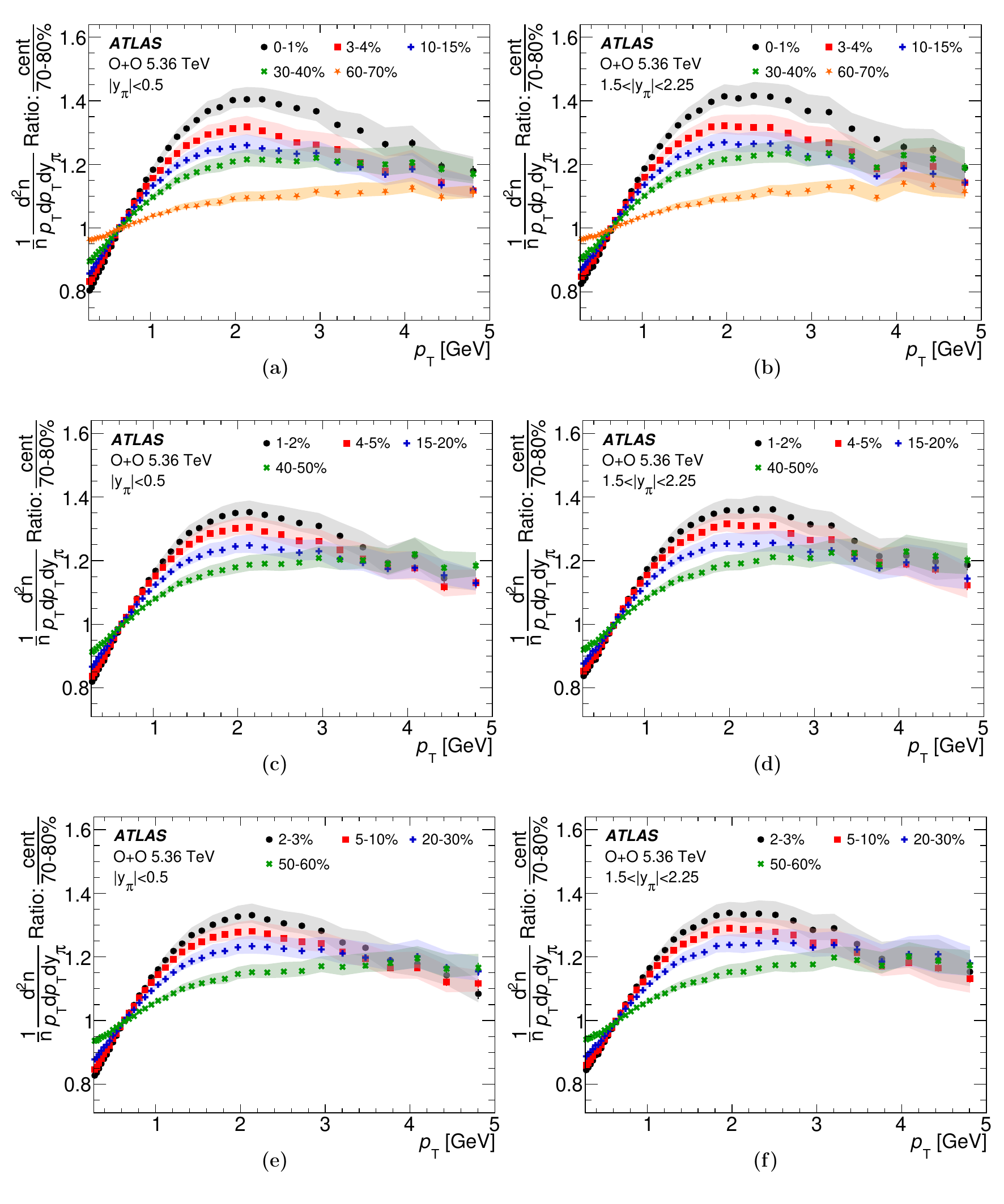}
\caption{
Comparisons between ratios of \pt\ spectra measured in \OO\ collisions at
various centralities to those measured in the 70--80\%
centrality interval.
Panels (a, c, e) show the measurements at mid-rapidity ($|\ypi|<0.5$), and
panels (b, d, f) show the measurements at forward rapidity ($1.5<|\ypi|<2.25$).
The vertical lines and shaded bands indicate statistical and
systematic uncertainties, respectively.
This figure is similar to Figure~\ref{fig:results_RatioPeriph_OO_aux},
but shows results in $\ypi$ intervals.
Panels (a) and (b) are identical to Figures~\ref{fig:results_RatioPeriph_OO}b
and~\ref{fig:results_RatioPeriph_OO}d, respectively.
}
\label{fig:results_RatioPeriphYpi_OO_aux}
\end{figure}

\begin{figure}[h]
\centering
\includegraphics[width=0.99\linewidth]{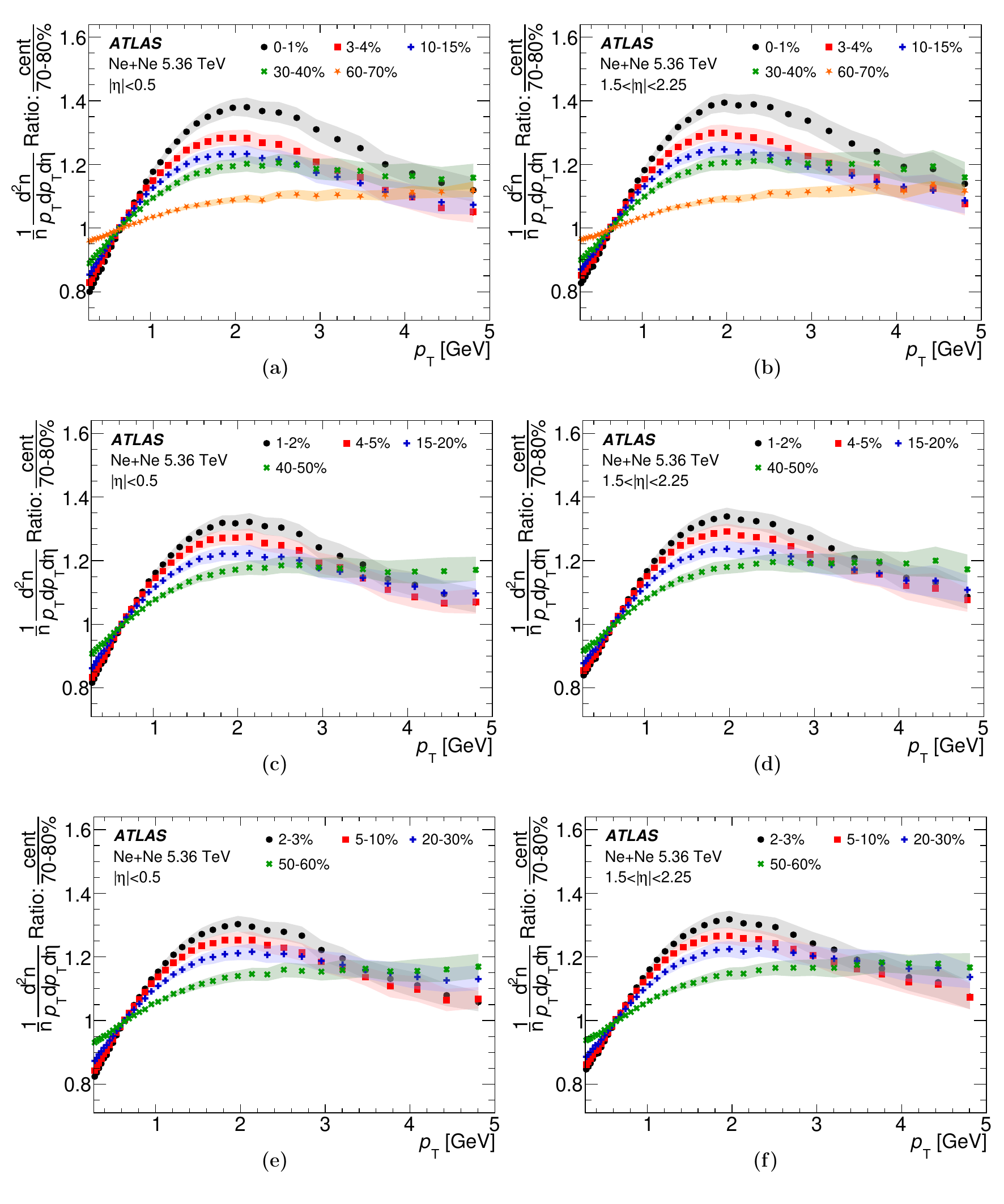}
\caption{
Same as Figure~\ref{fig:results_RatioPeriph_OO_aux}, but for
\NeNe\ collisions.
}
\label{fig:results_RatioPeriph_NeNe_aux}
\end{figure}

\begin{figure}[h]
\centering
\includegraphics[width=0.99\linewidth]{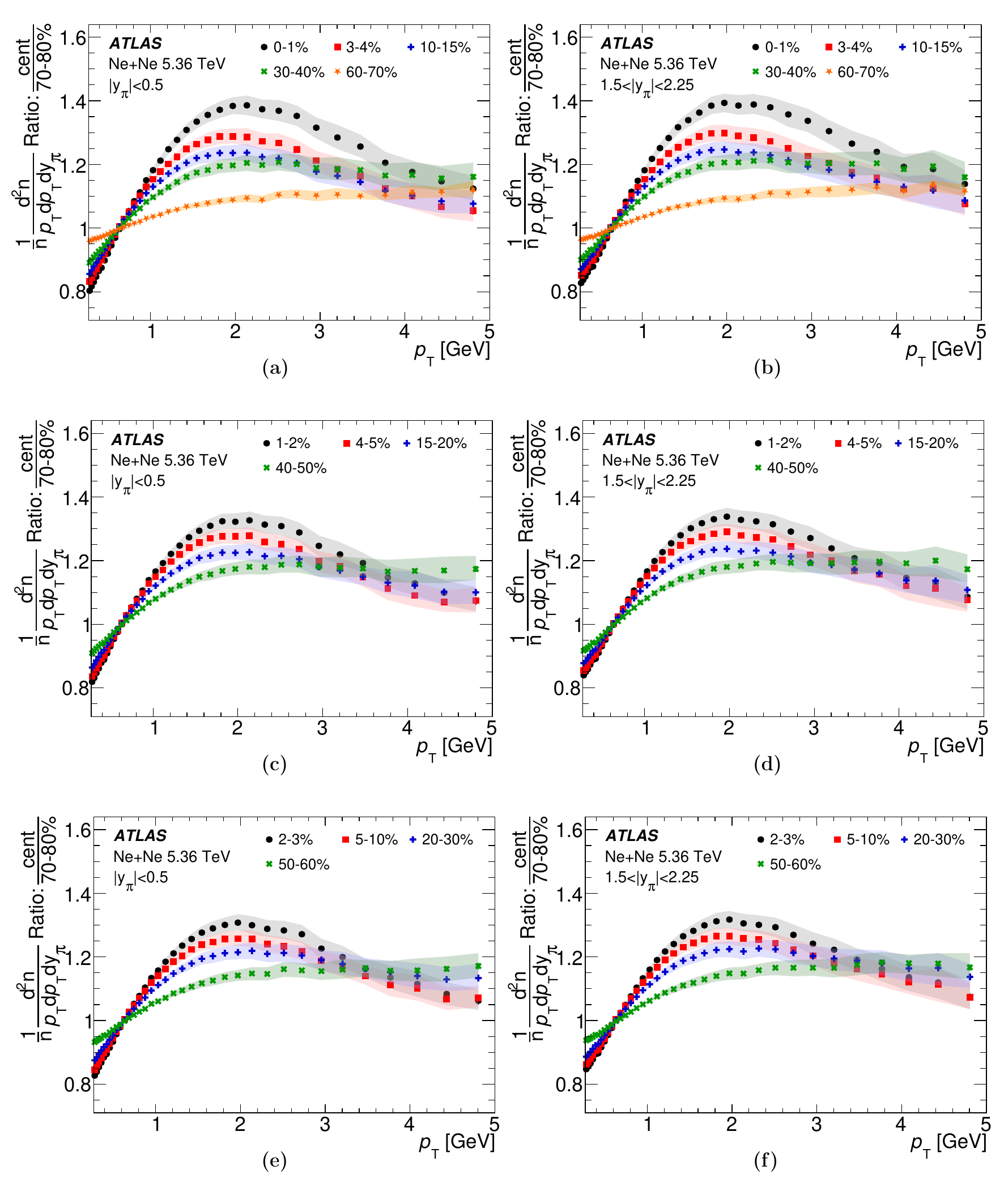}
\caption{
Same as Figure~\ref{fig:results_RatioPeriphYpi_OO_aux}, but for
\NeNe\ collisions.
}
\label{fig:results_RatioPeriphYpi_NeNe_aux}
\end{figure}
\begin{figure}[h]
\centering
\includegraphics[width=0.85\linewidth]{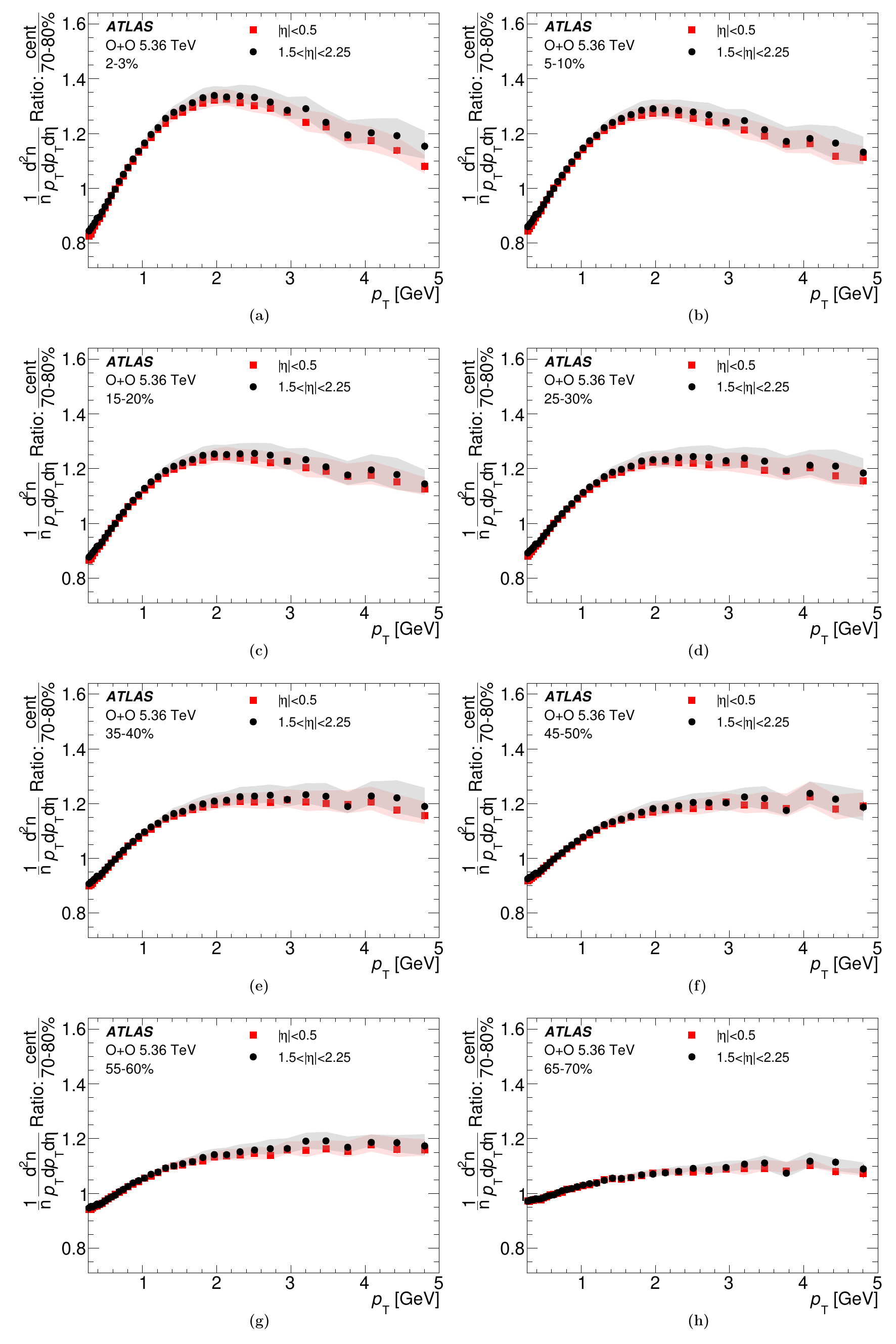}
\caption{
Comparisons between ratios of \pt\ spectra measured in
\OO\ collisions at various centralities to those
measured in the 70--80\% centrality interval.
Each panel corresponds to a different centrality interval.
This figure contains the same information as Figure~\ref{fig:results_RatioPeriph_OO_aux},
but compares the mid-rapidity and forward-rapidity results on the
same panel.
The vertical lines and shaded bands indicate statistical and
systematic uncertainties, respectively.
}
\label{fig:results_RatioPeriph_OO_etacompare_aux}
\end{figure}

\begin{figure}[h]
\centering
\includegraphics[width=0.85\linewidth]{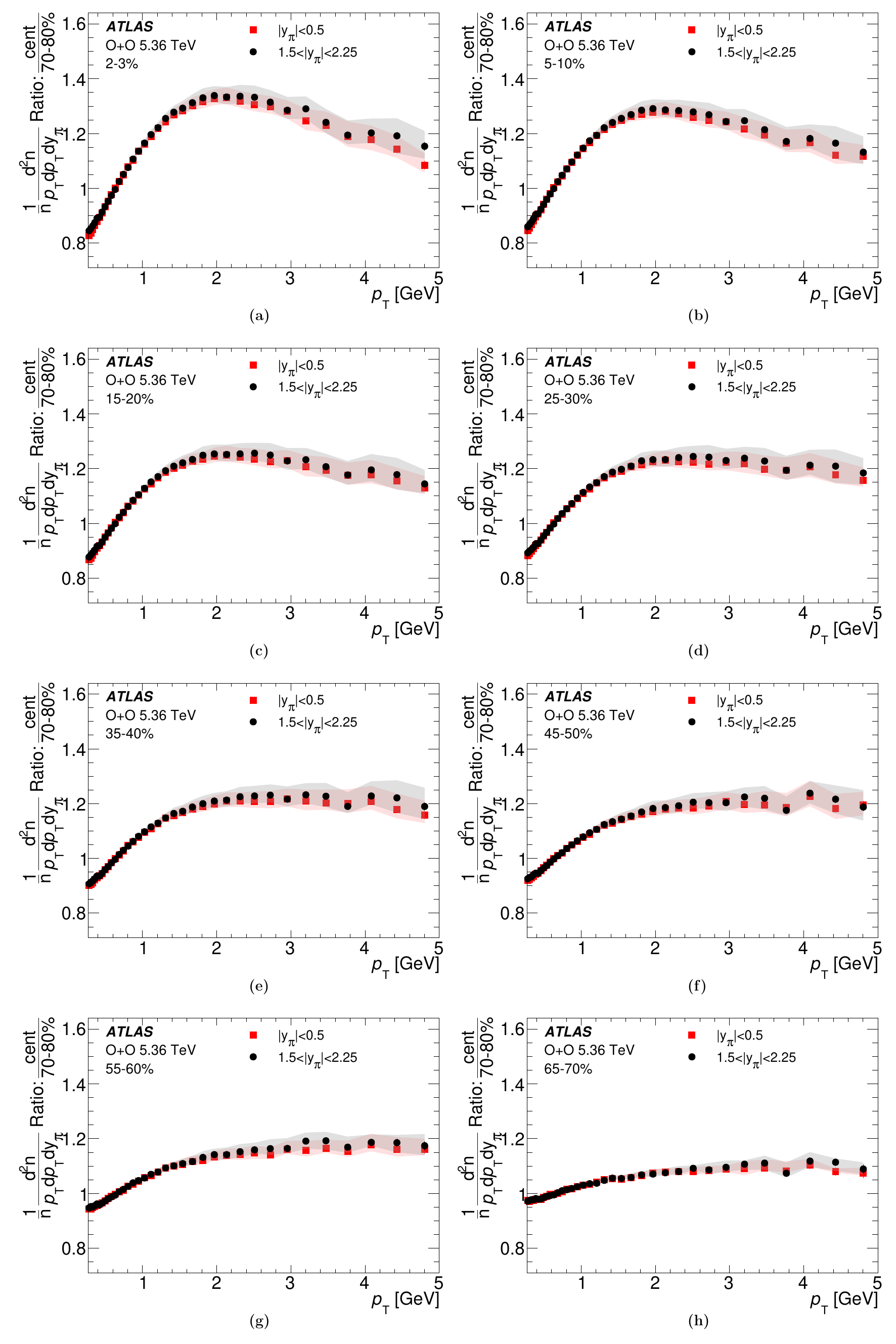}
\caption{
Comparisons between ratios of \pt\ spectra measured in \OO\ collisions at
various centralities to those measured in the 70--80\%
centrality interval.
Each panel corresponds to a different centrality interval.
This figure contains the same information as Figure~\ref{fig:results_RatioPeriphYpi_OO_aux},
but compares the mid-rapidity ($|\ypi|<0.5$) and forward-rapidity ($1.5<|\ypi|<2.25$) results on the
same panel.
The vertical lines and shaded bands indicate statistical and
systematic uncertainties, respectively.
}
\label{fig:results_RatioPeriphYpi_OO_etacompare_aux}
\end{figure}

\begin{figure}[h]
\centering
\includegraphics[width=0.85\linewidth]{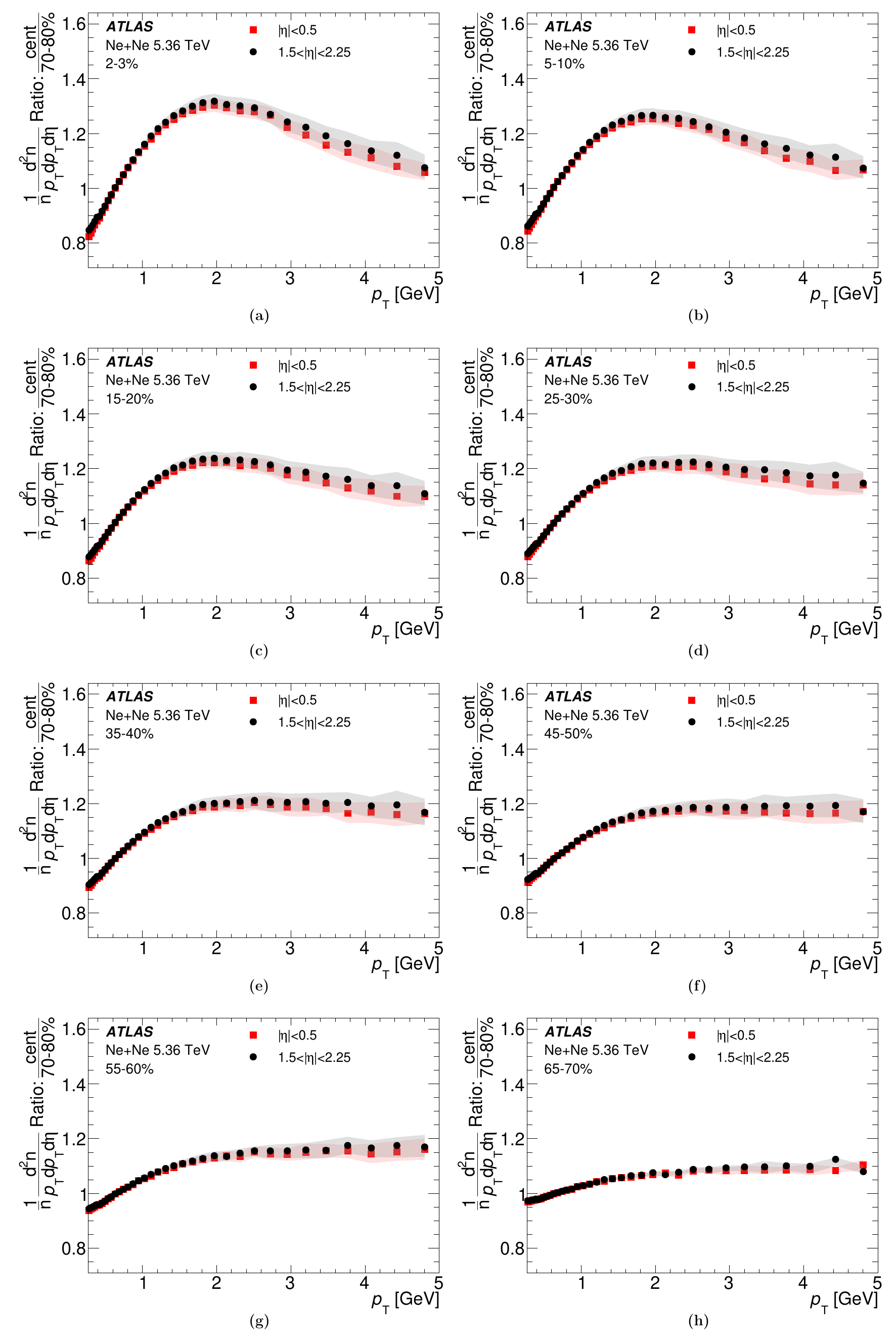}
\caption{
Same as Figure~\ref{fig:results_RatioPeriph_OO_etacompare_aux},
but for \NeNe\ collisions.
This figure contains the same information as Figure~\ref{fig:results_RatioPeriph_NeNe_aux},
but compares the mid-rapidity and forward-rapidity results on the
same panel.
}
\label{fig:results_RatioPeriph_NeNe_etacompare_aux}
\end{figure}

\begin{figure}[h]
\centering
\includegraphics[width=0.85\linewidth]{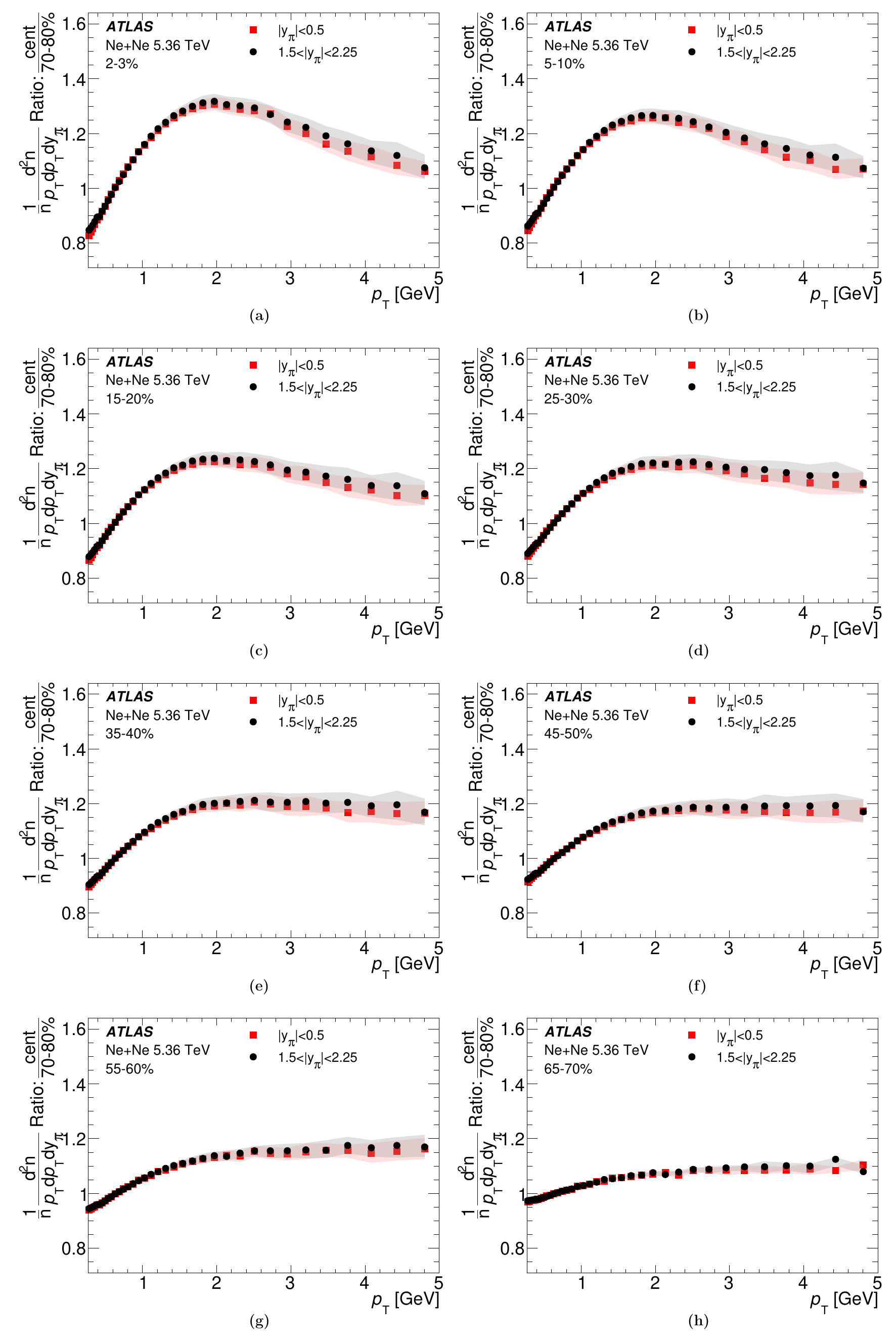}
\caption{
Same as Figure~\ref{fig:results_RatioPeriphYpi_OO_etacompare_aux},
but for \NeNe\ collisions.
This figure contains the same information as Figure~\ref{fig:results_RatioPeriphYpi_NeNe_aux},
but compares the mid-rapidity ($|\ypi|<0.5$) and forward-rapidity ($1.5<|\ypi|<2.25$) results on the
same panel.
}
\label{fig:results_RatioPeriphYpi_NeNe_etacompare_aux}
\end{figure}

\begin{figure}[h]
\centering
\includegraphics[width=0.85\linewidth]{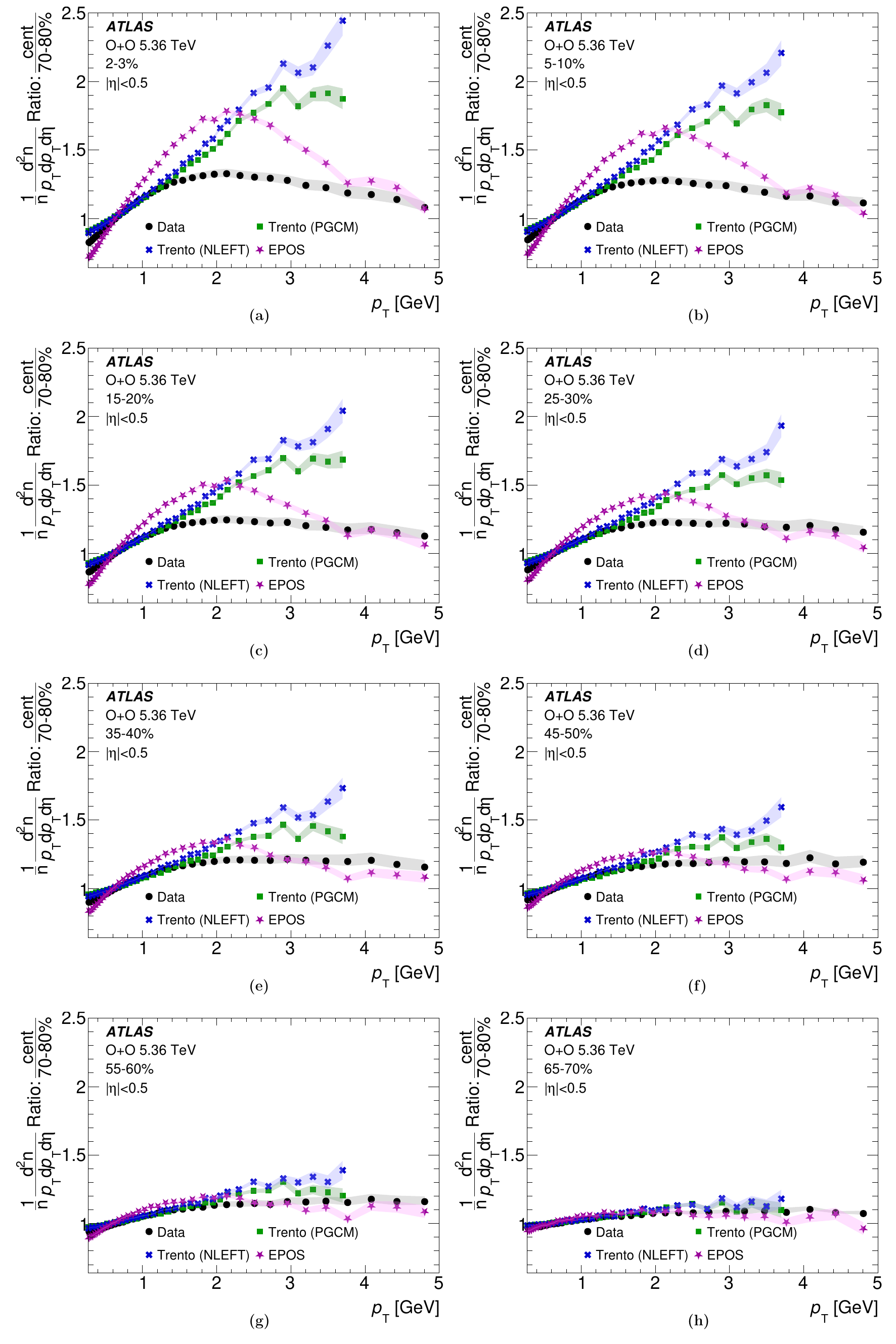}
\caption{
Ratios of the \pt\ spectra measured in \OO\ collisions at
various centralities to those measured in the 70--80\%
centrality interval.
Each panel corresponds to a different centrality interval.
Comparisons with the EPOS model and hydrodynamic models based
on IPGlasma and Trento calculations are also shown.
For the data, the vertical lines and shaded bands indicate
statistical uncertainties (always smaller than the size
of the data points) and systematic uncertainties, respectively.
For the theory points, the shaded bands indicate total uncertainties.
}
\label{fig:results_RatioPeriph_OO_TheoryCompare_aux}
\end{figure}

\begin{figure}[h]
\centering
\includegraphics[width=0.85\linewidth]{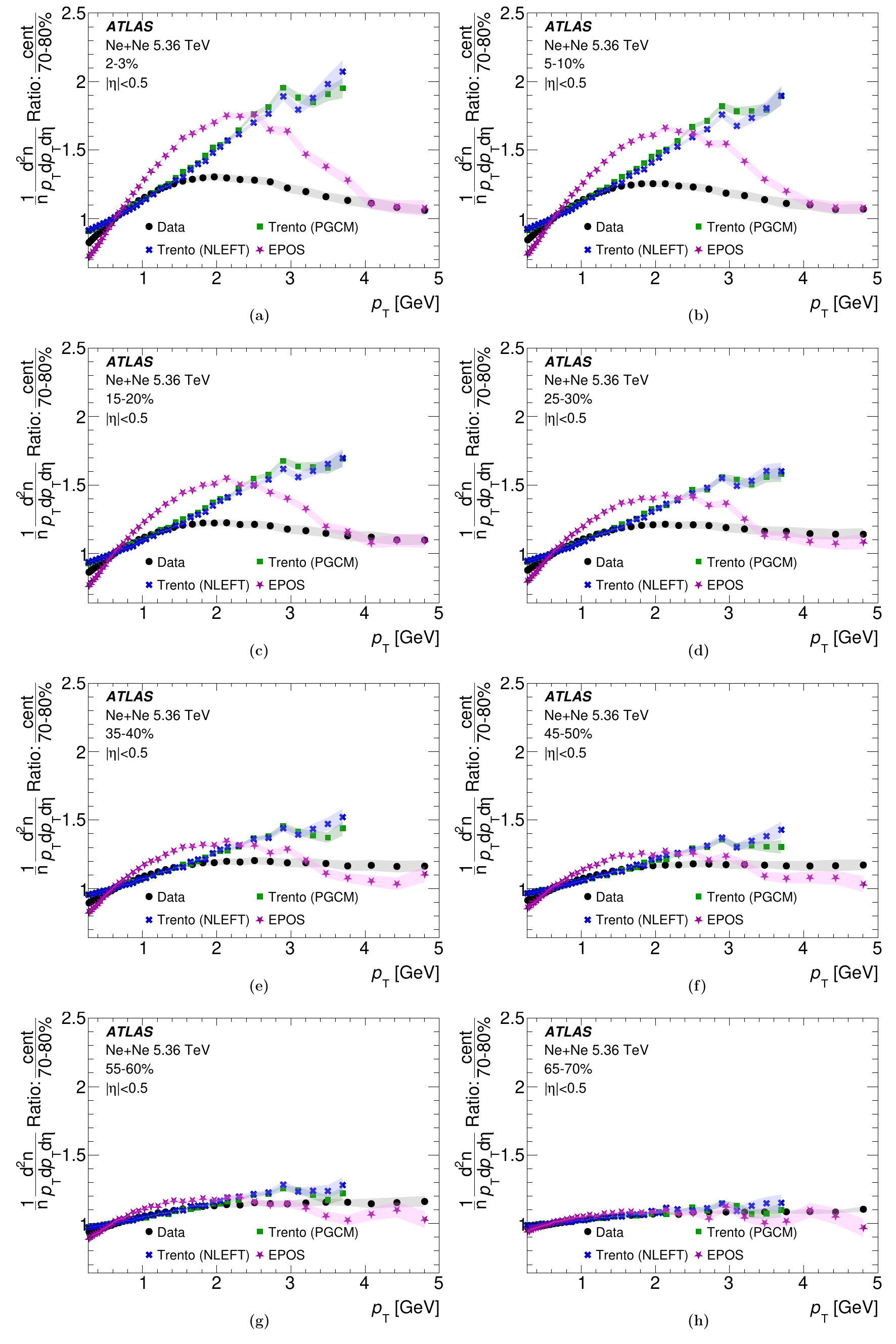}
\caption{
Same as Figure~\ref{fig:results_RatioPeriph_OO_TheoryCompare_aux}, but for
\NeNe\ collisions.
}
\label{fig:results_RatioPeriph_NeNe_TheoryCompare_aux}
\end{figure}

\begin{figure}[h]
\centering
\includegraphics[width=0.99\linewidth]{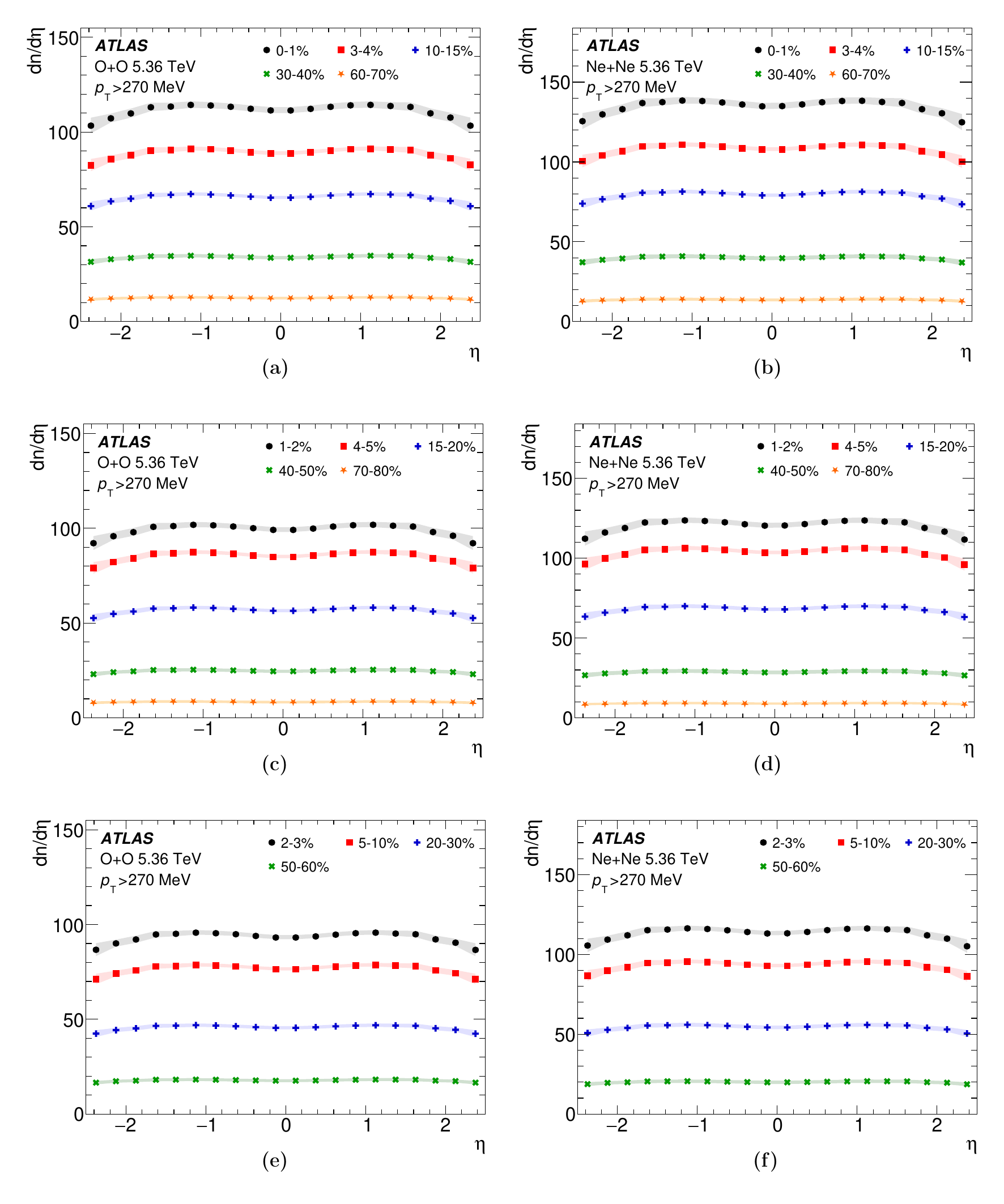}
\caption{
The \dndeta\ distributions in (a, c, e) \OO  and
(b, d, f) \NeNe collisions, as a function of $\eta$,
for several centrality intervals.
The vertical lines and shaded bands indicate statistical and
systematic uncertainties, respectively.
Statistical uncertainties in all cases are smaller than the size
of the data points.
The results correspond to the $\pT>270~\MeV$ fiducial measurements.
}
\label{fig:results_dndetaOO_aux}
\end{figure}

\begin{figure}[h]
\centering
\includegraphics[width=0.99\linewidth]{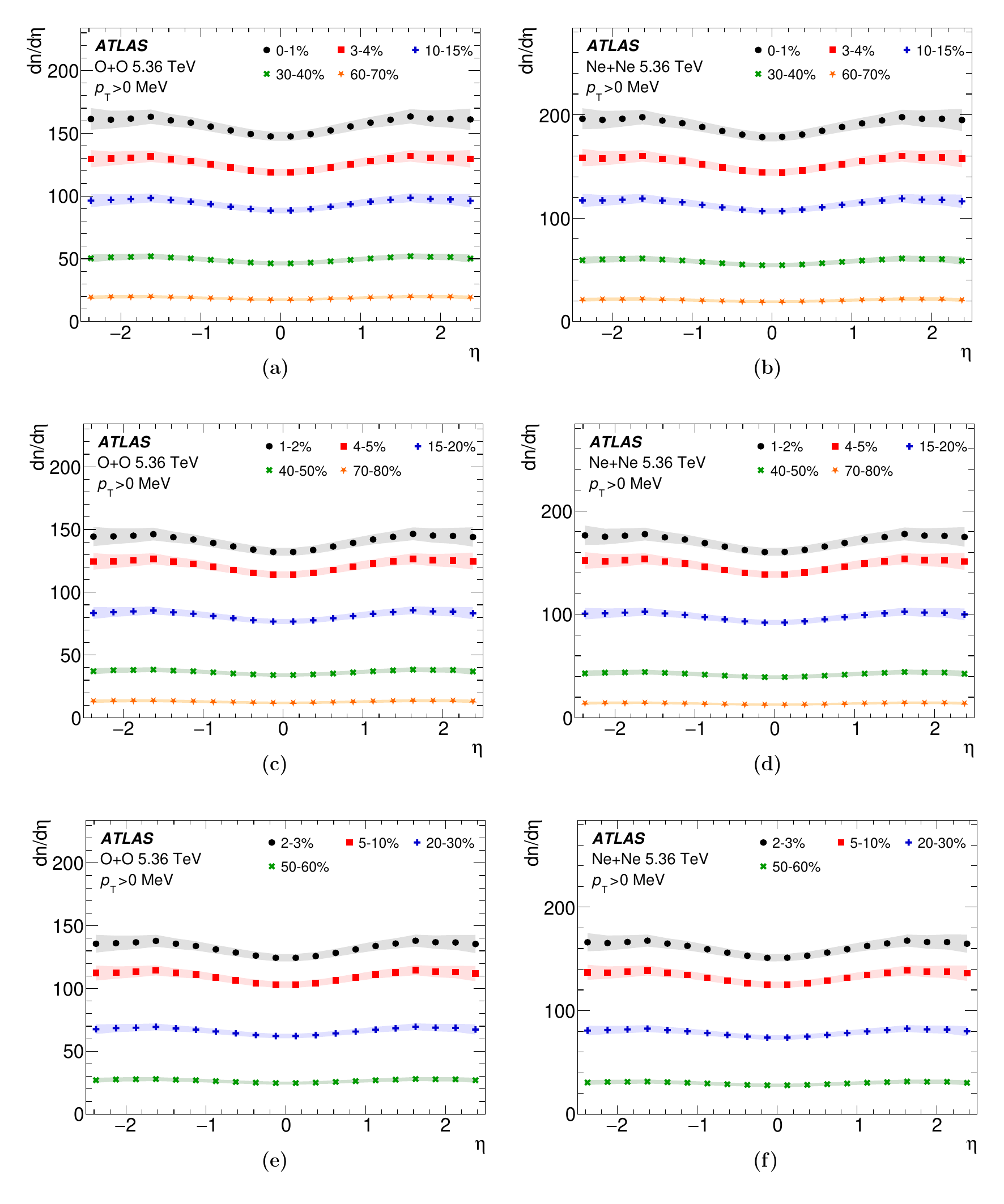}
\caption{
Same as Figure~\ref{fig:results_dndetaOO_aux}, but for the \pteqo\ case.
}
\label{fig:results_dndetaOO_ptExtrap0_aux}
\end{figure}

\begin{figure}[h]
\centering
\includegraphics[width=0.99\linewidth]{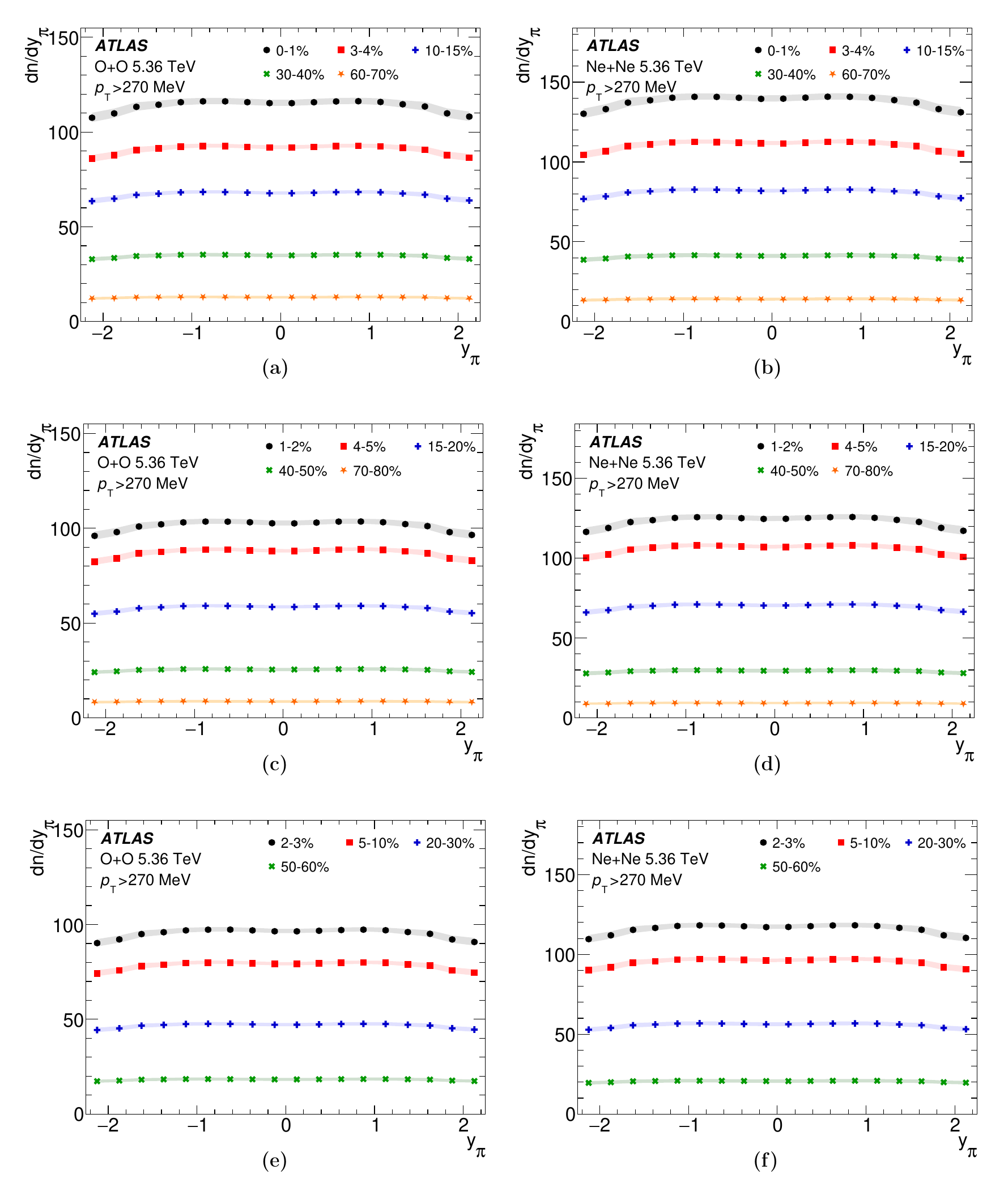}
\caption{
Same as Figure~\ref{fig:results_dndetaOO_aux}, but for
measurements as a function of \ypi.
}
\label{fig:results_dndypiOO_aux}
\end{figure}

\begin{figure}[h]
\centering
\includegraphics[width=0.99\linewidth]{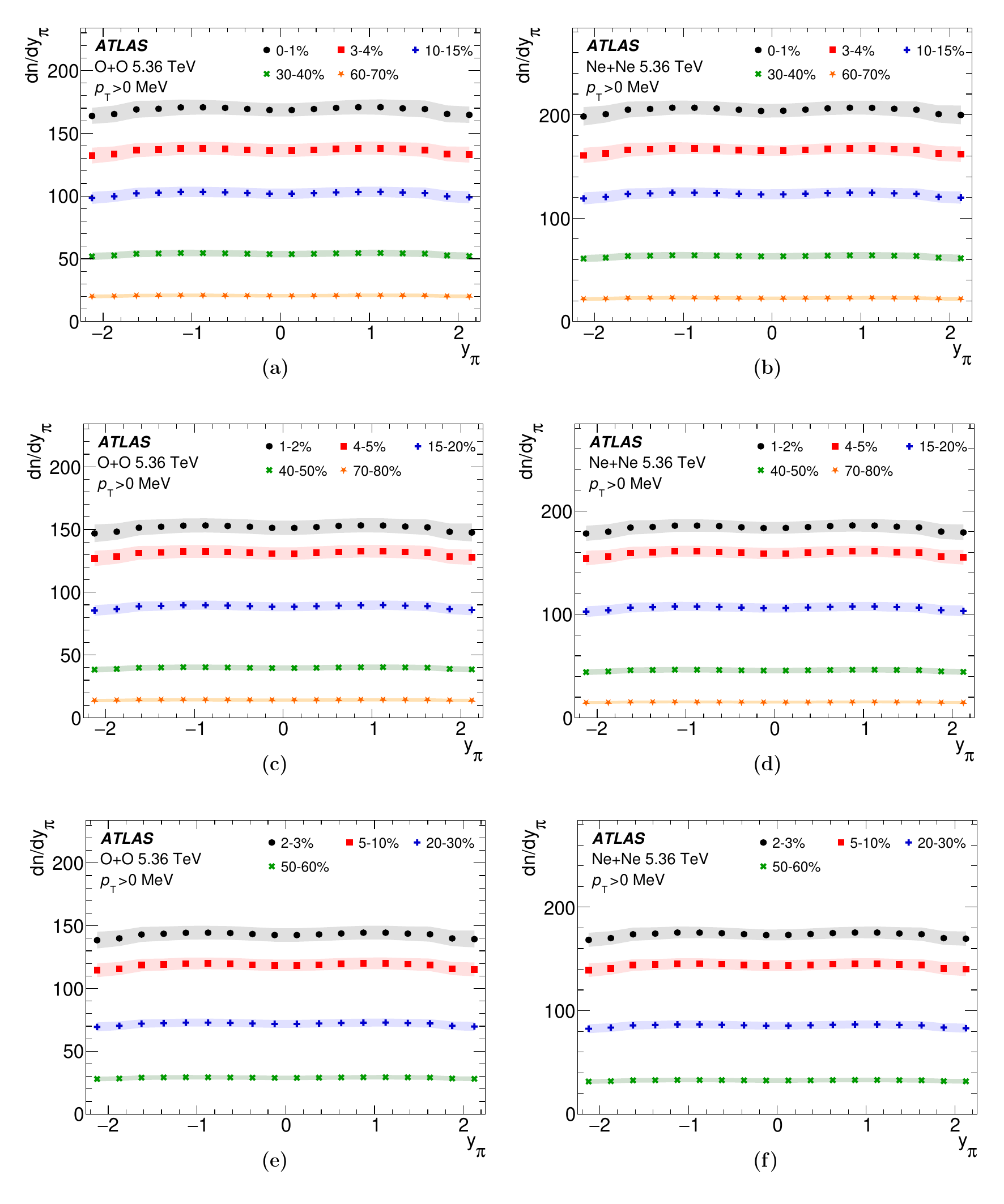}
\caption{
Same as Figure~\ref{fig:results_dndetaOO_aux}, but for
measurements as a function of \ypi\ and for the \pteqo\ case.
}
\label{fig:results_dndypiOO_ptExtrap0_aux}
\end{figure}
\begin{figure}[h]
\centering
\includegraphics[width=0.99\linewidth]{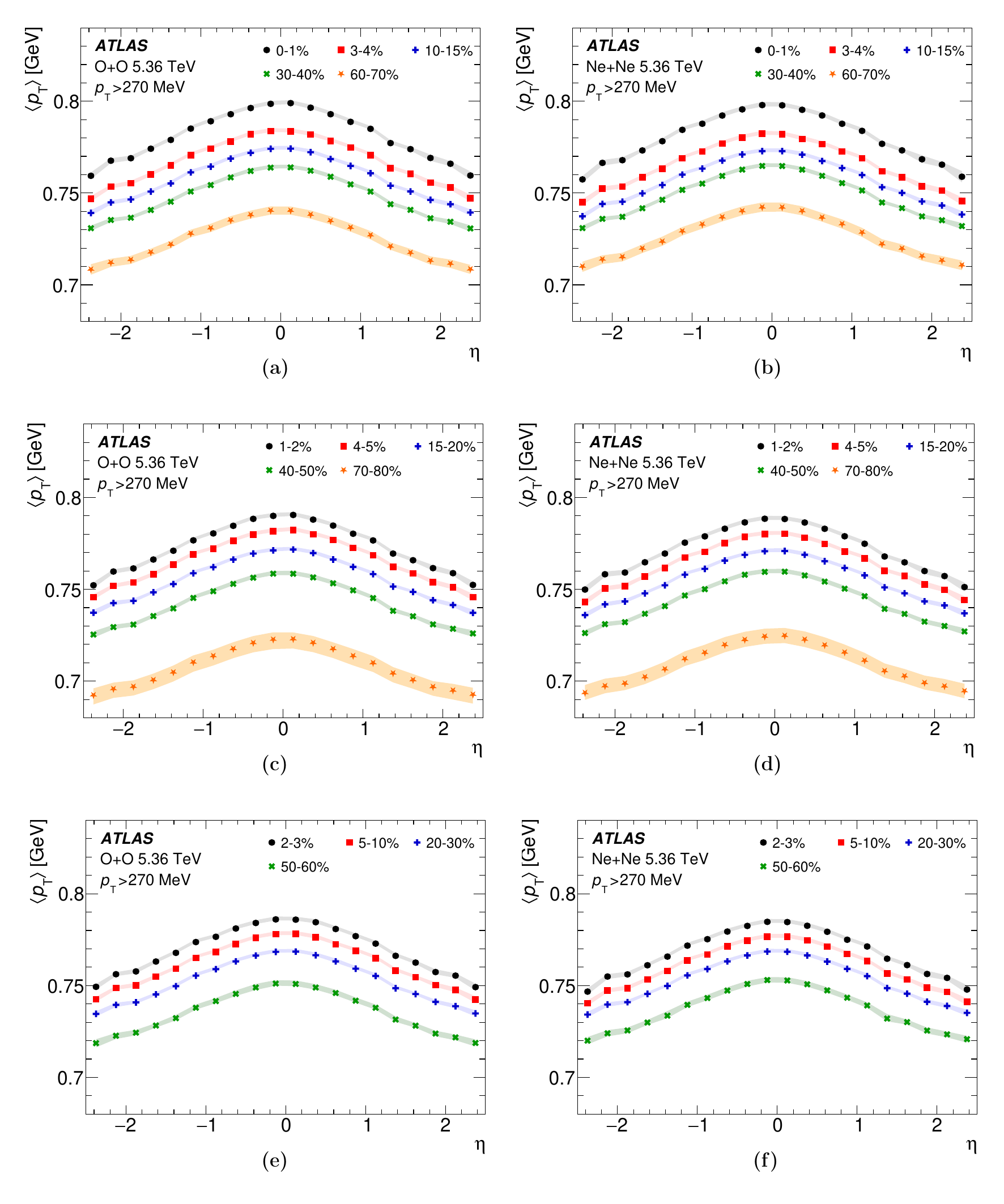}
\caption{
The \meanpt\ distributions in (a, c, e) \OO\ and
(b, d, f) \NeNe\ collisions, as a function of $\eta$,
for several centrality intervals.
The vertical lines and shaded bands indicate statistical and
systematic uncertainties, respectively.
Statistical uncertainties in all cases are smaller than the size
of the data points.
The results correspond to the $\pT>270~\MeV$ fiducial measurements.
}
\label{fig:results_MeanpTvsEtaOO_aux}
\end{figure}

\begin{figure}[h]
\centering
\includegraphics[width=0.99\linewidth]{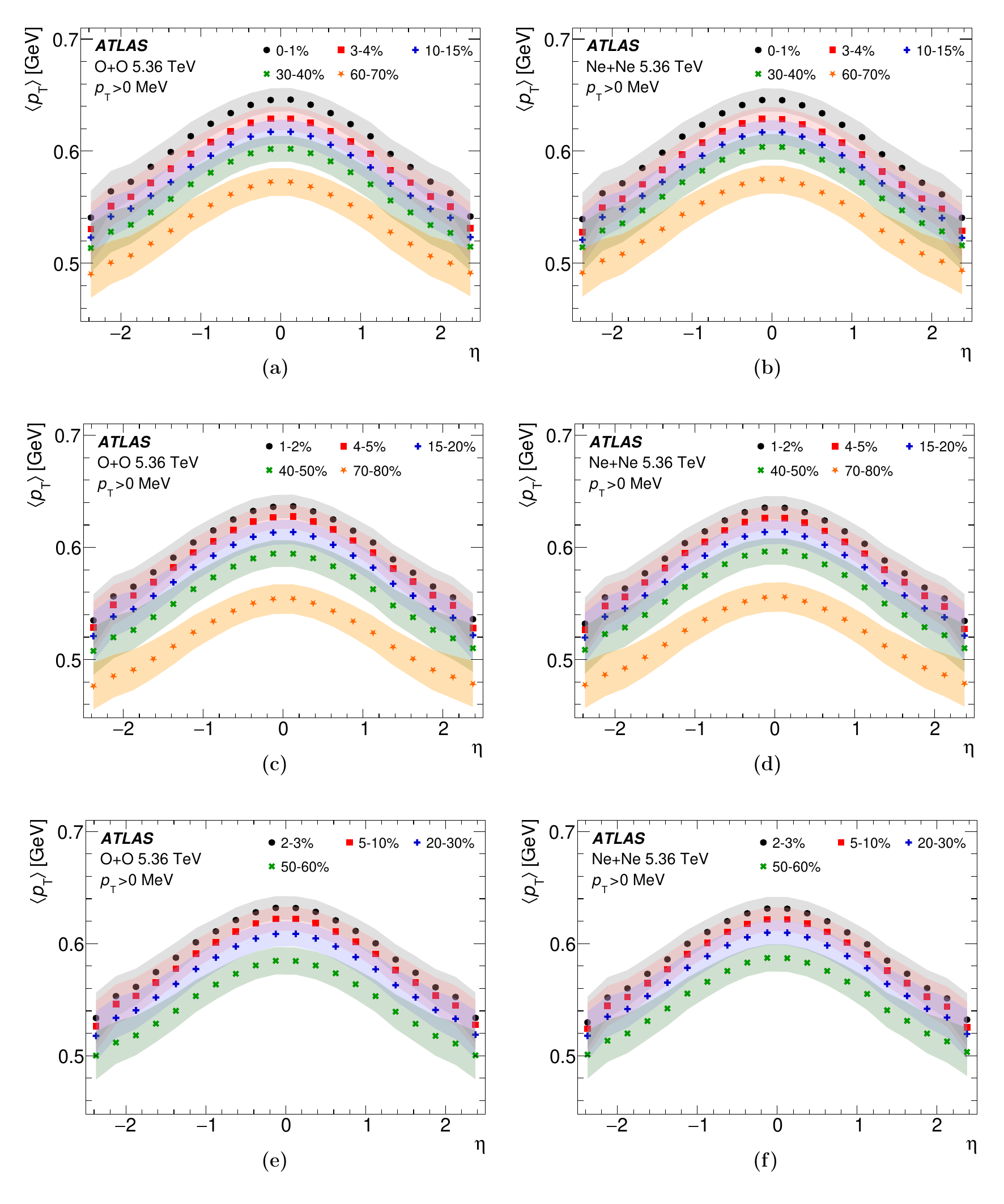}
\caption{
Same as Figure~\ref{fig:results_MeanpTvsEtaOO_aux}, but for the \pteqo\ measurements.
}
\label{fig:results_MeanpTvsEtaOO_ptExtrap0_aux}
\end{figure}

\begin{figure}[h]
\centering
\includegraphics[width=0.99\linewidth]{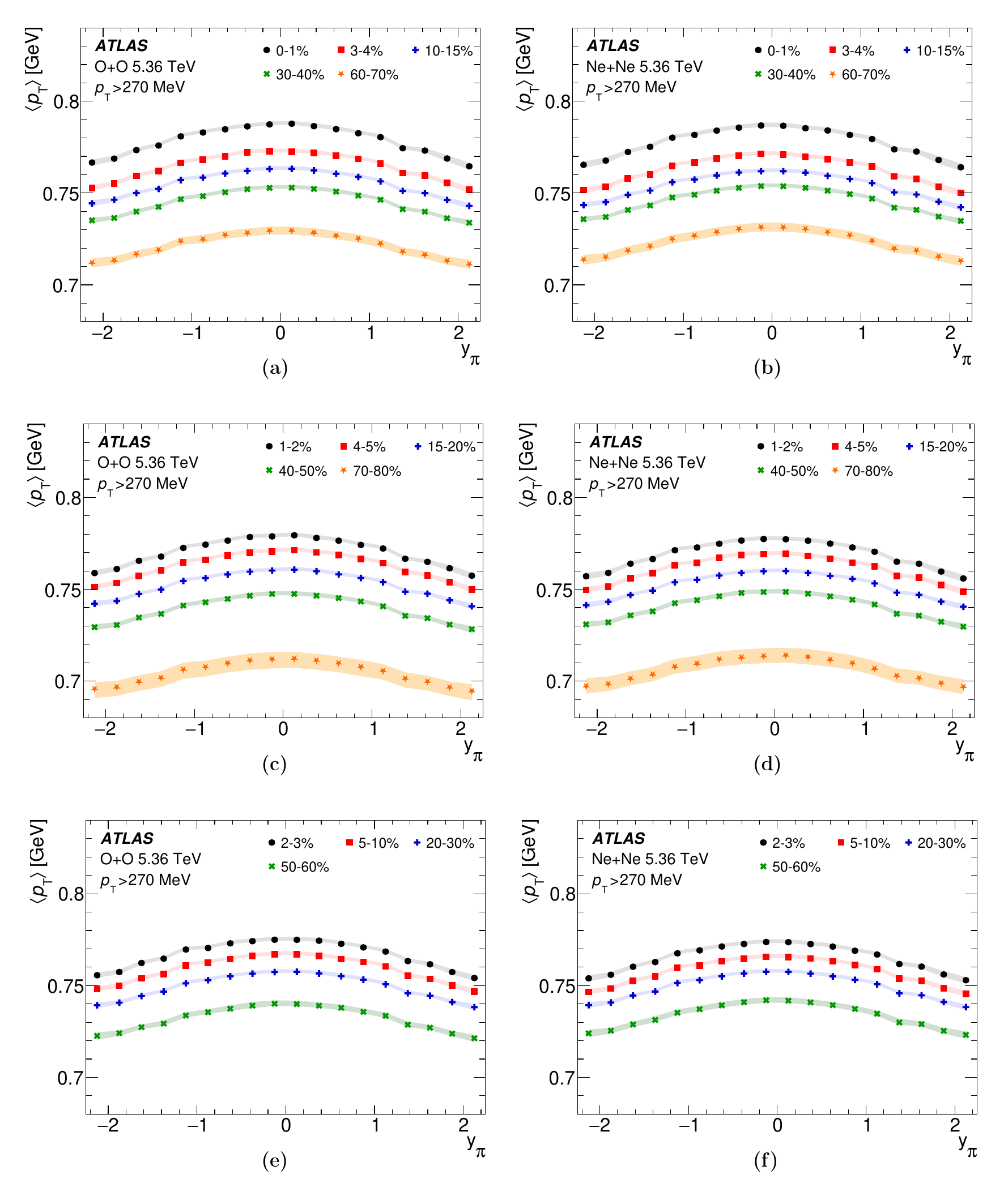}
\caption{
Same as Figure~\ref{fig:results_MeanpTvsEtaOO_aux}, but for measurements as a function of \ypi.
}
\label{fig:results_MeanpTvsypiOO_aux}
\end{figure}

\begin{figure}[h]
\centering
\includegraphics[width=0.99\linewidth]{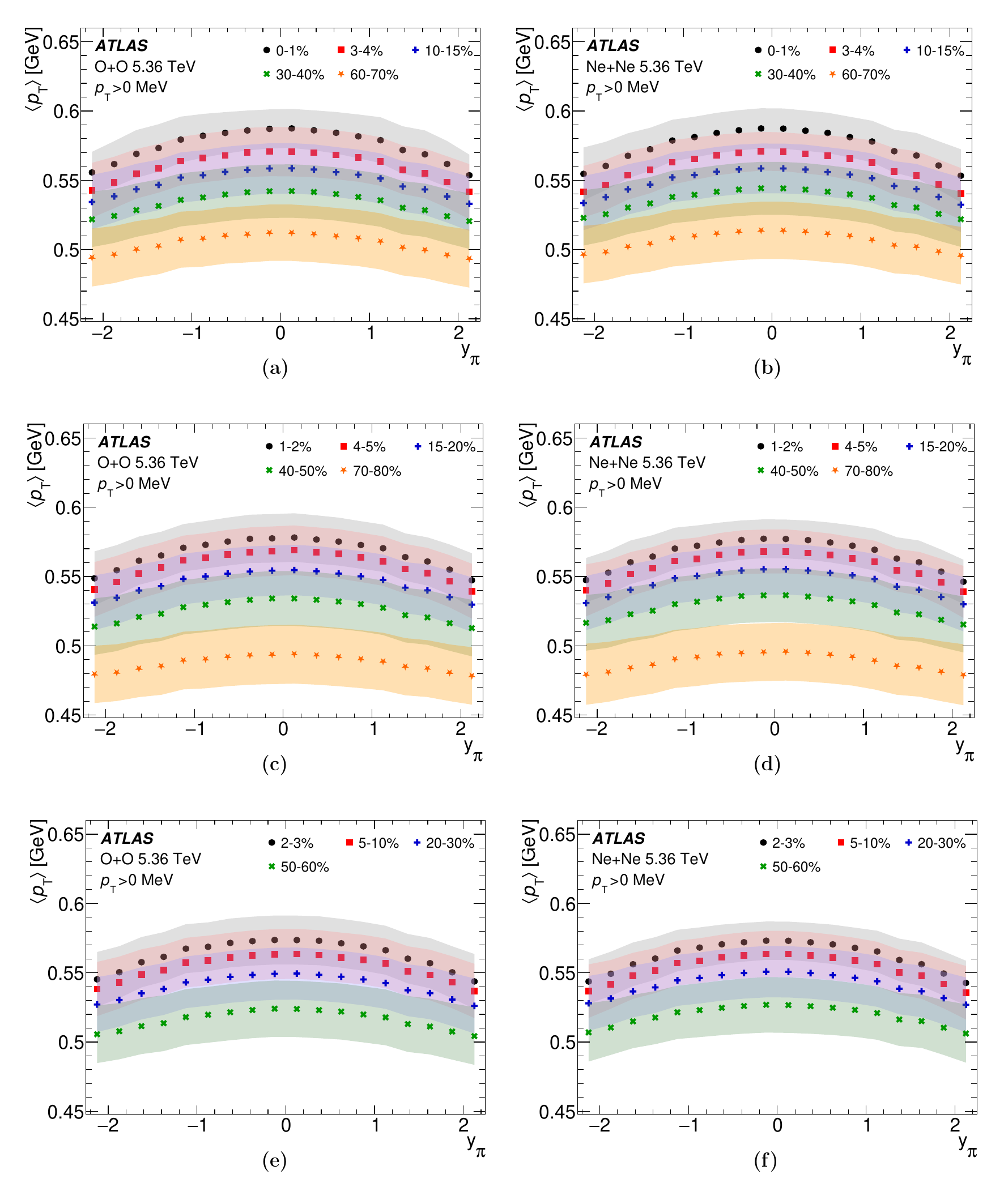}
\caption{
Same as Figure~\ref{fig:results_MeanpTvsEtaOO_aux}, but for
measurements as a function of \ypi\ and for \pteqo.
}
\label{fig:results_MeanpTvsypiOO_ptExtrap0_aux}
\end{figure}
\begin{figure}[h]
\centering
\includegraphics[width=0.90\linewidth]{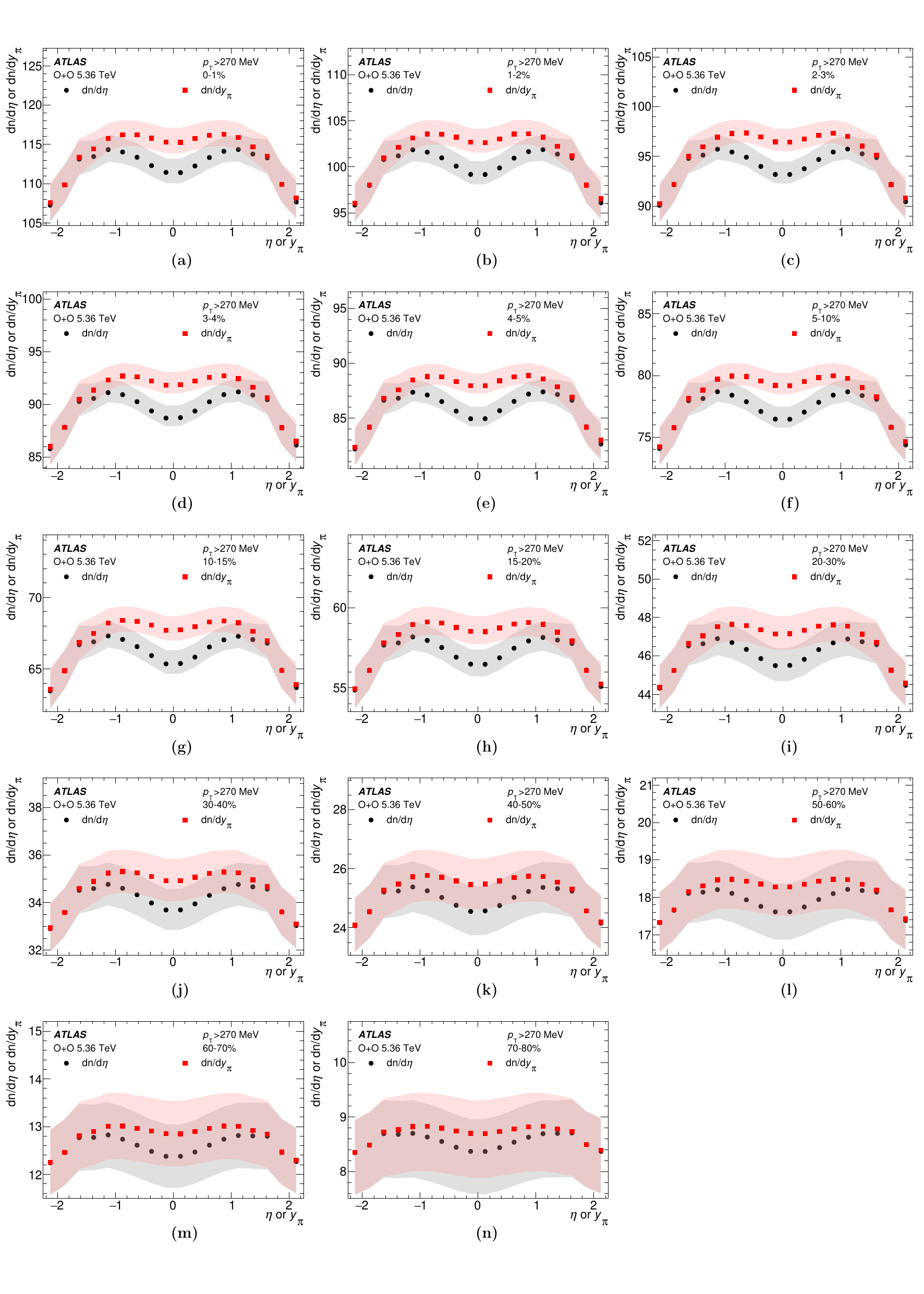}
\caption{
Comparisons of the \dndeta\ and \dndypi\ distributions in \OO\
collisions for several centrality intervals.
The vertical lines and shaded bands indicate statistical and
systematic uncertainties, respectively.
Statistical uncertainties in all cases are smaller than the size
of the data points.
The \dndypi\ data points are slightly shifted along the $x$-axis for clarity.
}
\label{fig:results_CompareYandEta_OO_dNdeta_aux}
\end{figure}

\begin{figure}[h]
\centering
\includegraphics[width=0.90\linewidth]{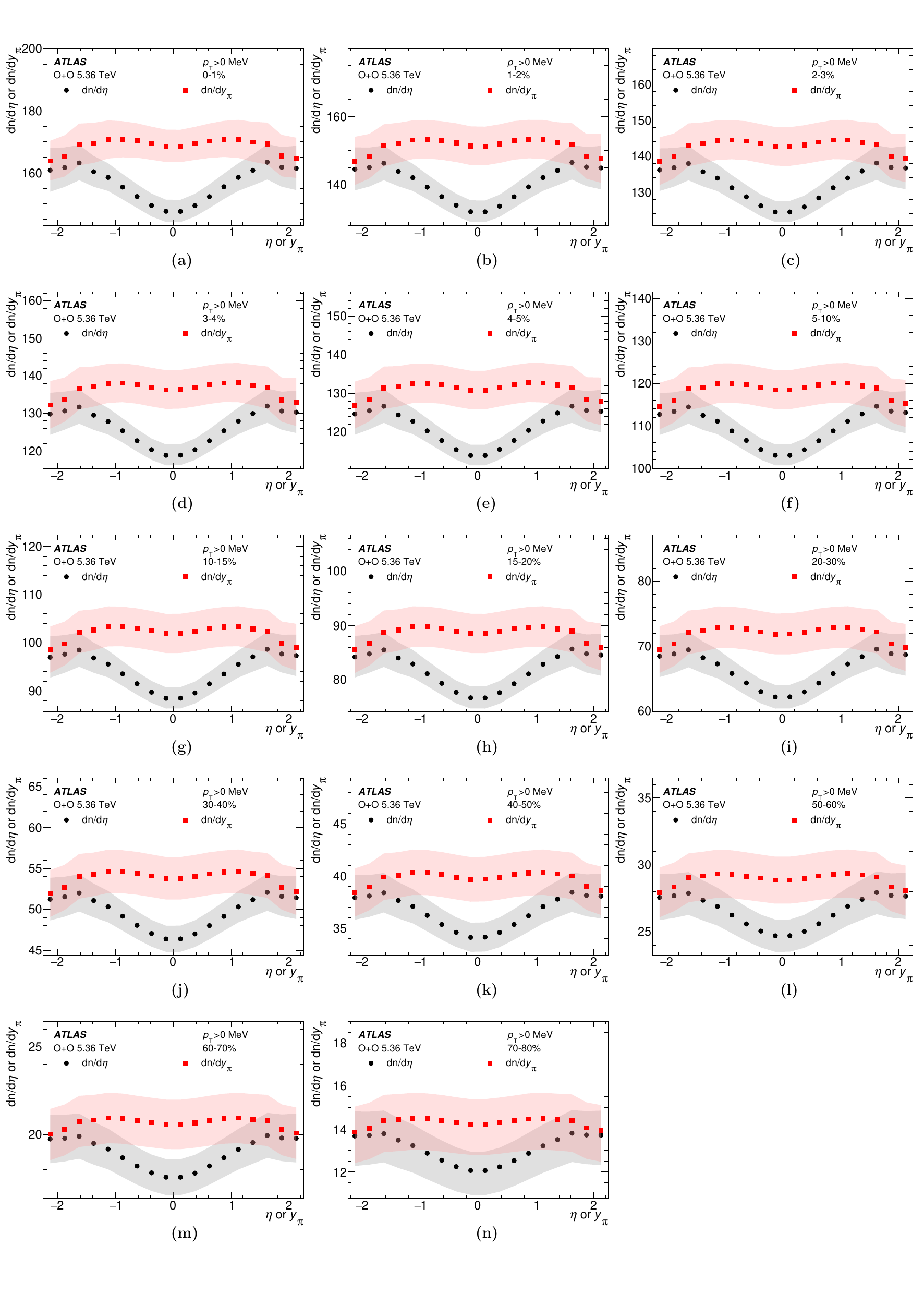}
\caption{
Same as Figure~\ref{fig:results_CompareYandEta_OO_dNdeta_aux} but for \pteqo.
}
\label{fig:results_CompareYandEta_OO_dNdeta_ptExtrap0_aux}
\end{figure}

\begin{figure}[h]
\centering
\includegraphics[width=0.90\linewidth]{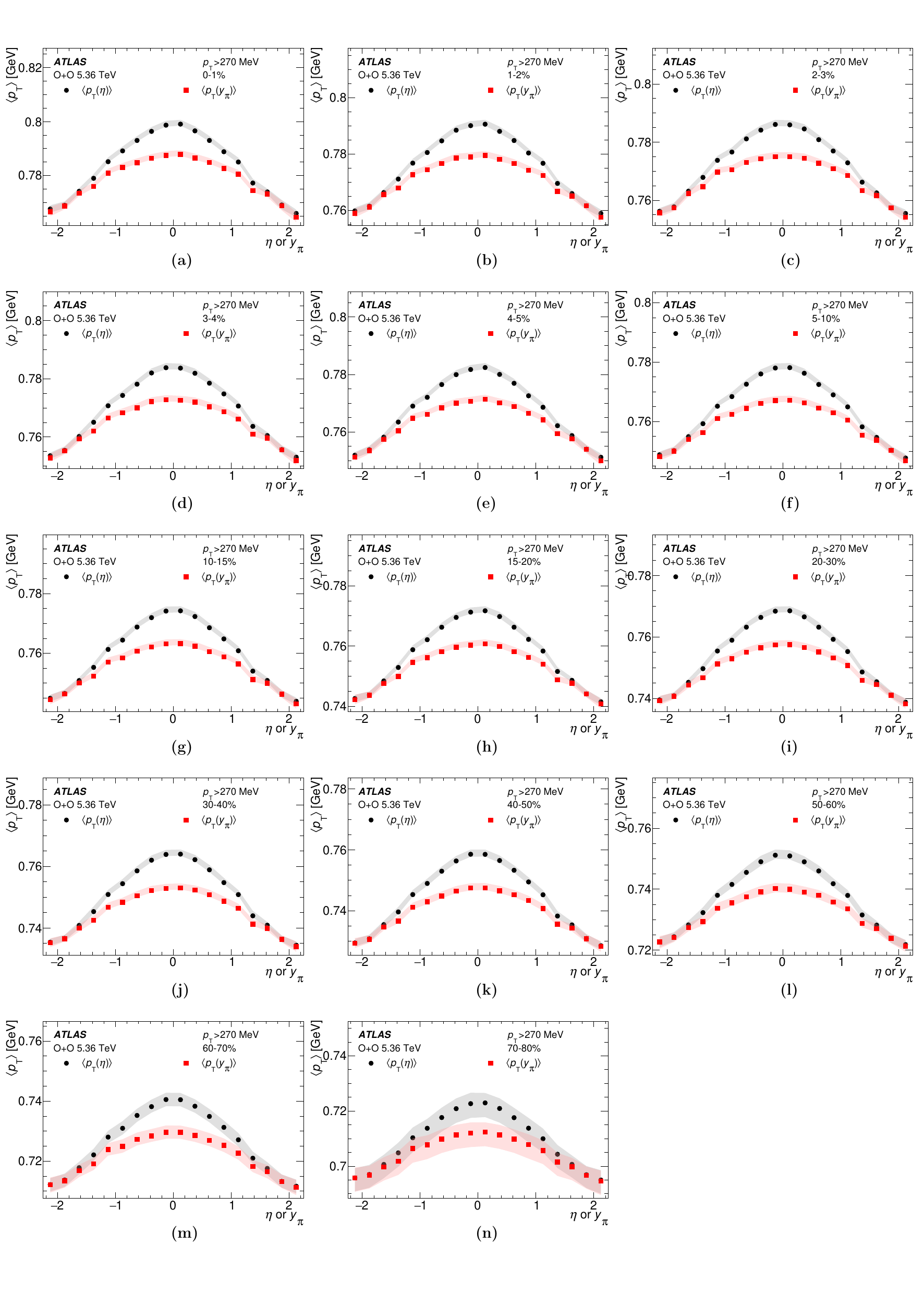}
\caption{
Comparisons of \meanpt\ vs.\ $\eta$ and \meanpt\  vs.\ $y_{\pi}$ in \OO\
collisions for several centrality intervals.
The vertical lines and shaded bands indicate statistical and
systematic uncertainties, respectively.
Statistical uncertainties in all cases are smaller than the size
of the data points.
The \dndypi\ data points are slightly shifted along the $x$-axis for clarity.
}
\label{fig:results_CompareYandEta_OO_meanpT_aux}
\end{figure}

\begin{figure}[h]
\centering
\includegraphics[width=0.99\linewidth]{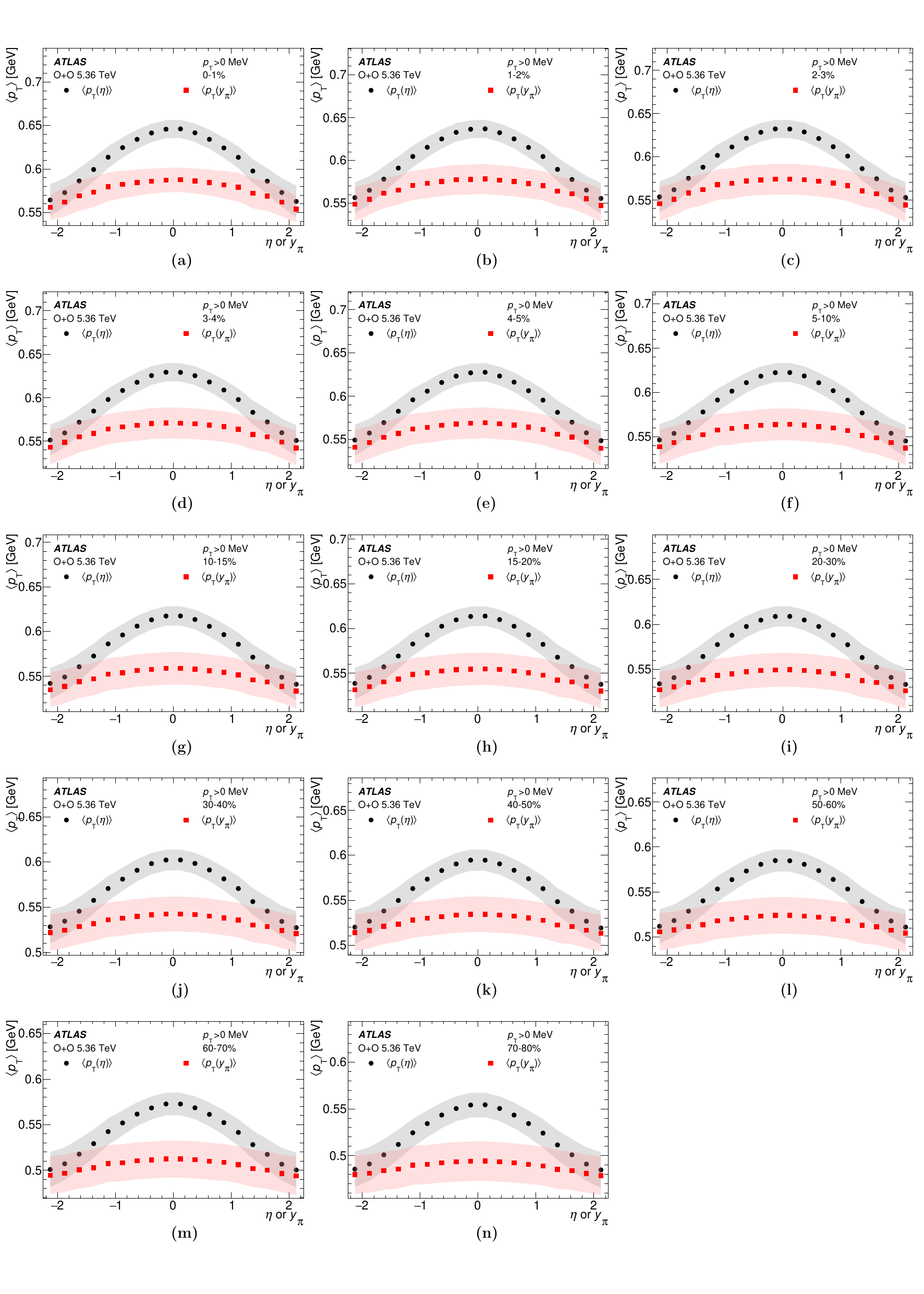}
\caption{
Same as Figure~\ref{fig:results_CompareYandEta_OO_meanpT_aux} but for \pteqo.
}
\label{fig:results_CompareYandEta_OO_meanpT_ptExtrap0_aux}
\end{figure}

\begin{figure}[h]
\centering
\includegraphics[width=0.90\linewidth]{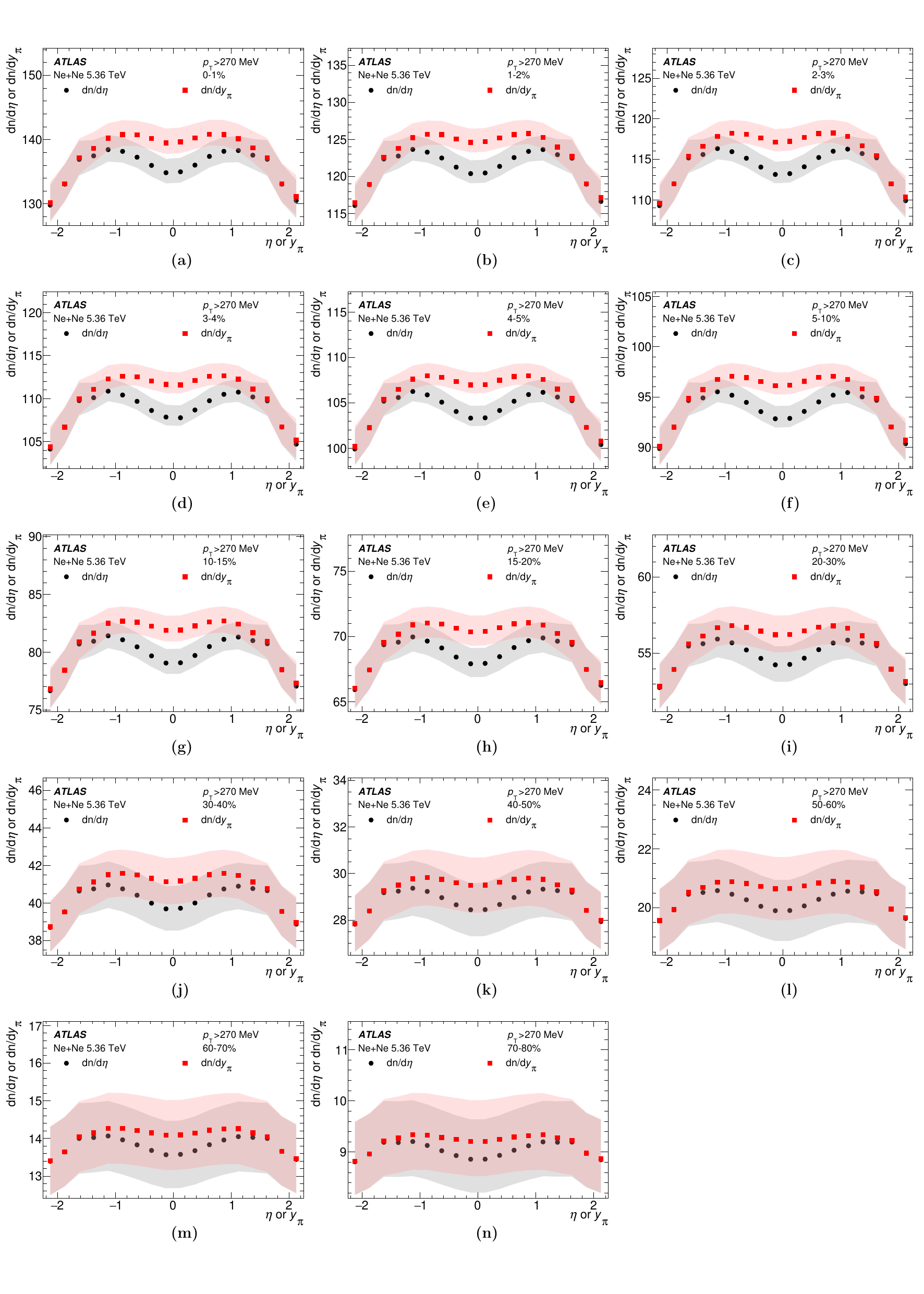}
\caption{
Same as Figure~\ref{fig:results_CompareYandEta_OO_dNdeta_aux} but for \NeNe\ collisions.
}
\label{fig:results_CompareYandEta_NeNe_dNdeta_aux}
\end{figure}

\begin{figure}[h]
\centering
\includegraphics[width=0.90\linewidth]{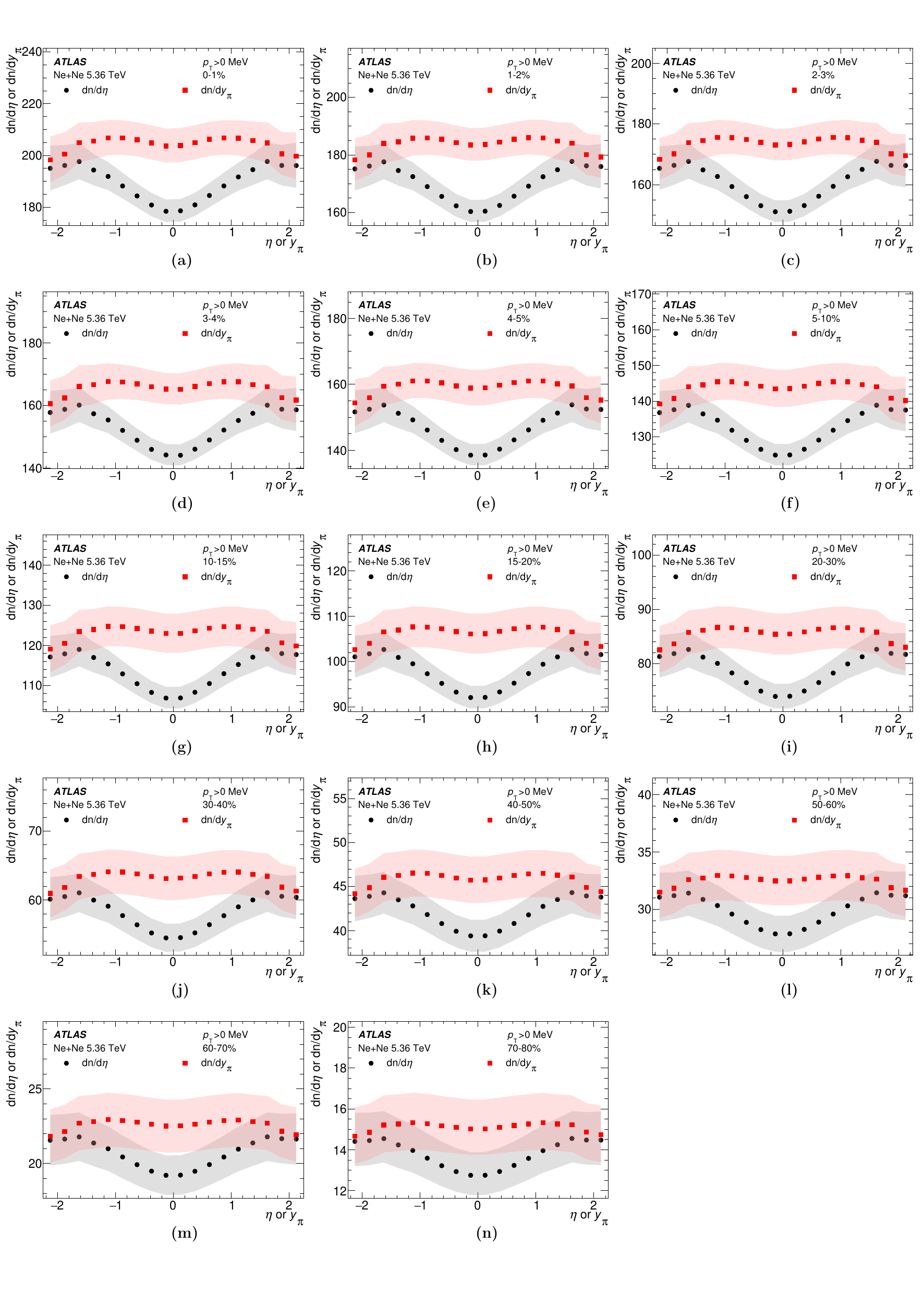}
\caption{
Same as Figure~\ref{fig:results_CompareYandEta_OO_dNdeta_aux} but for \pteqo\ and \NeNe\ collisions.
}
\label{fig:results_CompareYandEta_NeNe_dNdeta_ptExtrap0_aux}
\end{figure}

\begin{figure}[h]
\centering
\includegraphics[width=0.90\linewidth]{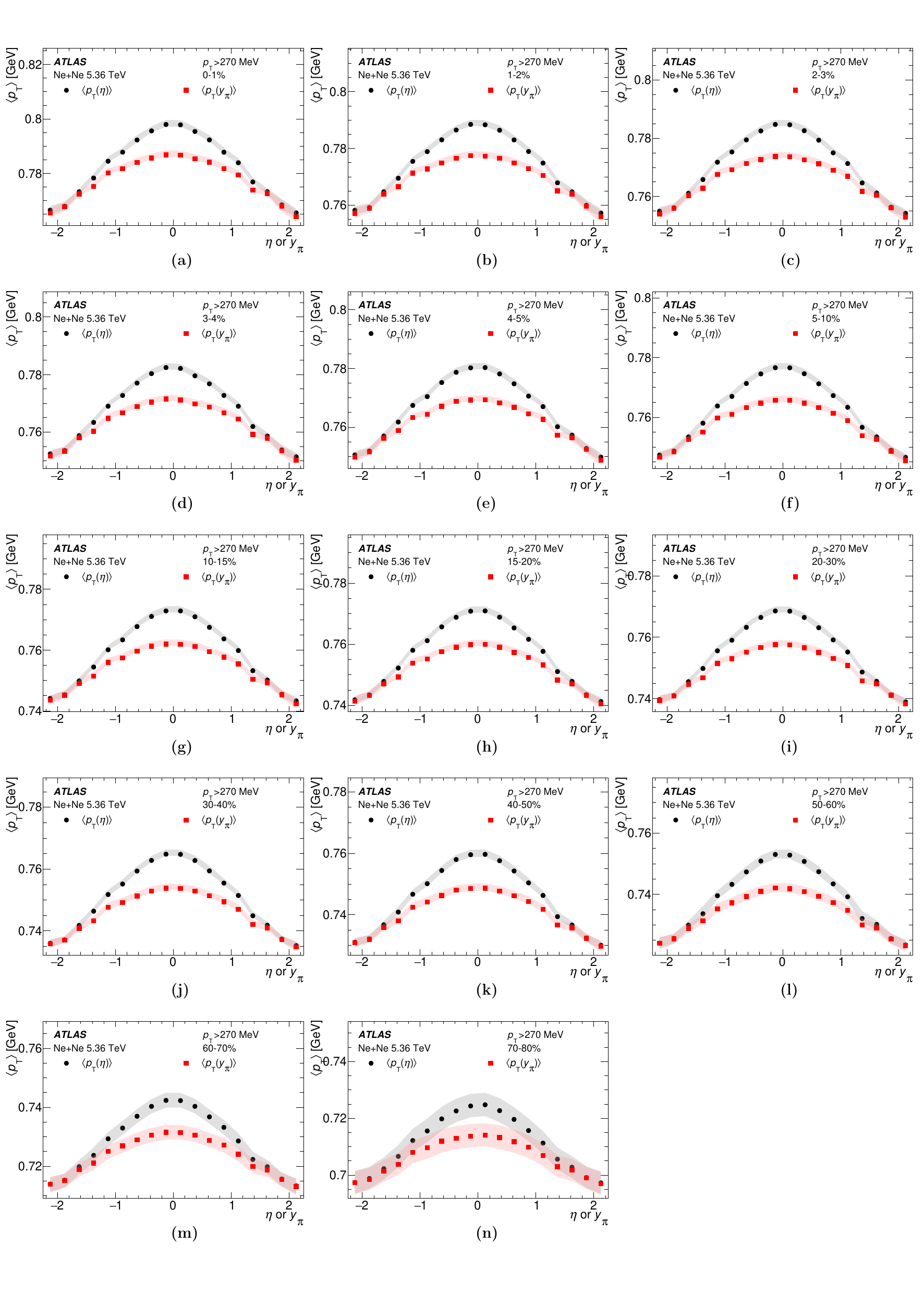}
\caption{
Same as Figure~\ref{fig:results_CompareYandEta_OO_meanpT_aux} but for \NeNe\ collisions.
}
\label{fig:results_CompareYandEta_NeNe_meanpT_aux}
\end{figure}

\begin{figure}[h]
\centering
\includegraphics[width=0.99\linewidth]{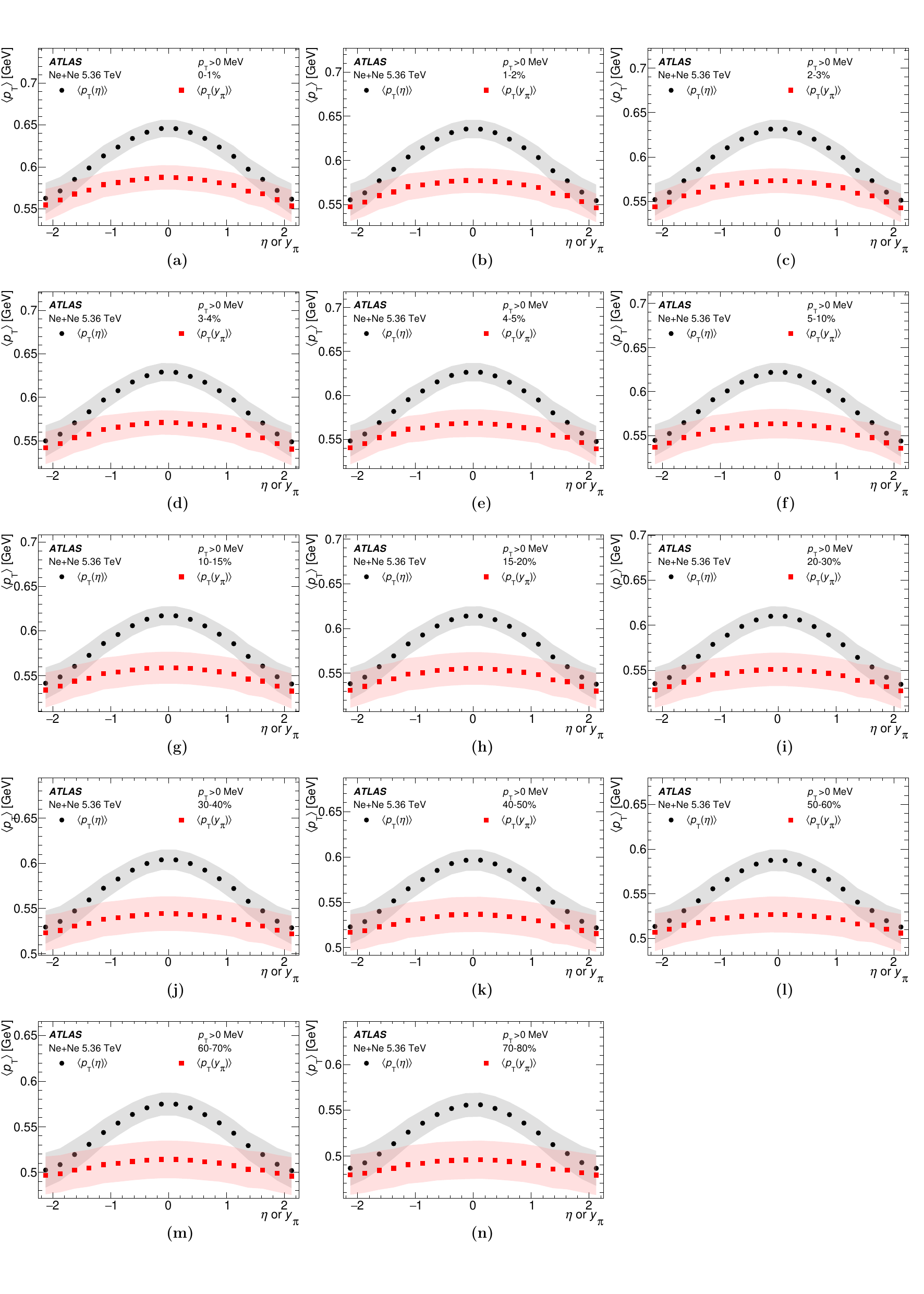}
\caption{
Same as Figure~\ref{fig:results_CompareYandEta_OO_meanpT_aux} but for \pteqo\ and \NeNe\ collisions.
}
\label{fig:results_CompareYandEta_NeNe_meanpT_ptExtrap0_aux}
\end{figure}
\begin{figure}[h]
\centering
\includegraphics[width=0.99\linewidth]{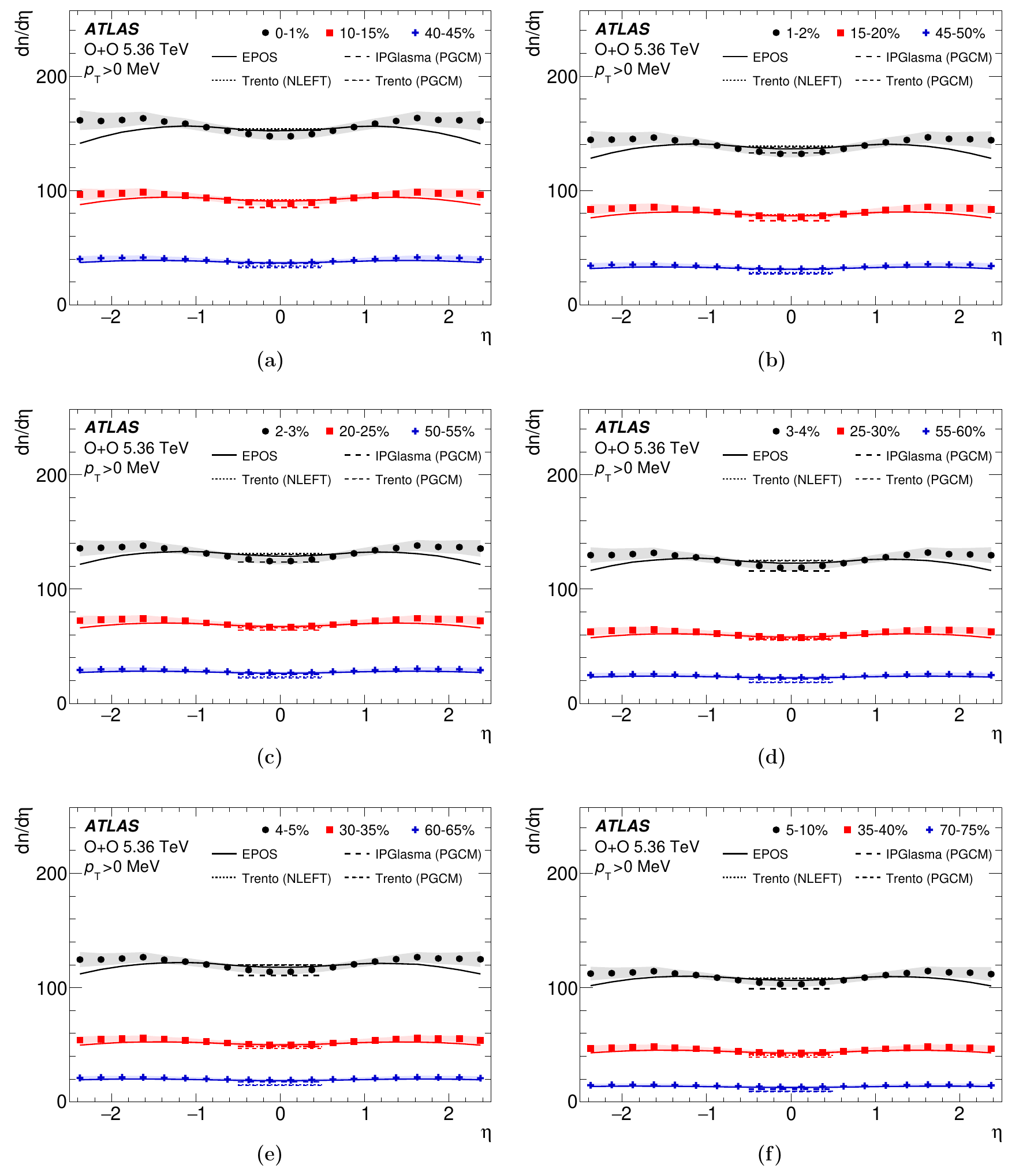}
\caption{
Comparisons between the extrapolated \dndeta measurements and the predictions from
theoretical calculations and the EPOS event generator for 5.36~\TeV\
\OO collisions in various centrality intervals.
}
\label{fig:results_OO_dNdeta_ptExtrap0_Theory_aux}
\end{figure}

\begin{figure}[h]
\centering
\includegraphics[width=0.99\linewidth]{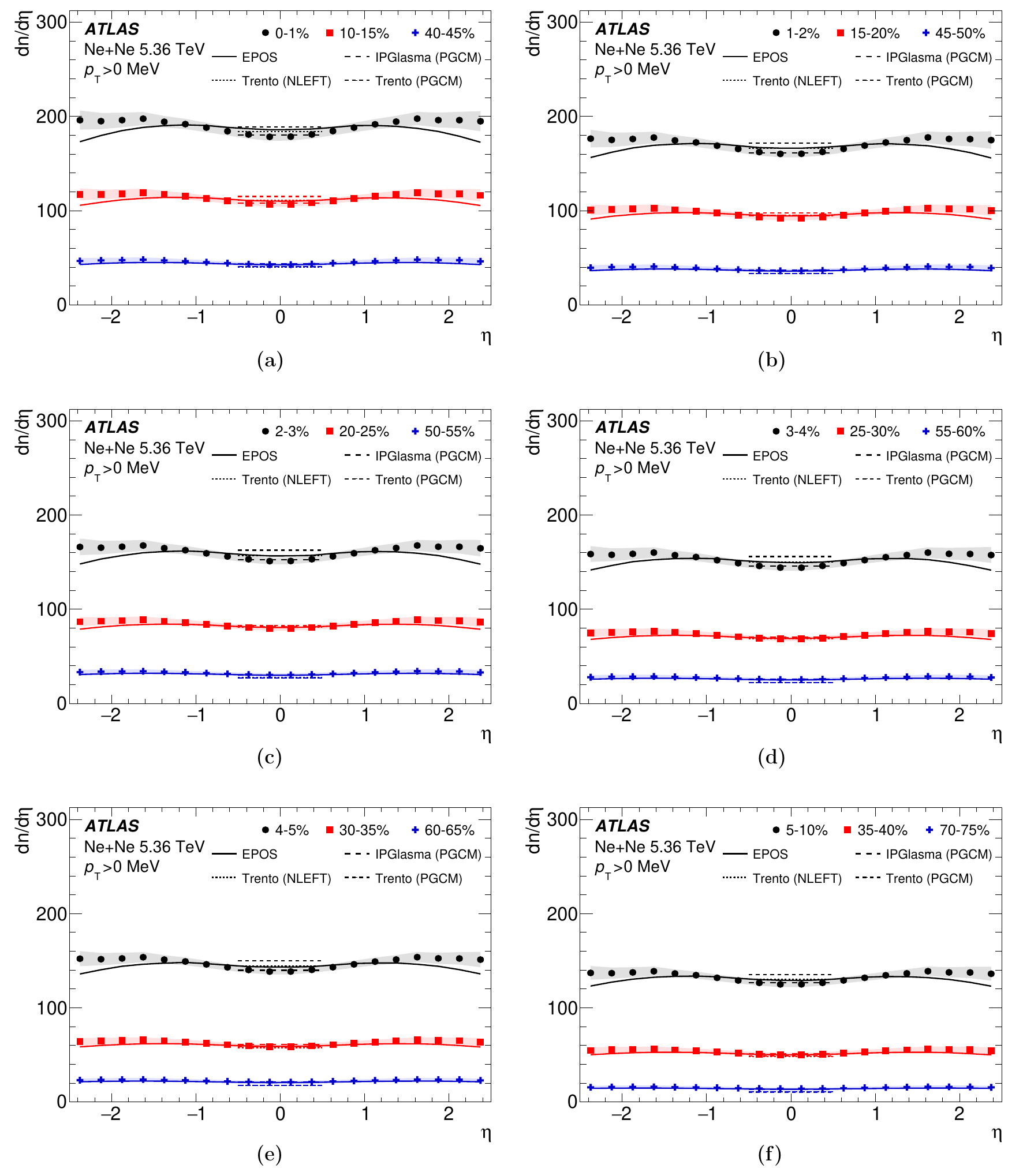}
\caption{
Comparisons between the extrapolated \dndeta measurements and the predictions from
theoretical calculations and the EPOS event generator for 5.36~\TeV\
\NeNe collisions in various centrality intervals.
}
\label{fig:results_NeNe_dNdeta_ptExtrap0_Theory_aux}
\end{figure}

\begin{figure}[h]
\centering
\includegraphics[width=0.99\linewidth]{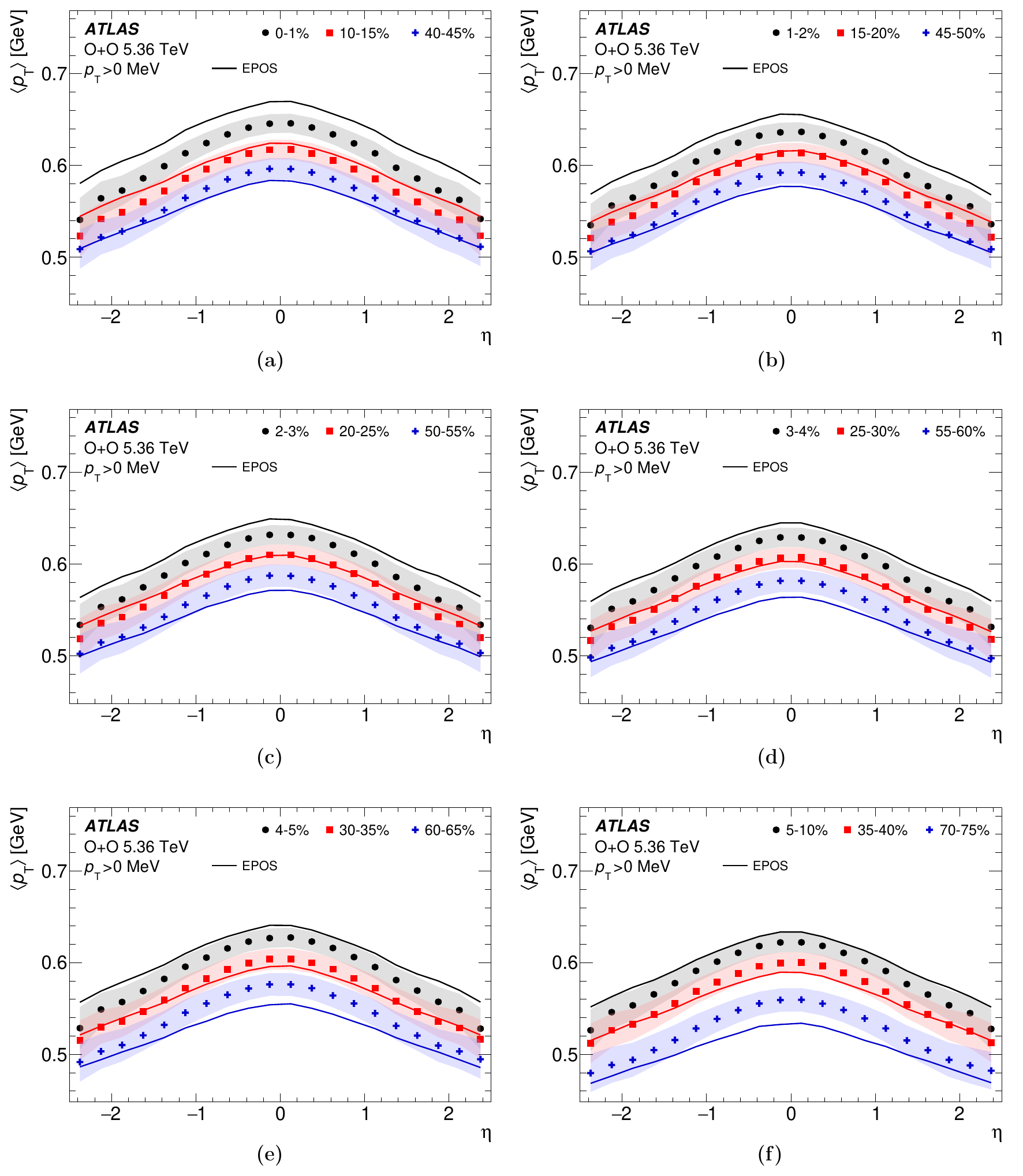}
\caption{
Comparisons between the extrapolated $\meanpt(\eta)$ measurements and the predictions from
the EPOS event generator for 5.36~\TeV\
\OO collisions in various centrality intervals.
}
\label{fig:results_OO_meanpT_ptExtrap0_Theory_aux}
\end{figure}

\begin{figure}[h]
\centering
\includegraphics[width=0.99\linewidth]{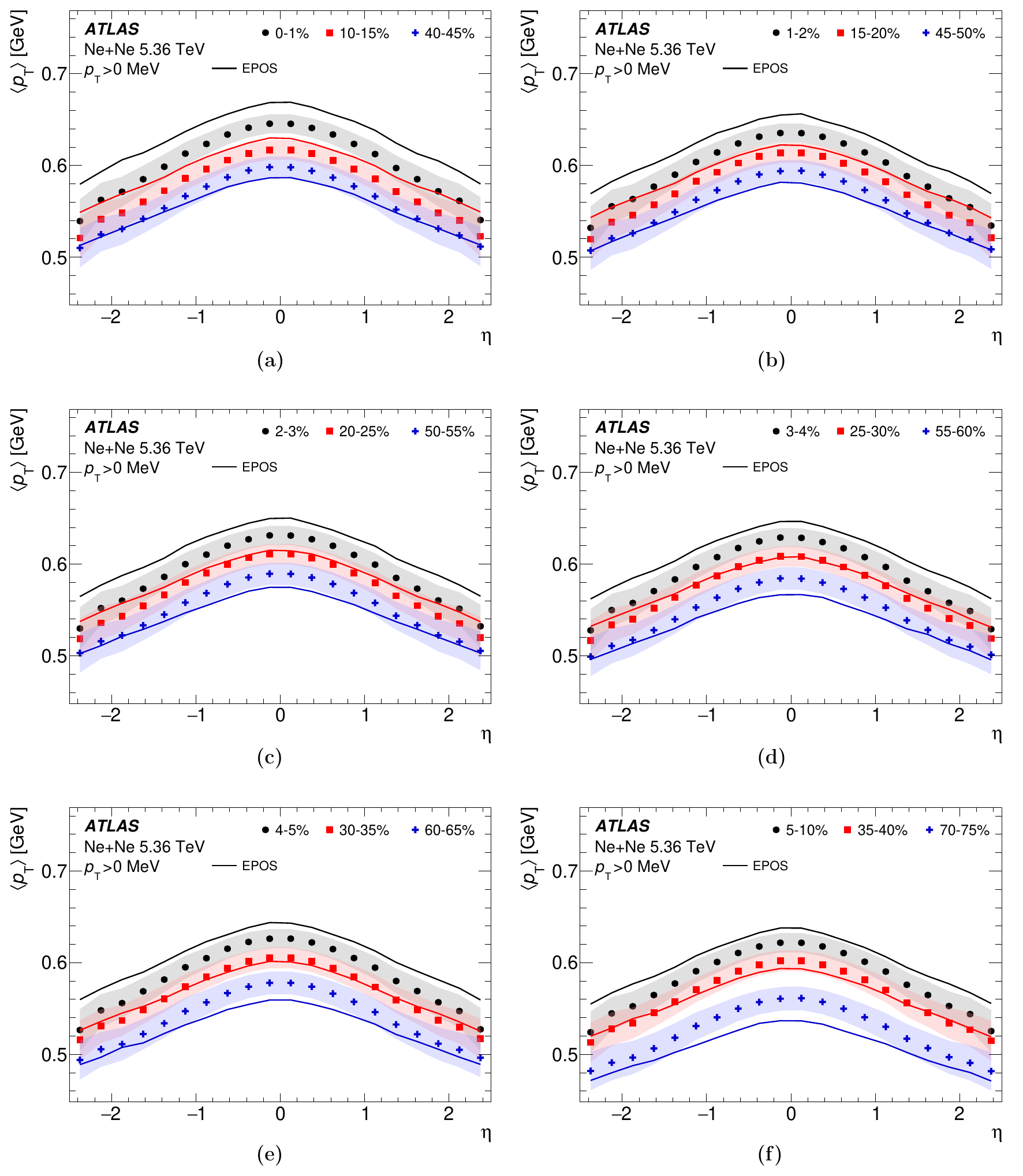}
\caption{
Comparisons between the extrapolated $\meanpt(\eta)$ measurements and the predictions from
the EPOS event generator for 5.36~\TeV\
\NeNe collisions in various centrality intervals.
}
\label{fig:results_NeNe_meanpT_ptExtrap0_Theory_aux}
\end{figure}
\begin{figure}[!h]
\centering
\includegraphics[width=0.99\linewidth]{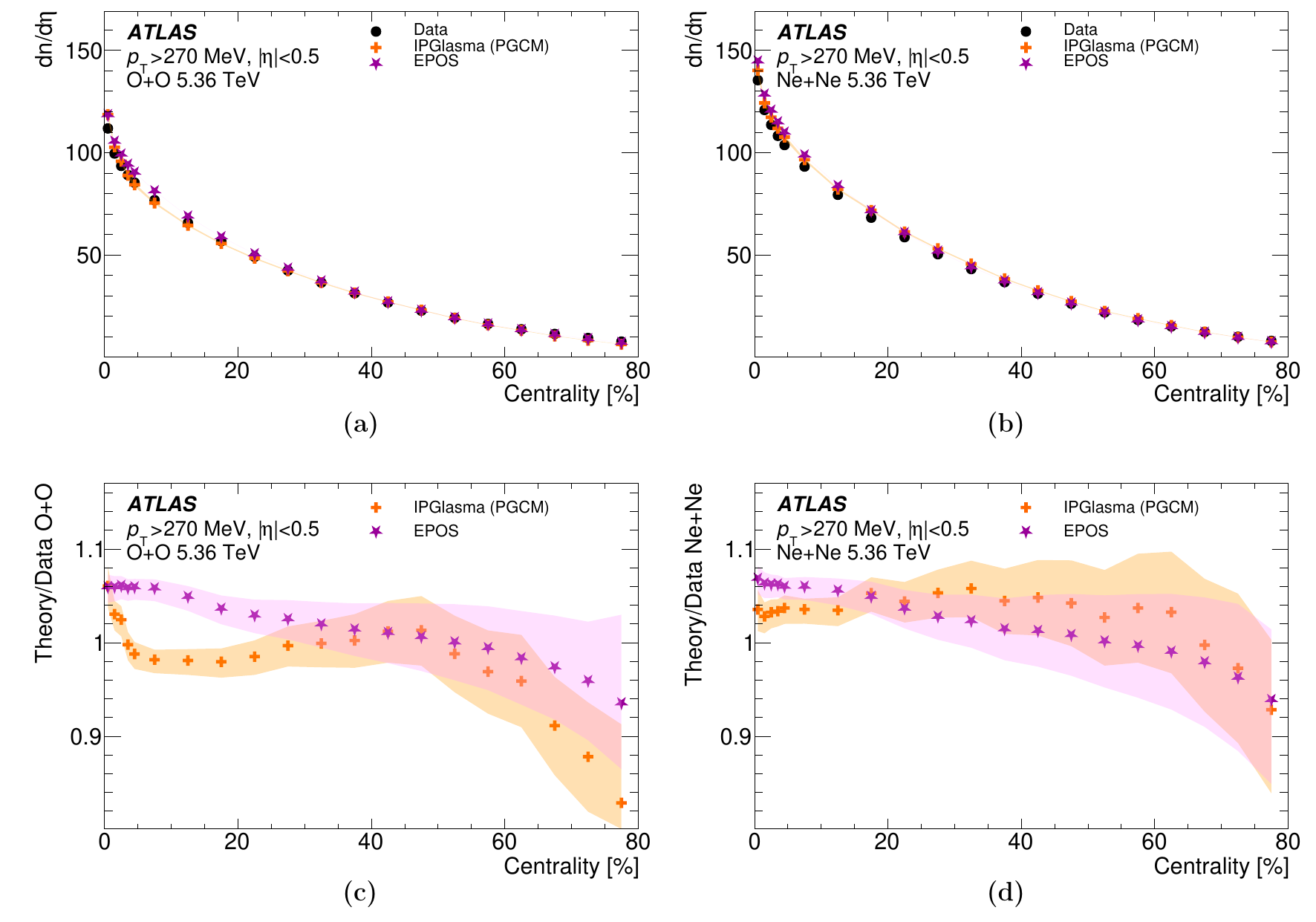}
\caption{
The centrality dependence of the \dndeta\ measured at mid-rapidity
($|\eta|<0.5$) in (a) \OO\ and (b) \NeNe\ collisions for
the $\pt>270~\MeV$ fiducial range.
Comparisons with EPOS and calculations from models described in
Ref.~\cite{Mantysaari:2025tcg} (labeled ``IPGlasma'') are also shown.
Theory-to-data ratios are shown for (c) \OO\ and (d) \NeNe\ collisions.
For the data in panels (a, b), the vertical lines and vertical bars
indicate statistical and systematic uncertainties, respectively.
Statistical uncertainties in all cases are smaller than the size
of the data points.
For the theory points shown in panels (a, b),
and the theory-to-data ratios shown in (c, d),
the shaded bands indicate combined (statistical + systematic) uncertainties.
}
\label{fig:results_dndetavsCent_aux}
\end{figure}

\begin{figure}[!h]
\centering
\includegraphics[width=0.99\linewidth]{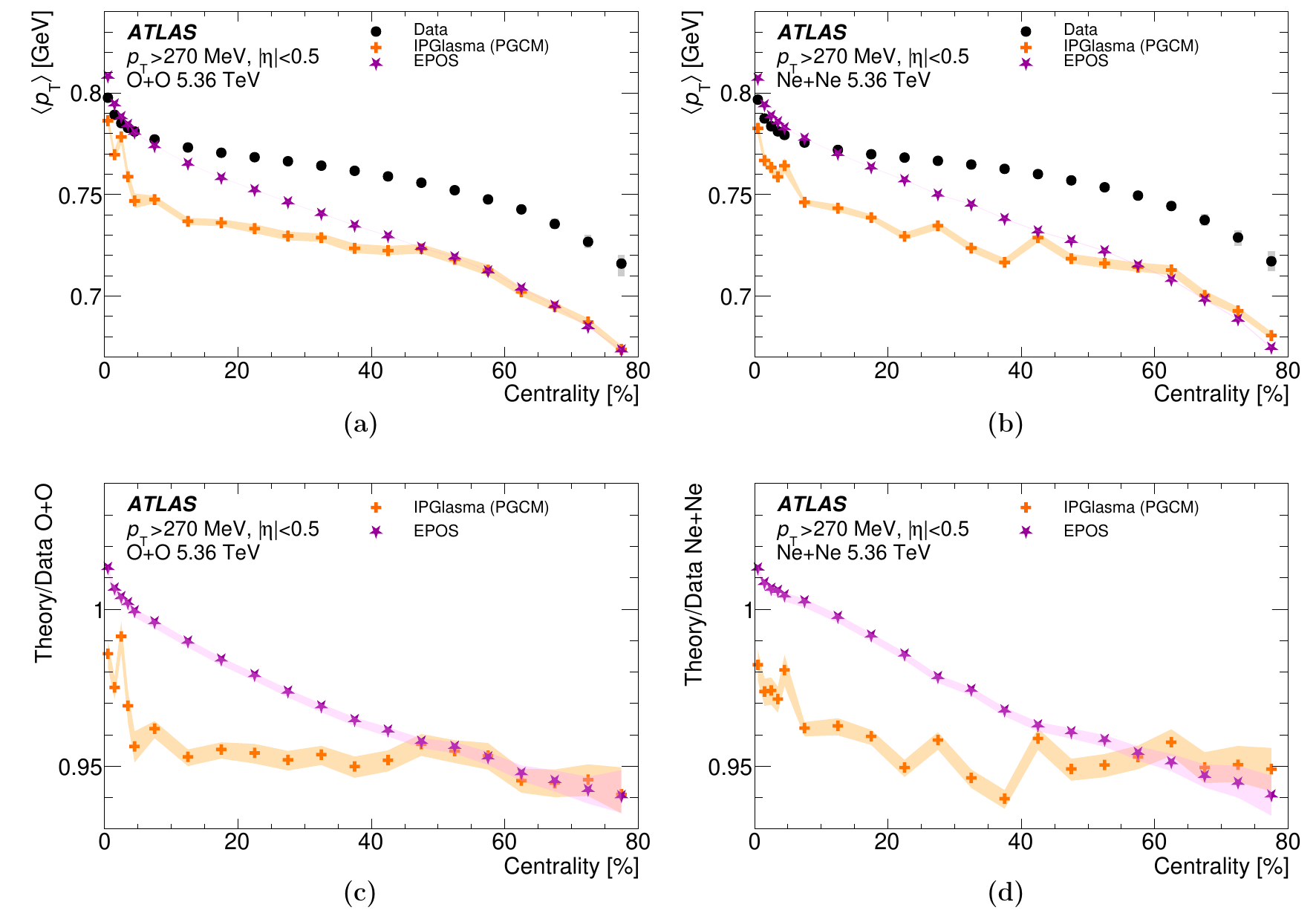}
\caption{
Same as Figure~\ref{fig:results_dndetavsCent_aux},
but for measurements of the \meanpt.
}
\label{fig:results_meanpTvsCent_aux}
\end{figure}

\begin{figure}[!tb]
\centering
\includegraphics[width=0.99\linewidth]{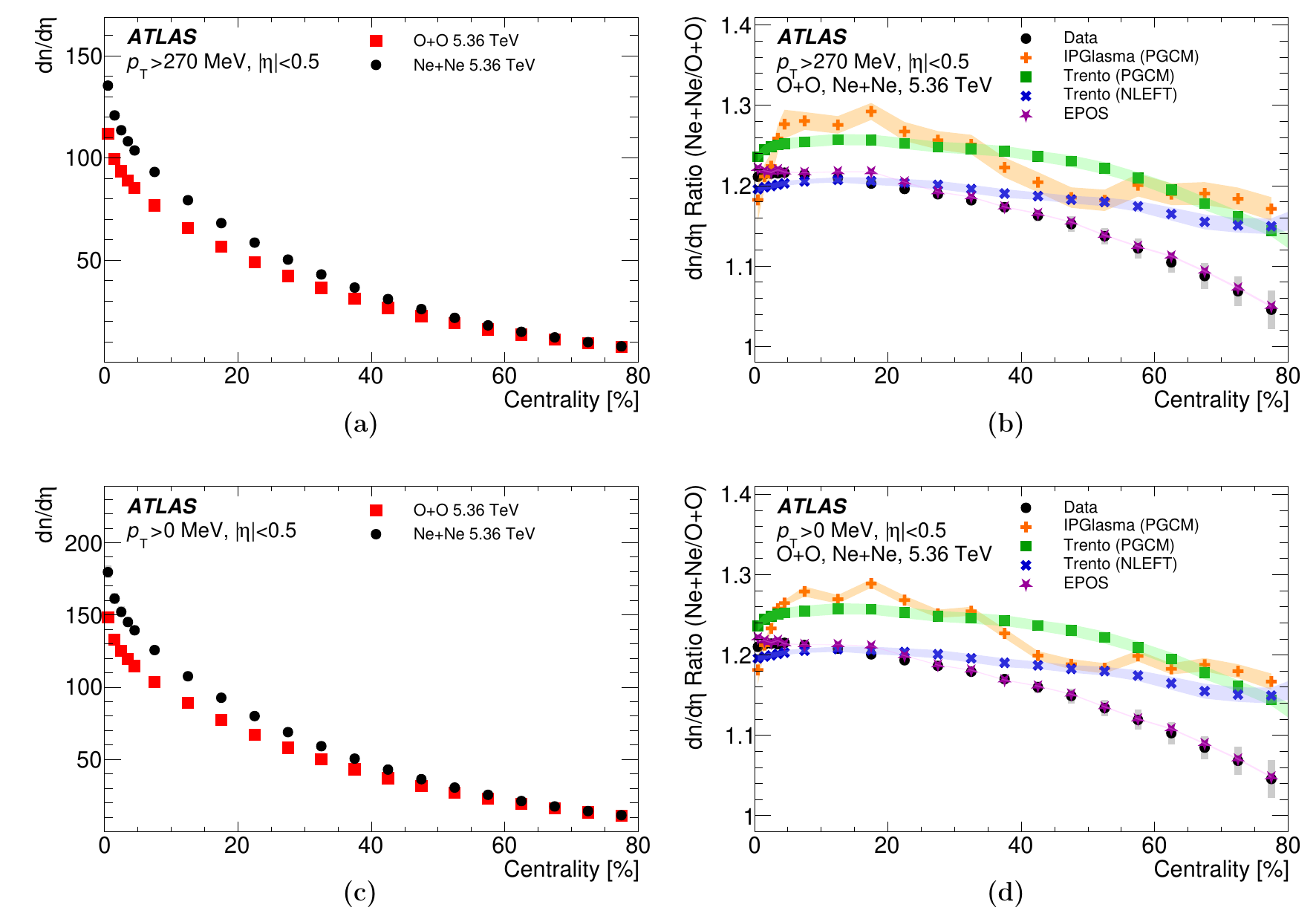}
\caption{
Comparison of the centrality dependence of \dndeta\ in \OO\ and \NeNe\
collisions for the (a) $\pt>270~\MeV$ fiducial region and the (c) \pteqo\ extrapolation.
Panels (b, d) show the centrality dependence of the \NeNe/\OO ratio for \dndeta.
Results are for measurements at mid-rapidity ($|\eta|<0.5$).
Panels (b, d) also show comparisons with EPOS and calculations from models described in
Ref.~\cite{Mantysaari:2025tcg} (labeled ``IPGlasma'')
and Ref.~\cite{Giacalone:2024luz} (labeled ``Trento'').
For the data, the vertical lines and vertical bars
indicate statistical and systematic uncertainties, respectively.
Statistical uncertainties in all cases are smaller than the size
of the data points.
For the theory points, the error bars indicate combined
(statistical+systematic) uncertainties.
The Trento model calculations are always for the \pteqo\ case,
and are identical in panels (b, d).
Panel (d) is identical to Figure~\ref{fig:results_ratio_ptExtrap0}a.
}
\label{fig:results_dndeta_ratio_aux}
\end{figure}

\begin{figure}[!tb]
\centering
\includegraphics[width=0.99\linewidth]{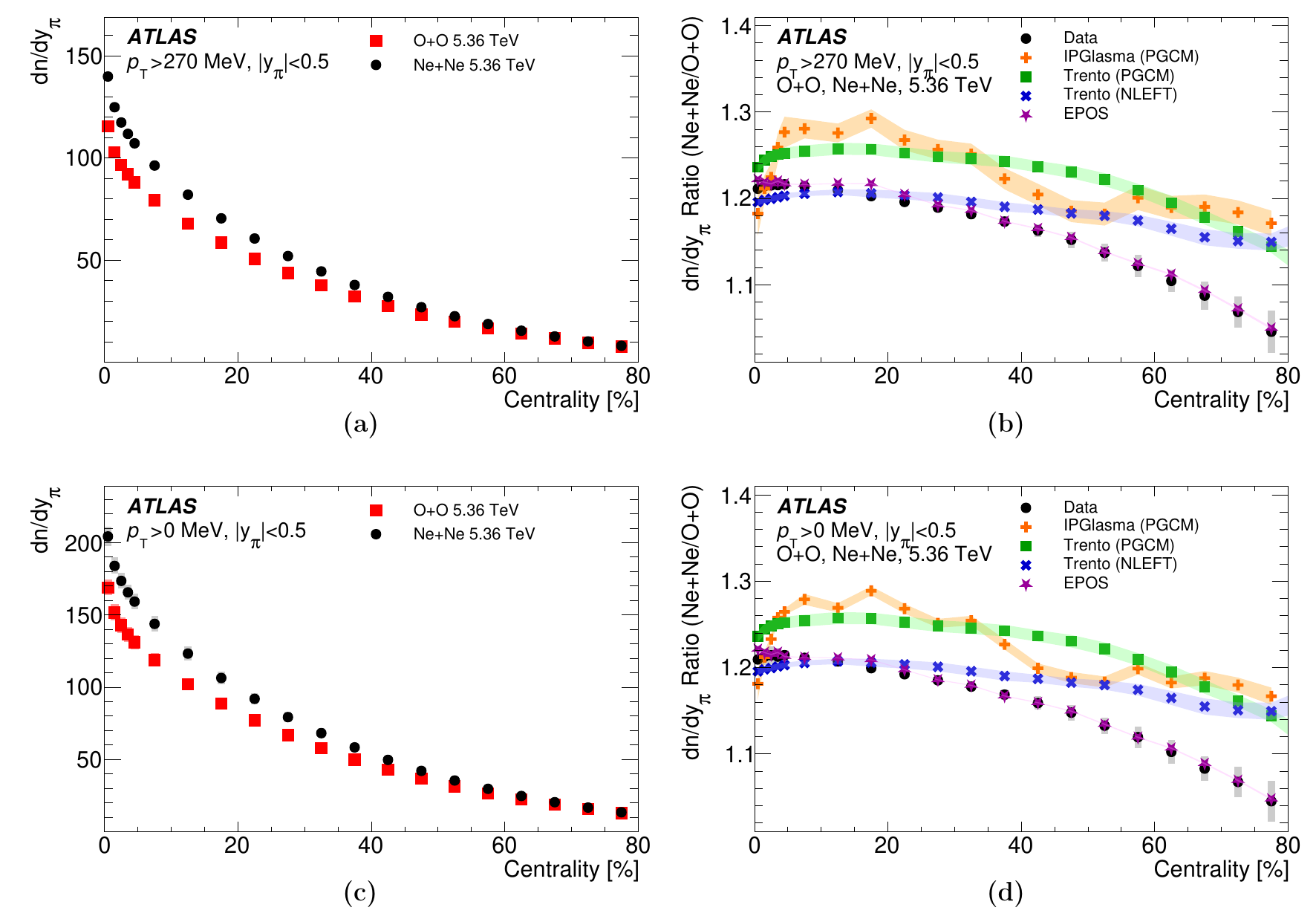}
\caption{
Same as Figure~\ref{fig:results_dndeta_ratio_aux}, but for $|\ypi|<0.5$,
(instead of $|\eta|<0.5$).
However, the theory calculations are for $|\eta|<0.5$, and in the case of the
Trento model, always for the \pteqo\ case.
Panel (d) is identical to Figure~\ref{fig:results_ratio_ptExtrap0}c.
}
\label{fig:results_dndypi_ratio_aux}
\end{figure}

\begin{figure}[!tb]
\centering
\includegraphics[width=0.99\linewidth]{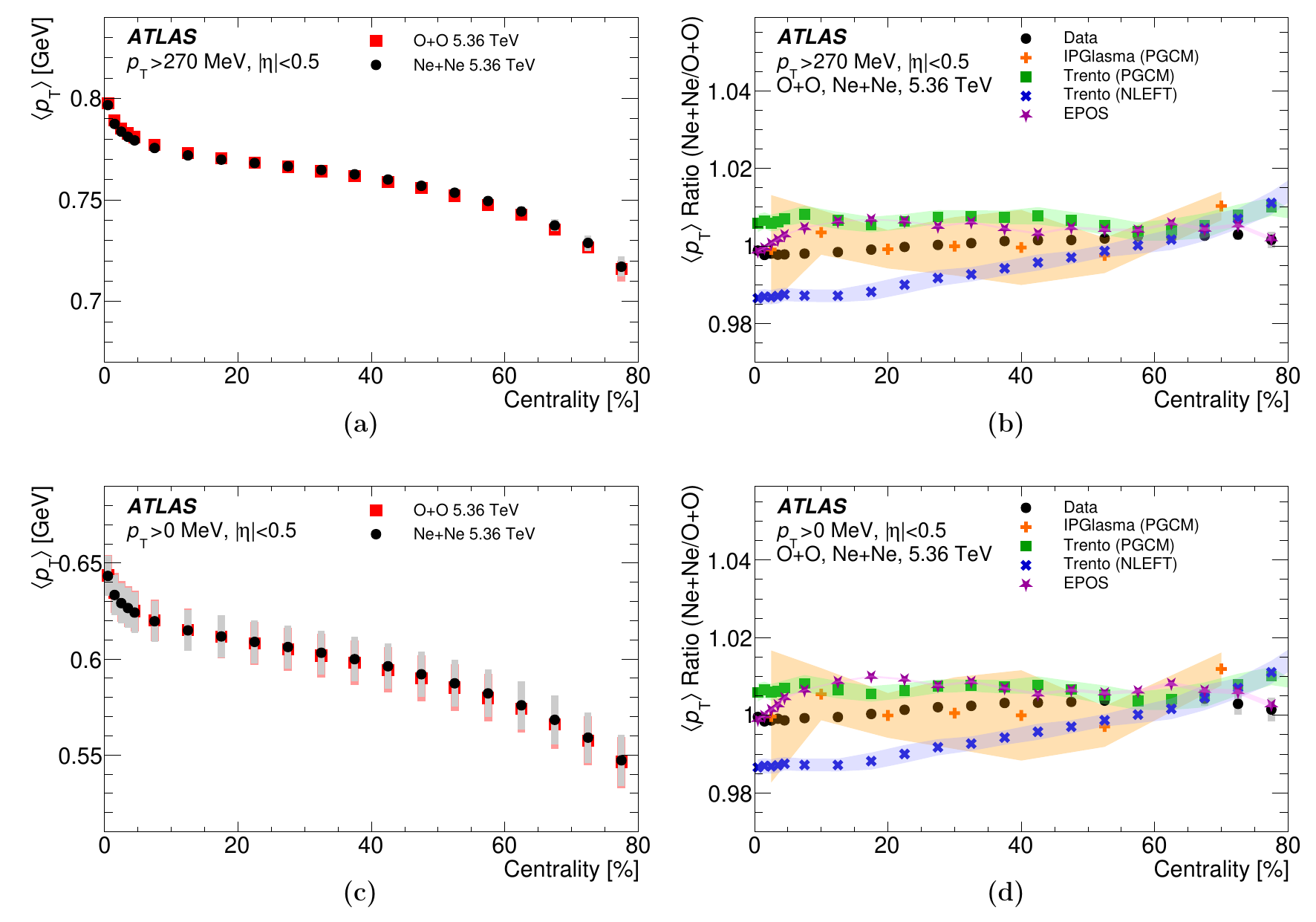}
\caption{
Comparison of the centrality dependence of \meanpt\ in \OO\ and \NeNe\
collisions for the (a) $\pt>270~\MeV$ fiducial region and the (c) \pteqo\ extrapolation.
Panels (b, d) show the centrality dependence of the \NeNe/\OO ratio for \dndeta.
Results are for measurements at mid-rapidity ($|\eta|<0.5$).
Panels (b, d) also show comparisons with EPOS and calculations from models described in
Ref.~\cite{Mantysaari:2025tcg} (labeled ``IPGlasma'')
and Ref.~\cite{Giacalone:2024luz} (labeled ``Trento'').
For the data, the vertical lines and vertical bars
indicate statistical and systematic uncertainties, respectively.
Statistical uncertainties in all cases are smaller than the size
of the data points.
For the theory points, the error bars indicate combined
(statistical + systematic) uncertainties.
The Trento model calculations are always for the \pteqo\ case,
and are identical in panels (b, d).
Panel (d) is identical to Figure~\ref{fig:results_ratio_ptExtrap0}b.
}
\label{fig:results_meanpT_ratio_aux}
\end{figure}

\begin{figure}[!tb]
\centering
\includegraphics[width=0.99\linewidth]{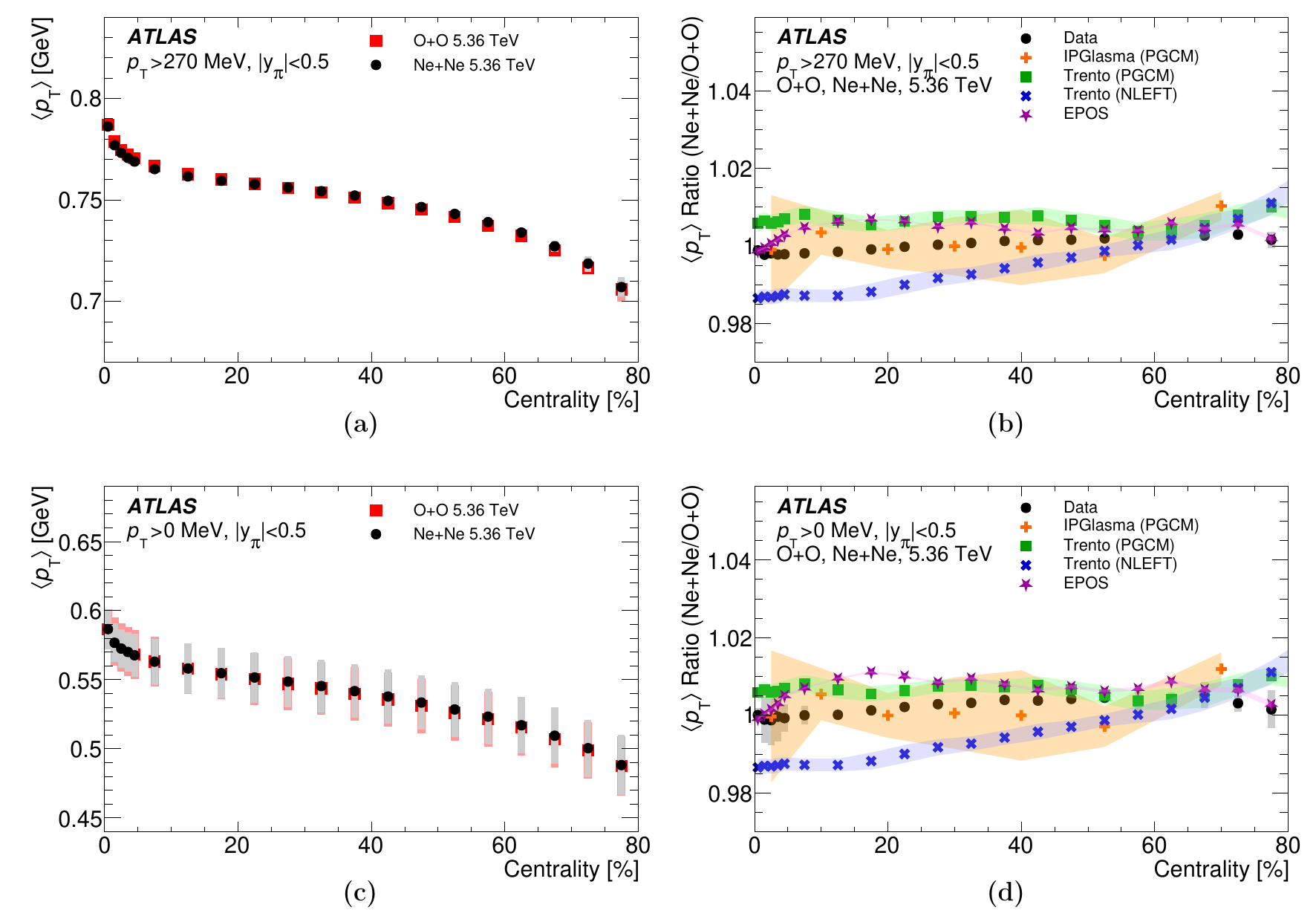}
\caption{
Same as Figure~\ref{fig:results_meanpT_ratio_aux}, but for $|\ypi|<0.5$,
(instead of $|\eta|<0.5$).
However, the theory calculations are for $|\eta|<0.5$, and in the case of the
Trento model, always for the \pteqo\ case.
Panel (d) is identical to Figure~\ref{fig:results_ratio_ptExtrap0}d.
}
\label{fig:results_meanpTvsypi_ratio_aux}
\end{figure}
%


%

%
%
\clearpage\FloatBarrier
\printbibliography
\clearpage
 
\begin{flushleft}
\hypersetup{urlcolor=black}
{\Large The ATLAS Collaboration}

\bigskip

\AtlasOrcid[0000-0002-6665-4934]{G.~Aad}$^\textrm{\scriptsize 102}$,
\AtlasOrcid[0000-0001-7616-1554]{E.~Aakvaag}$^\textrm{\scriptsize 17}$,
\AtlasOrcid[0000-0002-5888-2734]{B.~Abbott}$^\textrm{\scriptsize 121}$,
\AtlasOrcid[0000-0002-0287-5869]{S.~Abdelhameed}$^\textrm{\scriptsize 83b}$,
\AtlasOrcid[0000-0002-1002-1652]{K.~Abeling}$^\textrm{\scriptsize 54}$,
\AtlasOrcid[0000-0001-5763-2760]{N.J.~Abicht}$^\textrm{\scriptsize 48}$,
\AtlasOrcid[0000-0002-8496-9294]{S.H.~Abidi}$^\textrm{\scriptsize 30}$,
\AtlasOrcid[0009-0003-6578-220X]{M.~Aboelela}$^\textrm{\scriptsize 44}$,
\AtlasOrcid[0000-0002-9987-2292]{A.~Aboulhorma}$^\textrm{\scriptsize 36e}$,
\AtlasOrcid[0000-0001-5329-6640]{H.~Abramowicz}$^\textrm{\scriptsize 154}$
% ...
\bigskip
\\

% ...
$^{17}$Department for Physics and Technology, University of Bergen, Bergen; Norway.\\
% ...
$^{30}$Physics Department, Brookhaven National Laboratory, Upton NY; United States of America.\\
% ...
$^{36}$$^{(a)}$Facult\'e des Sciences Ain Chock, Universit\'e Hassan II de Casablanca;$^{(b)}$Facult\'{e} des Sciences, Universit\'{e} Ibn-Tofail, K\'{e}nitra;$^{(c)}$Facult\'e des Sciences Semlalia, Universit\'e Cadi Ayyad, LPHEA-Marrakech;$^{(d)}$LPMR, Facult\'e des Sciences, Universit\'e Mohamed Premier, Oujda;$^{(e)}$Facult\'e des sciences, Universit\'e Mohammed V, Rabat;$^{(f)}$Institute of Applied Physics, Mohammed VI Polytechnic University, Ben Guerir; Morocco.\\
% ...
$^{44}$Physics Department, Southern Methodist University, Dallas TX; United States of America.\\
% ...
$^{48}$Fakult\"{a}t Physik, Technische Universit{\"a}t Dortmund, Dortmund; Germany.\\
% ...
$^{54}$II. Physikalisches Institut, Georg-August-Universit\"{a}t G\"ottingen, G\"ottingen; Germany.\\
% ...
$^{83}$$^{(a)}$Khalifa University of Science and Technology, Abu Dhabi;$^{(b)}$New York University Abu Dhabi, Abu Dhabi;$^{(c)}$United Arab Emirates University, Al Ain;$^{(d)}$University of Sharjah, Sharjah; United Arab Emirates.\\
% ...
$^{102}$CPPM, Aix-Marseille Universit\'e, CNRS/IN2P3, Marseille; France.\\
% ...
$^{121}$Homer L. Dodge Department of Physics and Astronomy, University of Oklahoma, Norman OK; United States of America.\\
% ...
$^{154}$Raymond and Beverly Sackler School of Physics and Astronomy, Tel Aviv University, Tel Aviv; Israel.\\
% ...
$^{a}$ Also at Affiliated with an institute formerly covered by a cooperation agreement with CERN.\\
$^{b}$ Also at An-Najah National University, Nablus; Palestine.\\
$^{c}$ Also at Borough of Manhattan Community College, City University of New York, New York NY; United States of America.\\
$^{d}$ Also at Center for Interdisciplinary Research and Innovation (CIRI-AUTH), Thessaloniki; Greece.\\
$^{e}$ Also at Centre of Physics of the Universities of Minho and Porto (CF-UM-UP); Portugal.\\
$^{f}$ Also at CERN, Geneva; Switzerland.\\
% ...

\end{flushleft}


%

%
%
%
%

%
%
%
%

\end{document}